\documentclass[a4paper,12pt]{article}

\usepackage[T1]{fontenc} 

\usepackage{draft}
\usepackage{putex} 

\usepackage{cite}

\usepackage[
      colorlinks=true,
      linkcolor=black,
      urlcolor=blue,
      filecolor=black,
      citecolor=red,
      ]{hyperref}
      
\usepackage{graphicx}
\usepackage{xcolor}
\usepackage{amsfonts,amsmath,amssymb,bm}
\usepackage{verbatim}
\usepackage{array}
\usepackage{mathrsfs}
\usepackage{mathtools}
\usepackage{yfonts}
\usepackage{dsfont}
\usepackage{bbm}
\usepackage{colonequals}
\usepackage{amscd}
\usepackage{slashed}

\usepackage{float}
\usepackage{afterpage}

\usepackage{subcaption}

\usepackage{wrapfig}

\usepackage{suffix,mathtools,cancel,bbm}

\usepackage{multirow}

\makeatletter
\def\leftrightarrowsfill@{\arrowfill@\leftrarrows\Rrelbar\lrightarrows}
\newcommand{\xleftrightarrows}[2][]{\ext@arrow 3399\leftrightarrowsfill@{#1}{#2}}
\makeatother

\begin{document}

\preprint{PUPT-2642}

 \institution{PU}{Joseph Henry Laboratories, Princeton University, Princeton, NJ 08544, USA}

 \institution{IAS}{Institute for Advanced Study, Princeton, NJ 08540, USA}

 \institution{NYU}{Center for Cosmology and Particle Physics, New York University, New York, NY 10003, USA}

\title{ 
Probing Supersymmetric Black Holes \\
with Surface Defects
}

\authors{Yiming Chen,\worksat{\PU} Matthew Heydeman,\worksat{\IAS,\PU} Yifan Wang,\worksat{\NYU} and Mengyang Zhang\worksat{\PU}}

\abstract{

It has long been conjectured that the large $N$ deconfinement phase transition of $\cN=4$ ${\rm SU}(N)$ super-Yang-Mills corresponds via AdS/CFT to the Hawking-Page transition in which black holes dominate the thermal ensemble, and quantitative evidence of this has come through the recent matching of the superconformal index of ${1\over 16}$-BPS states to the supersymmetric black hole entropy. We introduce the half-BPS Gukov-Witten surface defect as a probe of the superconformal index, which also serves as an order parameter for the deconfinement transition. This can be studied directly in field theory as a modification of the usual unitary matrix model or in the dual description as a D3-brane probe in the background of a (complex) supersymmetric black hole. Using a saddle point approximation,
we determine our defect index in the large $N$ limit as a simple function of the chemical potentials and show independently that it is reproduced by the renormalized action of the brane in the black hole background. Along the way, we also comment on the Cardy limit and the thermodynamics of the D3-brane in the generalized ensemble. The defect index sharply distinguishes between the confining and the deconfining phases of the gauge theory and thus is a supersymmetric non-perturbative order parameter for these large $N$ phase transitions which deserves further investigation. Finally, our work provides an example where the properties of a black hole coupled to an external system can be analyzed precisely.

}
\date{}

\maketitle

\tableofcontents

\pagebreak

\setcounter{page}{1}

\section{Introduction and Summary}
\label{sec:intro}

\subsection{General motivation}
\label{ssec:generalmotivation}
 
Black holes are arguably the most extreme and most mysterious phases of matter known to mankind. First discovered as solutions in the classical theory of General Relativity,  it soon became clear that  black holes have many key features that are intrinsically quantum mechanical and provide stringent tests of quantum gravity. To note a few, 
black holes
 behave as thermodynamic objects with a temperature $T_{\rm BH}={\kappa\over 2\pi}$ (known as Hawking temperature)  proportional to the surface gravity $\kappa$, and emit elementary particles via  Hawking radiation, 
which is a purely quantum mechanical process. Black holes also possess an entropy determined by the  event horizon area $A$ via the famous Bekenstein-Hawking formula $S_{\rm BH}={A\over 4 G_N \hbar}$. The black hole entropy implies an underlying statistical ensemble of microstates, which are necessarily quantum due to the classical no-hair theorem. Finally, the black hole singularity pushes all physical theories to their limits, and its resolution demands a consistent theory of quantum gravity.

The study of black holes have inspired many developments in quantum gravity over the past half century. One revolutionary idea  that was born in this period is the AdS/CFT correspondence \cite{Maldacena:1997re}, which conjectures a non-perturbative definition of quantum gravity on $d+1$-dimensional anti-de-Sitter (AdS$_{d+1}$) space by a dual conformal field theory (CFT) in $d$ spacetime dimensions. In turn, the AdS/CFT framework provides an indispensable platform to study the quantum natures of black holes using field theory tools. One of the best understood AdS/CFT dual pairs is between type IIB string theory on ${\rm AdS}_5\times S^5$ and four-dimensional $\cN=4$ ${\rm SU}(N)$ super-Yang-Mills (SYM) \cite{Maldacena:1997re,Gubser:1998bc,Witten:1998qj}, which we summarize below, including the dictionary between the bulk data (string length $\ell_s$, string coupling $g_s$ and axion $C_0$) and boundary data (gauge group rank $N$, Yang-Mills coupling $g_{\rm YM}$ and theta angle $\theta$),
\ie 
\text{Type IIB string theory on ${\rm AdS}_5\times S^5$}&\xleftrightarrow[\tau={i\over g_s} +C_0 =\,{4\pi i \over g_{\rm {
 YM}}^2}+{\theta\over 2\pi }]{\left(L\over \ell_s\right)^4=\,g_{\rm YM}^2 N }{\cN=4}~\textrm{SU} (N)~\text{super-Yang-Mills} 
 \label{AdSCFT}
\fe
Here in the 't Hooft large $N$ limit with large 't Hooft coupling $\lambda=g_{\rm YM}^2 N$, the bulk string theory admits a semiclassical description by 5d gauged supergravity on ${\rm AdS}_5$, which has a large family of charged, rotating, black hole solutions with a $S^3$ event horizon of finite area and consequently macroscopic entropy \cite{Gutowski:2004ez, Gutowski:2004yv,Chong:2005hr, Chen:2005zj, Kunduri:2006ek, Chong:2005da,Gauntlett:1998fz,Cvetic:2005zi}. Via the AdS/CFT correspondence, they are expected to encode the spectrum of certain highly excited states in the Hilbert space $\cH (S^3)$ of the SYM on $S^3$,  and equivalently heavy
local (point-like) operators by the operator-state correspondence.
Furthermore, the Hawking-Page transition from the thermal AdS to the large black hole has long been conjectured to be the holographic dual of the non-abelian confinement/deconfinement phase transition in the large $N$ SYM theory\cite{Witten:1998qj,Witten:1998zw,Sundborg:1999ue,Aharony:2003sx}. Therefore, it is imperative to investigate this dictionary in detail,  in order to gain insights on both the black holes and non-abelian gauge theory dynamics.

In view of \eqref{AdSCFT}, heavy CFT local operators contain the quantum microstates for the black holes in the AdS. The immediate challenge is to account for the black hole entropy from enumerating these CFT operators. This is difficult due to the strong coupling effects on the SYM side. Nonetheless, one can first focus on black holes that preserve a small fraction of the supersymmetry, and optimistically this symmetry is sufficiently strong to allow a counting of the microstates at weak coupling which can then safely be extrapolated to the strong coupling regime. It turns out that the 5d supergravity hosts a four-parameter family of asymptotically ${\rm AdS}_5$ black hole solutions preserving 2 out of the 32 supercharges, and thus they are often referred to as ${1\over 16}$-BPS black holes \cite{Gutowski:2004ez, Gutowski:2004yv,Chong:2005hr, Chen:2005zj, Kunduri:2006ek, Chong:2005da,Gauntlett:1998fz,Cvetic:2005zi}. By AdS/CFT, properties of these supersymmetric black holes are naturally encoded in supersymmetric  observables in the dual SYM involving a class of ${1\over 16}$-BPS local operators with large degeneracy.
This provides a natural arena to 
investigate the quantum nature of black holes and perform 
nontrivial tests of quantum gravity, using well-established tools in supersymmetric field theories. 

Even after restricting to the supersymmetric operators, the exact degeneracies (as they contribute to a thermal partition function) are difficult to determine. However, there are examples of supersymmetric partition functions which can be reliably computed at weak coupling, because they are protected due to the insertion of a $(-1)^F$ operator (with $F$ the fermion number) which ensures boson/fermion cancellation when supersymmetry is preserved irrespective of the coupling \cite{Witten:1982im,Witten:1982df}. The simplest such quantity appropriate for the counting of BPS operators in conformal field theory is the superconformal index (SCI)\cite{Kinney:2005ej,Romelsberger:2005eg}, which has recently been shown to capture the degeneracies of the dual black holes and exhibit the confinement/deconfinement phase transition as one tunes supersymmetry preserving potentials as we review below.
The main goal of this paper is to  study a further refinement of the superconformal index by inserting an extra supersymmetric surface defect operator. This extended defect is analogous to the usual Polyakov loop \cite{Polyakov:1975rs} and turns out to be a natural supersymmetry preserving order parameter for the deconfinement phase transition in the large $N$ SYM on $S^1\times S^3$.\footnote{On the contrary, the naive Polyakov loop does not preserve the desired supersymmety and thus is not an order parameter for the deconfinement phase transition in the supersymmetric ensemble. See Section~\ref{sec:PLcomment} for more discussions.} Furthermore, this surface defect is described by a probe D3-brane in the holographic dual and captures finer properties of the supersymmetric black hole beyond the bare index.

Before describing our setup that involves the surface defect in more detail,  it is useful to start with a review of the superconformal index itself.  The superconformal index of $\cN=4$ ${\rm SU}(N)$ SYM in the grand canonical ensemble is a kind of graded partition function which counts states (and thus CFT operators) weighted by their charges under the maximal commuting subalgebra of the $\mf{psu}(2,2|4)$ superconformal algebra. More explicitly, it is defined as,
\ie 
&\cI_{\rm SCI}(p,q,y_i)
=\Tr_{\cH({S^3})}\left( (-1)^F e^{-\B\{\cQ,\cQ^\dagger\} }
p^{ J_1+{1\over 2}R_3}q^{ J_2+{1\over 2}R_3} y_1^{{1\over 2}(R_1-R_3)}y_2^{ {1\over 2}(R_2-R_3)}
\right)\,,
\label{N4SCI}
\fe
with the chemical potentials
\ie 
\label{eq:potentialfugacity}
p\equiv e^{2\pi i \sigma}\,,\quad 
q\equiv e^{2\pi i \tau}\,, \quad 
y_i\equiv e^{2\pi i \Delta_i}\,, \quad i = 1, 2\,,
\fe
 which counts ${1\over 16}$-BPS local operators annihilated by a chosen pair of supercharges $\cQ,\cQ^\dagger$ in SYM, satisfying the anticommutation relation
 \ie 
\{\cQ,\cQ^\dagger\}=E-J_1-J_2-{R_1+R_2+R_3\over 2}\,.
\label{QSac}
\fe
The counting is
refined by chemical potentials $\sigma, \, \tau,\, \Delta_{1,2}$ for linear combinations of conserved charges that commute with  $\cQ,\cQ^\dagger$. Here $J_1,J_2$ are Cartan generators of the $\mf{so}(4)_{\rm rot}$ isometry on $S^3$ and $R_1,R_2,R_3$ are Cartan generators of the $\mf{su}(4)_R$ R-symmetry, both of which are bosonic subgroups of the full $\mf{psu}(2,2|4)$ superconformal symmetry.\footnote{See Appendix~\ref{sec:superalgebra} for more details on the superconformal algebra.} By standard argument, the index doesn't depend on the regularization parameter $\B$~\cite{Romelsberger:2005eg, Kinney:2005ej}.\footnote{Besides being independent of $\beta$, the index is also independent of the 't Hooft coupling (and the $\theta$-angle), so we can compute it in the free field theory limit (where the bulk is very stringy), but extrapolate to the large 't Hooft coupling limit where the bulk is semiclassical.}

Recently there has been exciting progress in the microstate counting for ${1\over 16}$-BPS black holes using the $\cN=4$ superconformal index, written as a unitary matrix model that can be derived in the path integral approach from localization \cite{Nawata:2011un,Assel:2014paa}, 
\ie 
\cI_{\rm SCI}(p,q,y_i)=\int_{SU(N)} [dU] 
\exp\left[\sum_{n=1}^{\infty} {1\over n} f(p^n,q^n,y_1^n,y_2^n)(\Tr U^n \Tr U^{\dagger n}-1)\right]\,,
\label{4dUmm}
\fe
where $f(p,q,y_1,y_2)$ is the so-called single letter index and coincides with the character for the $\mf{psu}(1,2|3)\subset \mf{psu}(2,2|4)$ subgroup which (super)commutes with $\cQ,\cQ^\dagger$ (see \cite{Nawata:2011un} for details).
Since the integrand is a class function on ${\rm SU}(N)$, it is standard to write the above matrix integral over the ${\rm SU}(N)$ gauge group as an integral over the maximal torus $T$ parametrized by $\text{diag}(e^{2\pi i u_1},...,e^{2\pi i u_N})$ with $u_i\sim u_i+1$ subject to the constraint $\sum_{i=1}^N u_i \in \mZ$, 
where an additional Vandermonde determinant is introduced. After taking  into account the contributions from all fields in \eqref{4dUmm}, the 4d index of $\mathcal{N}=4$ SYM takes the following explicit form,
\begin{equation}
\label{4dSYMUmm}
    \mathcal{I}_{\text{SCI}} (p,q,y_i) = \frac{\kappa^{N-1}}{N!}\oint_T \left(\prod_{i=1}^{N-1} du_i\right)\, \prod_{i\neq j}\frac{\prod_{a=1}^3 \Gamma(\Delta_a+u_{ij};\sigma,\tau)}{\Gamma(u_{ij};\sigma,\tau)}\,,
\end{equation}
where
$u_{ij} \equiv u_i-u_j$ and we have defined $\Delta_3$ such that the chemical potentials satisfy the constraint
\begin{equation}\label{constraintsigma}
    \sigma + \tau - \Delta_1 - \Delta_2 - \Delta_3 = 1\,.
\end{equation}
The elliptic gamma functions $\Gamma(u;\sigma,\tau)$ (whose precise form will not be needed here) in the numerator of (\ref{4dSYMUmm}) come from the three $\mathcal{N}=1$ chiral multiplets in $\mathcal{N}=4$ SYM and the one in the denominator is from the $\mathcal{N}=1$ vector multiplet. The $\frac{1}{N!}$ factor in front of the integral in (\ref{4dSYMUmm}) is from modding out the Weyl group of ${\rm SU}(N)$. The prefactor $\kappa^{N-1}$ comes from the multiplets in the Cartan directions and does not depend on $u_i$, 
\ie 
\kappa = (p;p)(q;q) \prod_{a=1}^3\Gamma(\Delta_a;\sigma,\tau)\,,
\fe 
where $(p;q)\equiv \prod_{k=0}^\infty (1-pq^k)$ is the Pochammer symbol.
Evaluating the remaining integral over $u_i$ at large $N$ has been the subject of much recent study.

To better understand the role of the chemical potentials in the superconformal index as well as the relationship to the dual black holes, we first note that the superconformal index (\ref{N4SCI}) can be related to the ordinary thermal partition function\footnote{Here in both the superconformal index and the partition function, we normalize the contribution of the vacuum state to be one. }
\begin{equation}
  \cI_{\rm SCI}(p,q,y_i) =   Z = \textrm{Tr}_{\mathcal{H} (S^3)}\left( e^{-\beta H + \sum_{i} \beta \Omega_i J_i + \frac{1}{2} \sum_{a=1}^3 \beta\Phi_a R_a }  \right)\,,
  \label{eq:grandcanonicalindex}
\end{equation}
with chemical potentials $\{\Omega_1, \Omega_2 , \Phi_1 , \Phi_2 , \Phi_3\}$ determined by the following relations, 
\begin{equation}
    \sigma = \frac{\beta (\Omega_1 - 1)}{2\pi i }\,, \quad \tau = \frac{\beta (\Omega_2 - 1)}{2\pi i }\,, \quad \Delta_a = \frac{\beta (\Phi_a - 1)}{2\pi i }\,, \quad a = 1, 2,3\,,
\end{equation}
which satisfy the constraint \cite{Hosseini:2017mds,Cabo-Bizet:2018ehj,Choi:2018hmj,Benini:2018ywd} 
\begin{equation}\label{complexconstraint}
    \beta \left(1 + \Omega_1 + \Omega_2 - \Phi_1 - \Phi_2 - \Phi_3 \right) =2\pi i \,.
\end{equation}
The constraint (\ref{complexconstraint}), which is equivalent to (\ref{constraintsigma}), implies that at least one of the chemical potentials has to be complex in order to enforce the supersymmetric boundary condition, where the $(-1)^F$ twist is equivalently implemented by $e^{i\pi R_3}$ in the SYM theory \cite{Choi:2018hmj,Cabo-Bizet:2018ehj}.\footnote{\label{ft:spincharge}Note that in $\cN=4$ SYM, the quantum numbers of all the fields satisfy $F=2J_{1,2}=R_{1,2,3}\mod 2$.} These chemical potentials are related to the asymptotic values of various off-diagonal elements of the metric (up to possible shifts that we will describe) in the gravity description, implying the necessity of considering \emph{complex} solutions in the Euclidean gravitational path integral that computes \eqref{eq:grandcanonicalindex} from the holographic dual. 
Conversely the supersymmetric black holes in AdS$_5$ with large entropy necessarily have nonzero angular momentum and thus become complex solutions after Wick rotation.

Since the superconformal index can be interpreted as a statistical ensemble which is dominated by the black hole solutions in the large $N$ dual (for a range of the chemical potentials), knowing the large $N$ limit of the index \eqref{eq:4dindexaction} and performing a Legendre transformation produces immediately the black hole entropy that matches via the Bekenstein-Hawking formula with  the ${1\over 16}$-BPS solution in supergravity \cite{Benini:2015eyy,Cabo-Bizet:2018ehj,Choi:2018hmj,Choi:2018vbz, Benini:2018ywd, Honda:2019cio, ArabiArdehali:2019tdm, Kim:2019yrz, Cabo-Bizet:2019osg, GonzalezLezcano:2019nca, Cabo-Bizet:2019eaf, ArabiArdehali:2019orz, Murthy:2020rbd, Agarwal:2020zwm, Benini:2020gjh, Cabo-Bizet:2020nkr, GonzalezLezcano:2020yeb, Copetti:2020dil,   Cabo-Bizet:2020ewf, Lezcano:2021qbj,ArabiArdehali:2021nsx, Cassani:2021fyv, Aharony:2021zkr, David:2021qaa, Goldstein:2020yvj,Choi:2021rxi,Boruch:2022tno,Mamroud:2022msu}. We will now proceed to review some progress in this direction. 

One main breakthrough comes from a careful analysis of \eqref{4dSYMUmm} in the large $N$ limit, by picking out the relevant
saddle point that produces the dominant contribution with complex chemical potentials. 
Previous attempts to study the large $N$ limit of the index \cite{Kinney:2005ej,Romelsberger:2005eg,Janik:2007pm,Grant:2008sk,Chang:2013fba} revealed an $\mathcal{O}(1)$ growth of the degeneracies rather than the $\mathcal{O}(N^2)$ behavior predicted by the Bekenstein-Hawking formula for AdS$_5$ black holes. This slower growth is consistent with a gas of bulk supergravitons, signifying the dual conformal field theory is in the confining phase. It is believed that the absence of the large degeneracies in the index was due to unnecessary boson/fermion cancellation, and we will briefly review below how this is avoided by the appropriate complex saddle point and complex chemical potentials, which allow us to explore the deconfining regime. From the point of view of fixed charges, one expects the large charge sector of the CFT to be dominated by black hole states, but it turns out that the relevant BPS black holes obey a nonlinear relation among their charges which is not manifest in the field theory calculation\cite{Cabo-Bizet:2018ehj,Hosseini:2017mds}.

For any value of $N$, in principle \eqref{4dSYMUmm} provides an exact answer for the index. In order to produce the semiclassical black hole entropy at large $N$, there are several different strategies. The first is a Cardy-like limit \cite{DiPietro:2014bca} in which $N$ is kept fixed while the chemical potentials are tuned so that the index is dominated by large charge states. 
 This dramatically simplifies the expression of the unitary matrix integral even before the large $N$ limit is taken (so one is effectively counting large AdS black holes)\cite{Choi:2018hmj,Choi:2018vbz,Honda:2019cio,Kim:2019yrz,ArabiArdehali:2019tdm,Cassani:2021fyv,ArabiArdehali:2021nsx,Amariti:2020jyx, Amariti:2021ubd}. A second approach known as the Thermodynamic Bethe Ansatz \cite{Closset:2017bse,Benini:2018mlo,Benini:2018ywd,Hong:2018viz,GonzalezLezcano:2019nca,Benini:2020gjh,Lezcano:2021qbj,Benini:2021ano,Mamroud:2022msu} is appropriate when the potentials obey a certain relation (and most useful in the large $N$ limit). Here, one may write the exact large $N$ index as a formal sum over the solution set of the Bethe Ansatz equations (see \cite{Benini:2018mlo,Aharony:2021zkr} for details). Not all of these solutions are known, but many of the isolated solutions can be given a direct gravity interpretation in terms of Euclidean black hole solutions decorated by wrapped D3-branes \cite{Aharony:2021zkr}.\footnote{These Bethe Ansatz equations also admit continuous families of solutions for $N\geq 3$ \cite{ArabiArdehali:2019orz,Lezcano:2021qbj,Benini:2021ano}. Their gravitational interpretation is yet to be understood.} The final approach and the one most relevant for this work involves the direct saddle point evaluation of \eqref{4dUmm} as a large $N$ matrix integral of certain special elliptic functions \cite{Cabo-Bizet:2019eaf,Cabo-Bizet:2020nkr,Cabo-Bizet:2020ewf}. A comparison of these approaches as well as studies of subleading contributions is given in \cite{GonzalezLezcano:2020yeb}. Further studies of subleading saddle points, phase transitions in gauge theory, and the relationship to small AdS black holes are given in \cite{Choi:2018vbz,Copetti:2020dil,Choi:2021lbk}. Finally, it is possible to study the finite $N$ cohomology problem in order to directly detect the growth of the index and enumerate the operators which can be dual to black holes\cite{Murthy:2020rbd,Agarwal:2020zwm,Chang:2022mjp,Choi:2022caq,Choi:2023znd}.

Focusing on the large $N$ evaluation of the matrix integral by saddle point approximation, we will eventually make use of a configuration in which the two angular potentials $\sigma$ and $\tau$ in \eqref{eq:potentialfugacity} are \emph{unequal}. An appropriate formulation of the saddle point problem for this non-collinear configuration is found in \cite{Choi:2021rxi}. There \eqref{4dSYMUmm} is evaluated on a family of saddles in which the matrix eigenvalues are distributed uniformly over a complex parallelogram whose shape depends on $\sigma,\tau$, rather than the more familiar linear eigenvalue distribution. The final result for the free energy of the leading saddle is
\begin{equation}
\label{eq:4dindexaction}
    \textrm{log}~\mathcal{I}_{\text{SCI}} \approx - i \pi N^2 \frac{ \Delta_1 \Delta_2 \Delta_3}{\sigma \tau} \, ,
\end{equation}
which matches the on-shell action of the semiclassical black hole solution and thus reproduces the Bekenstein-Hawking entropy of the generically complex black hole.

However there is much more to these  black hole solutions than just the entropy, which is simply determined by the bulk on-shell action. Indeed, they are complicated geometries that involve nontrivial topologies and nontrivial fibrations. A natural question is how one can probe such refined properties in the dual SYM theory. Since the field theory
provides the unambiguous non-perturbative completion for the semiclassical expansion around these gravitational saddle points, understanding this connection will also be indispensable for elucidating the structure of the
gravitational path integral.

In general, features of semiclassical saddles in a quantum theory can be detected by analyzing certain light observables whose backreaction is suppressed. A well-studied setup concerns the correlation functions of light operators on a semiclassical background. In particular, the thermal correlation functions in holographic CFTs above the Hawking-Page transition are known to encode subtle and salient properties of black holes in the bulk dual, including the structure of the black hole singularity by one-point \cite{Grinberg:2020fdj} and two-point functions \cite{Fidkowski:2003nf,Festuccia:2005pi,Rodriguez-Gomez:2021pfh}, and the fast scrambling and quantum chaos by the out-of-time-order four-point functions \cite{Shenker:2013pqa,Maldacena:2015waa}, and the black hole information loss 
via the so-called ``heavy-heavy-light-light'' four-point functions which provide the non-perturbative quantum gravity information beyond the semiclassical limit 
\cite{Fitzpatrick:2014vua,Fitzpatrick:2016ive}. 

It is important to note, however, that in some cases backreaction cannot be ignored\cite{Almheiri:2014cka}, even when the black hole is large compared to the Planck or string scales. This is the near extremal or near BPS limit in which $\beta \rightarrow \infty$ and the black hole geometry approaches a product of AdS$_2$ and a compact manifold. For a wide class of supersymmetric black holes, it has been argued that quantum fluctuations of the near horizon metric become strongly coupled in this limit, and their dynamics are captured by a supersymmetric version of JT gravity in AdS$_2$\cite{Heydeman:2020hhw,Boruch:2022tno}. This is an exactly solvable model of gravity which captures the universal low temperature corrections to the partition function, and the large $N$ index (being independent of $\beta$), is ultimately recovered from this theory\cite{Boruch:2022tno,Iliesiu:2022kny} for AdS$_5$ or flat space black holes. In the context of correlation functions on black hole backgrounds, the super-JT description was used in \cite{Lin:2022rzw,Lin:2022zxd} to study the late time behavior of these correlators for BPS black holes. We offer further comments on the near BPS limit in Section \ref{sec:conclusion}.

While these works shed light on the universal properties of black holes, here we are interested in the more sophisticated features of the special 
${1\over 16}$-BPS black holes in type IIB string theory on ${\rm AdS}_5\times S^5$. The benefit, however, is that the observables we identify in this work can in principle be computed exactly, despite being on a black hole background, as long as the residual ${1\over 16}$ supersymmetry is preserved, which helps to tame the strong interactions in the field theory. 
This motivates us to look for a further refinement of the superconformal index \eqref{N4SCI} by additional insertions that are compatible with the supercharges $\cQ,\cQ^\dagger$. Due to the anti-commutation relation (\ref{QSac}), it is immediate that local operator insertions cannot preserve $\cQ,\cQ^\dagger$ since they would transform nontrivially under the bosonic generators from the anti-commutator. Instead, one can consider extended operators, and in this work we study a class of two-dimensional operators known as surface defects in $\cN=4$ SYM that are compatible with the desired supercharges.\footnote{See Section \ref{sec:PLcomment} for further comments.} Their explicit construction was given in \cite{Gukov:2006jk} and are commonly referred to as Gukov-Witten (GW) surface operators. In the following subsection we will provide a brief review of the GW surface operators.

\subsection{Review of the Gukov-Witten surface operator}
\label{sec:GWrev}

The GW surface operator \cite{Gukov:2006jk} is a two spacetime dimensional extended operator that is defined by a codimension-two singularity in the SYM fields. To describe the profile of the fields, we first split the six adjoint scalars $\Phi^I$ in the SYM theory into $\Phi^{1,2,3,4}$ and $\Phi^{5,6}$ which transform as $(\bm 2,\bm 2)_0$ and $(\bm 1,\bm 1)_\pm $ respectively under the R-symmetry subalgebra $\mf{su}(2)_1\times \mf{su}(2)_2\times \mf{u}(1)_R \subset \mf{su}(4)_R$.
We identify the Cartan generators of $\mf{su}(2)_1$ and $\mf{su}(2)_2$ with $R_2+R_3$ and $R_2-R_3$ respectively. The  $\mf{u}(1)_R$ is generated by $R_1$ and we define the complex scalar ${\bm \Phi}=\Phi^{5}+i\Phi^{6}$ which has charge $R_1=2$. 
The GW surface defect on a two-dimensional submanifold $\Sigma$ is then described by the following scale invariant singular configuration for the complex scalar $\bm \Phi$ and the gauge field $A$, 
\ie
{\bf \Phi}={\B+i\C\over z}\,,\quad A=\A d\psi\,,
\label{GWconfig}
\fe
where $z=re^{i\psi}$ is the local normal holomorphic coordinate to $\Sigma$ with $r$ the transverse distance and $\psi$ the polar angle.  The coefficients $\A,\B,\C$ take values in the Lie algebra $\mf{g}$ of the gauge group $G$ and commute with one another as a consequence of the supersymmetry conditions (i.e. BPS equations),
\ie
F_{z\bar z}+[{\bf \Phi},{\bf \Phi}^\dag]=2\pi \A \D^2(z,\bar z)\,,\quad D_{\bar z}{\bf \Phi}=0\,,
\label{BPSeq}
\fe 
known as the Hitchin equations.  
Furthermore, the configuration \eqref{GWconfig} can be superimposed with topological terms
\ie
\exp \left(
i \int_{\Sigma} \Tr \eta F
\right)
\label{2dthetaterm}
\fe
with $d=2$ theta angles $\eta\in \mfg$. These surface operators are closely related to the Wilson-'t Hooft line operators constructed in \cite{Kapustin:2005py}. 
Physically, the configuration \eqref{GWconfig} together with \eqref{2dthetaterm} describes the worldsheet (along $\Sigma$) of the Dirac string emanating from a dyon with fractional electric and magnetic charges $(\eta,\A)$. Equivalently, this means the open GW surface operator along  $\Sigma$ with a non-empty boundary $\pa\Sigma$ gives the gauge invariant completion of a Wilson-'t Hooft loop along $\pa\Sigma$ with these unquantized charges. 
On the other hand, the closed GW surface defect  (which will be the focus of this work)  is a supersymmetric generalization of 
the Abrikosov-Nielsen-Olesen (ANO) vortex \cite{ABRIKOSOV1957199,Nielsen:1973cs} in four spacetime dimensions that supports a localized flux (equivalently nontrivial holonomy)  as in \eqref{GWconfig} and \eqref{BPSeq}.

The GW surface operator is half-BPS. When positioned along $\Sigma=\mR^2$ at $x^2=x^3=0$ in the flat spacetime $\mR^{1,3}$ with coordinates $x^\m$, it preserves the following subalgebra of the $\cN=4$ superconformal algebra,\footnote{This is also known as the small $\cN=(4,4)$ superconformal algebra centrally extended by $\mf{u}(1)_C$. The embedding of this algebra into 4d $\mathcal{N}=4$ superconformal algebra is reviewed in Appendix \ref{sec:superalgebra}.}
\ie 
\left( \mf{psu}(1,1|2)\times \mf{psu}(1,1|2) 
\right ) \ltimes \mf{u}(1)_C\,,
\label{defectalg}
\fe
which contains the conformal symmetry $\mf{so}(2,2)$ along the $\mR^2$ and the R-symmetry subalgebra $\mf{su}(2)_1\times \mf{su}(2)_2$ mentioned above, and is centrally extended by $\mf{u}(1)_C$ which is generated by a combination of the transverse rotation $J_{23}$ and the $\mf{u}(1)_R$ R-symmetry,
\ie 
C= J_{23} -{R_1\over 2}\,.
\label{centralc}
\fe

The distinct GW surface operators are classified by gauge inequivalent configurations of \eqref{GWconfig} together with \eqref{2dthetaterm}. Taking into account the global discrete identifications, each equivalent class is labelled by a Levi subgroup $L \subset G$ which specifies the gauge subgroup preserving \eqref{GWconfig} and
a four-tuple of continuous parameters \cite{Gukov:2006jk},
\ie 
(\A,\B,\C,\eta)\in (T\times \mf{t} \times \mf{t} \times T^\vee) / \cW_{L} \,,
\label{contdefectparam}
\fe 
where $T\subset G$ and $T^\vee \subset  G^\vee $ are the maximal torus subgroups of $G$ and its Langlands dual $G^\vee$ respectively, $\mf{t}\subset \mf{g}$ is the Cartan subalgebra, and $\cW_L$ is the Weyl group of the Levi subgroup $L$ that implements the residual gauge identifications. We denote the corresponding GW surface operator as $\cD_L[G]$.
The continuous parameters in \eqref{contdefectparam} are then coordinates on the conformal manifold of $\cD_L[G]$.  Conformal surface defects in general has a defect Weyl anomaly (also known as defect central charge) which we denote as $b_{2d}$ that is similar to the central charge of 2d CFTs. In particular, $b_{2d}$ counts the additional degrees of freedom introduced by the defect and monotonically decreases under defect renormalization group flows \cite{Jensen:2015swa,Casini:2018nym,Wang:2020xkc,Shachar:2022fqk,Casini:2023kyj}. At special loci on the defect conformal manifold, there is symmetry enhancement. In particular, at $\B=\C=0$, the surface defect preserves an extra bosonic symmetry generated by (see also Appendix~\ref{sec:superalgebra}),
\ie 
J_A=R_1-R_2\,.
\label{extrasym}
\fe

Instead of the definition in \eqref{GWconfig} as a disorder operator (also known as a mondromy defect), 
the GW surface operator on $\Sigma$ has an alternative description in terms of $\cN=4$ SYM coupled to 
a 2d $\cN=(4,4)$ non-linear sigma model localized on $\Sigma$ with hyper-K\"ahler target space $T^*(G/L)$ which is the space of solutions to \eqref{GWconfig} known as the Hitchin moduli space
\cite{Gukov:2006jk}. From the dimension of the target space $\dim T^*(G/L)=4(\dim G-\dim L)$, one can infer the defect central charge $b_{2d}=6(\dim G-\dim L)$ \cite{Chalabi:2020iie,Wang:2020xkc}.
The sigma model has a natural $G$ global symmetry which couples to the SYM gauge fields. This 2d-4d description 
turns out to be extremely useful for studying the physics of the GW surface operators. This is further facilitated by 2d $\cN=(4,4)$ gauge theories that provide simple UV gauged linear sigma model (GLSM) descriptions of the more complicated non-linear sigma model \cite{Gukov:2006jk,Gadde:2013dda}.\footnote{In the 2d-4d description, the defect conformal manifold \eqref{contdefectparam} comes from the complex structure and complex K\"ahler moduli of the target space $T^*(G/L)$, equivalently from the complex FI terms and the superpotential deformations in the GLSM.} It is this gauged linear sigma model description which will be useful for us in making contact with the saddle point analysis of the original index \eqref{4dSYMUmm} and studying its refinement due to the surface defect insertion. 

Here we focus on $G={\rm SU}(N)$ SYM, in which case the Levi subgroups are in one-to-one correspondence with integer partitions $[k_1,\dots,k_n]$ with  $0<k_1\leq \dots \leq k_n<N$,
\ie 
L={\rm S}[{\rm U}(k_1)\times \cdots \times {\rm U}(k_n)]\,,\quad N=\sum_{i=1}^n k_i\,,
\label{Levi}
\fe 
and the corresponding 2d gauge theory describing the GW surface defect is defined by a $\cN=(4,4)$ linear quiver gauge theory with gauge groups $\oplus_{i=1}^{n-1} U(p_i)$ where $p_i=\sum_{j=1}^i k_j$ and bifundamental hypermultiplets between adjacent quiver nodes (including the bulk ${\rm SU}(N)$ as the last node). Importantly, it is the Higgs branch of the 2d gauge theory that is relevant for the 2d-4d description of the GW surface operator \cite{Witten:1997yu,Gukov:2006jk}. The defect central charge for $\cD_{[k_1,\dots,k_n]}({\rm SU}(N))$ is \cite{Chalabi:2020iie,Wang:2020xkc}
\ie 
b_{2d}=3\left(N^2-\sum_{i=1}^n  k_i^2\right)\,.
\fe

Via AdS/CFT, these GW surface defects have different dual descriptions in type IIB string theory depending on the defect central charge $b_{2d}$. For $b_{2d}\sim \cO(N)$, the surface defect insertion in the vacuum is described in the bulk by probe D3-branes wrapping ${\rm AdS}_3\times {\rm S}^1$ submanifolds in the ${\rm AdS}_5\times {\rm S}^5$ where the specific locations are determined by the continuous parameters $(\B_i,\C_i)$ and the other periodic parameters $(\A_i,\eta_i)$ come from holonomies of the worldvolume gauge field along  S$^1$ and its electromagnetic dual on the probe D3-branes \cite{Drukker:2008wr,Koh:2008kt}. The worldvolume gauge fields on the D3 branes also have components along the AdS$_3$ directions. The GW surface defect corresponds to choosing the Neumann boundary condition for these gauge fields at infinity and therefore we should integrate over their boundary components.\footnote{One way to see this is that the GW surface defect do not introduce any extra global symmetry on the defect worldvolume.} For $b_{\rm 2d}\sim \cO(N^2)$, the backreaction from the D3-branes cannot be neglected and the surface defect is described by ``bubbling'' geometries that are asymptotically ${\rm AdS}_5\times S^5$ \cite{Gomis:2007fi,Lin:2004nb,Lin:2005nh}.

Here we are particularly interested in the simplest  GW surface defect $\cD_{[1,N-1]}(SU(N))$ in ${\rm SU}(N)$ SYM,\footnote{We comment on the general GW surface defects in Section~\ref{sec:genOtherDCFT}.} with defect central charge
\begin{equation}\label{c2d}
   b_{2d} = 6(N-1)\,.
\end{equation}
The defect conformal manifold is parametrized by four real parameters $(\A,\B,\C,\eta)$ where $\A\sim \A+1\,,\eta\sim \eta+1$. Below we focus on the case $\B=\C=0$ where the extra symmetry \eqref{extrasym} is the preserved and which enables us to study the fully refined surface defect index (with all the fugacities as in the bare index \eqref{N4SCI}).\footnote{Even though we keep $\A,\eta$ general, the defect index will be independent of these parameters. On the gravity side, this corresponds to the independence of the probe D3 brane action on the holonomy of the worldvolume gauge field (and its dual) along the $S^1\subset S^5$, which we further comment on in Section~\ref{sec:gravaction}.}

On the field theory side, the corresponding 4d-2d description involves a 2d $\cN=(4,4)$ SQED with $N$ hypermultiplets of charge one. 
On the bulk side (before taking the near horizon limit), the surface defect describes the two dimensional intersection of a single probe D3-brane extending along $0,1,4,5$ directions in $\mathbb{R}^{1,9}$ with a stack of $N$ D3-branes extending along $0,1,2,3$ directions, see Table~\ref{table:D3}. 
The properties of this intersecting D3-brane system were studied carefully in \cite{Constable:2002xt}. In the supergravity description, we have a single probe D3-brane wrapping an ${\rm AdS}_3\times S^1$ submanifold inside ${\rm AdS}_5\times S^5$, if we consider the system in the vacuum \cite{Constable:2002xt,Drukker:2008wr,Koh:2008kt}. 
In the vacuum, the surface operator in the absence of other insertions has a simple expectation value that is fixed by its conformal anomalies (such as $b_{2d}$) through the supersymmetric Casimir energy \cite{Chalabi:2020iie}. However in a (generalized) thermal ensemble, as we will see, the surface operator becomes a powerful probe for the state of the system. Here we would like to use the surface operator to probe the geometry of the ${1\over 16}$-BPS black hole states in the generalized thermal ensemble given by the superconformal index.  

\begin{table}[h!]
\begin{center}
\begin{tabular}{|c|c|c|c|c|c|c|c|c|c|c|}
\hline
 &  0 & 1 & 2 & 3 & 4 & 5 & 6 & 7 & 8 & 9 \\ \hline
\quad $N$ \textrm{D3}-branes & $\times$ & $\times$ & $\times$ & $\times$ &  &  &  &  &  &  \\ \hline
Probe D3-brane & $\times$ & $\times$ &  &  & $\times$ & $\times$ &  &  &  &  \\ \hline
\end{tabular}
\end{center}
\caption{The supersymmetric configuration of the $N$ D3-branes and a single probe D3-brane.}
\label{table:D3}
\end{table}

To preserve the desired supercharges $\cQ,\cQ^\dagger$ on $S^1\times S^3$, the surface operator worldvolume $\Sigma$ extends along the $S^1$ and wraps a great circle on $S^3$ that is generated by the Killing vector in \eqref{QSac}. In the presence of general chemical potentials as in \eqref{N4SCI}, $\Sigma$ is a torus with $\tau_{T^2}=\sigma$ as its complex structure modulus.\footnote{To avoid confusion with the other complex parameters of the problem, we will rarely write the complex structure of the defect $T^2$ explicitly. Throughout, $\tau$ and $\sigma$ will always refer to the background angular potentials on the rotating $S^3$, as in \eqref{eq:potentialfugacity}.}
 In the Lorentzian picture, since the surface defect extends in the time direction, it modifies the Hilbert space on $S^3$ to $\cH_{\cD}(S^3)$ which is also known as the defect Hilbert space, and gives rise to the defect superconformal index\footnote{Here we take $F=2J_1~{\rm mod}\,2$ which coincides with the usual fermion number on the 2d defect worldvolume. Note that the defect degrees of freedom in general do not obey the same spin-charge relation as the bulk degrees of freedom (see also Footnote~\ref{ft:spincharge}). 
 } 
\ie 
\cI_{\cD}=\Tr_{\cH_{\cD}({\rm S}^3)}\left( (-1)^F e^{-\B\{\cQ,\cQ^\dagger\} }
p^{ J_1+{1\over 2}R_3}q^{ J_2+{1\over 2}R_3} y_1^{{1\over 2}(R_1-R_3)}y_2^{ {1\over 2}(R_2-R_3)}
\right)\,.
\label{dSCI}
\fe
The usual state-operator correspondence in CFT extends readily to the case with a conformal defect and here $\cH_{\cD}(S^3)$ contains states that are in one-to-one correspondence with local operators on the surface defect $\cD$. Note that while some defect local operators come from the decomposition of
 bulk operators (namely those in $\cH(S^3)$)  via the bulk-defect operator-product-expansion (OPE), there are genuine defect local operators that do not have a bulk origin. Furthermore, the bulk-defect OPE is typically non-trivial, and consequently the defect local operators that come from bringing bulk operators to the defect in general have scaling dimensions that differ from their bulk parents. 

 There is a simple generalization of the matrix model formula for the  superconformal index \eqref{4dUmm} that holds for the defect index \eqref{dSCI}, which can either be derived from state counting or from localization of the coupled 4d-2d system. The inclusion of the surface defect amounts to inserting the superconformal index $\cI_{2d}(p,q,y_i;U)$ of the 2d defect fields (also known as the NSNS elliptic genus) in the unitary matrix integral \eqref{4dUmm}, which we refer to as the defect index,
 \ie 
\cI_{\cD}(p,q,y_i)=\int_{SU(N)} [dU]\, 
\cI_{4d}(p,q,y_i;U)\cI_{2d}(p,q,y_i;U)\,,\quad 
\label{defectmm}
 \fe
where $\cI_{4d}(p,q,y_i;U)$ denotes the integrand from the 4d fields in \eqref{4dUmm} and the $U$ dependence of the 2d index descends from the 2d-4d coupling by gauging the ${\rm SU}(N)$ global symmetry of the 2d defect theory \cite{Nakayama:2011pa,Gaiotto:2012xa,Iqbal:2012xm,Gadde:2013dda,Bullimore:2014nla,Gukov:2014gja}. 
Our proposal is that this expression provides the microscopic definition of a dual gravity system which includes black holes interacting with a D3-brane, and we will provide a precision check of this proposal in the large $N$ limit.

\subsection{Summary of main results}

\begin{figure}[t!]
    \begin{center}
   \includegraphics[scale=.3]{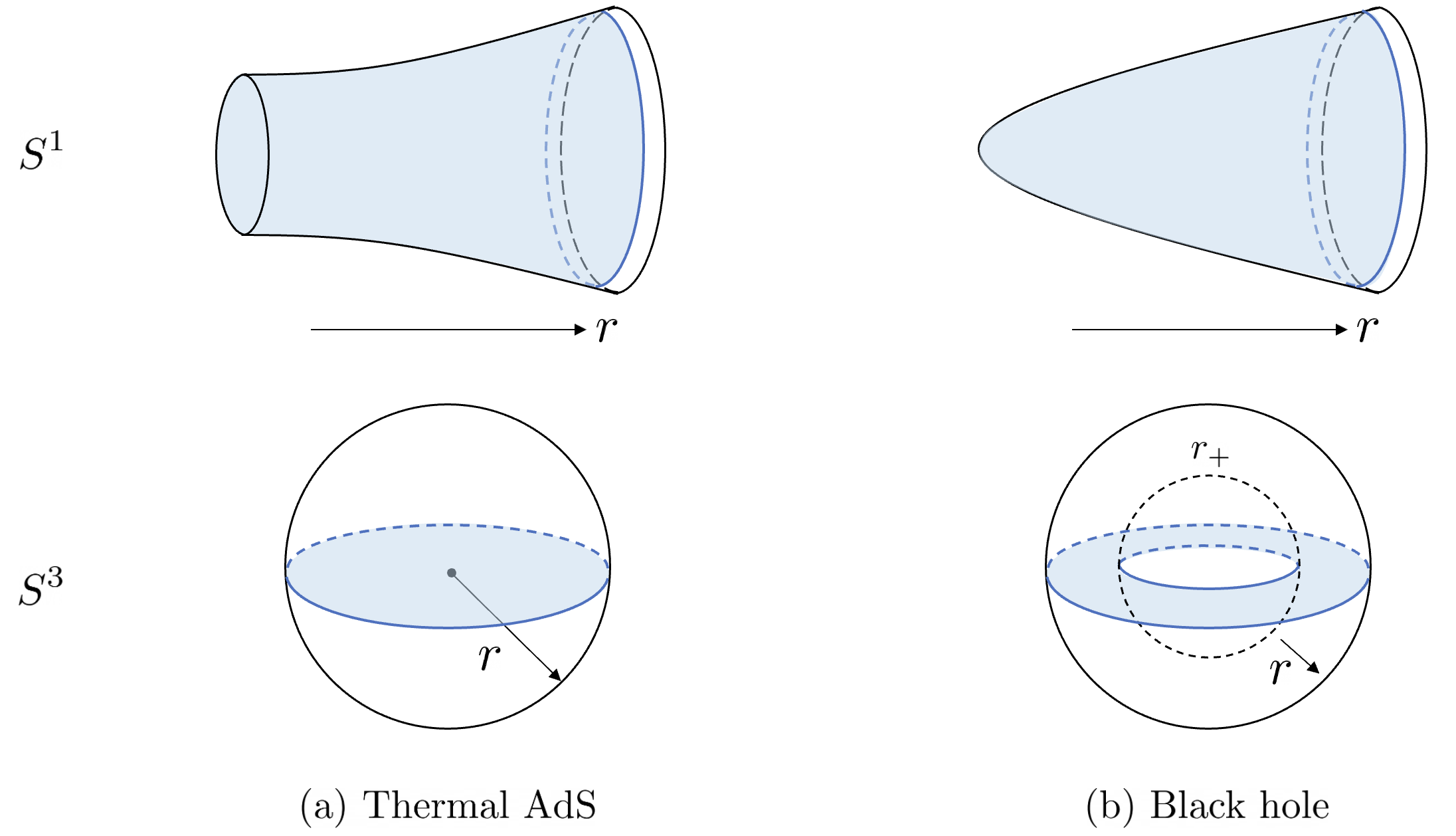}
    \end{center}
    \caption{An illustration of the different topologies of the bulk solutions and how the D3-brane (shaded in blue) probes them. The D3-brane fills the Euclidean time circle ($S^1$) and a circle in the spatial $S^3$. In the thermal AdS solution, the time circle is non-contractible while the spatial circle is contractible; the situation is reversed in the black hole solution. An extra circle in $S^5$ that the D3-brane is wrapping is omitted in the figure.} 
    \label{fig:topology}
\end{figure}

A main goal of this work is to analyze the matrix integral \eqref{defectmm} in the large $N$ limit where saddle point approximation is applicable. Importantly, since $b_{2d}=6(N-1) \ll N^2$ in this case, we can ignore the backreaction of the defect on the original matrix integral potential without the defect. Therefore, to leading order in the $1/N$ expansion, we find 
\ie \label{ID}
\cI_{\cD}(p,q,y_i)=\sum_{{\rm saddles}~U_*} \cI_{4d}(p,q,y_i;U_*) \cI_{2d}(p,q,y_i;U_*)+\dots \,,
\fe
where $U_*$ labels a saddle point in the unitary matrix model \eqref{4dUmm}. The saddle points of \eqref{4dUmm} are not completely understood due to complications from branch cuts of the effective action in the matrix model. Nonetheless, a large class of saddles have been identified for the special chemical potentials $\sigma=\tau$ in \cite{Cabo-Bizet:2019eaf} and more recently for the general $\sigma\neq \tau$ case in \cite{Choi:2021rxi}. In particular, the saddle points where the holonomies distribute along some complex directions are proposed to be the field theory avatars of the ${1\over 16}$-BPS black hole in the bulk. Indeed, the black hole entropy can be computed this way from the effective action of the leading saddle in the matrix model. 
As noted before, consideration of defects opens up a finer perspective on the structure of these saddles, and equivalently the black holes.  
Indeed, using the AdS/CFT dictionary, the normalized expectation value of the defect $\langle \mathcal{D}\rangle$ can be computed, in the large $N$ limit, by the on-shell action of a probe D3-brane on the black hole geometry,
\ie 
\langle \mathcal{D}\rangle = \cI_{2d}(p,q,y_i;U_*)\sim e^{-I_{D3, \textrm{E}} }\,.
\fe
The location of the D3-brane is fixed by the boundary conditions and a calibration condition coming from supersymmetry. Despite the complicated black hole solution, we identify this supersymmetric probe D3-brane and evaluate its on-shell action.\footnote{As mentioned before, here we focus on the simplest GW surface defect satisfying $\B,\C=0$ and preserving the extra symmetry \eqref{extrasym}. Correspondingly we identify explicitly the calibrated probe D3-brane on the black hole geometry that preserves the same symmetry. It would be interesting to generalize our gravity analysis for the case $\B,\C\neq 0$.} After taking into account the regularization of infinities from the asymptotic boundary, we find the following simple result,\footnote{The action is not a pure phase since $\sigma,\tau$ are in general complex. }
\ie 
I_{D3, \textrm{E, finite}} = -2\pi i N \frac{(\sigma + \tau - 1)^2}{9 \sigma} \,,
\label{SD3}
\fe
for general angular chemical potentials $\sigma,\tau$ and identical R-symmetry chemical potentials $\Delta_a =\Delta={\sigma+\tau-1\over 3}$. 
For the most general case we expect the answer will be
\ie \label{D3Egen}
I_{D3, \textrm{E, finite}} = -2\pi i N \frac{\Delta_2\Delta_3}{ \sigma} \,.
\fe 
To confirm (\ref{D3Egen}) we would need to extend our analysis to a more general family of black hole solutions with different electric charges \cite{Cvetic:2004hs,Cvetic:2005zi,Cassani:2019mms}. We leave this to future study. The order $N$ classical action we find in the black hole phase is distinct from the zero action configuration of the D3-brane in empty AdS space \cite{Drukker:2008wr}. This difference can be attributed to the fact that our D3-brane configuration is sensitive to the different topologies in the bulk, see Figure \ref{fig:topology}. 

On the field theory side, the integral expression of the 2d index for the corresponding GW surface defect has been previously derived in \cite{Gadde:2013dda}. We solve the coupled matrix model \eqref{defectmm} in the large $N$ limit by evaluating the 2d index on the proposed saddles of \eqref{4dUmm}, which gives the expectation value $\langle\mathcal{D}\rangle$ of the defect. We find precise agreement with \eqref{SD3}. In addition to achieving this match between field theory and gravity quantities, this result has a pleasingly simple form analogous to \eqref{eq:4dindexaction} for the superconformal index without the defect insertion. The order $N$ behavior of the defect expectation value signals that we are in the deconfined phase of the gauge theory, rather than the confined phase for which we will get an order one answer. 
This is much like the case for the Polyakov loop as a probe for the deconfinement phase transition of large $N$ SYM in the conventional (non-supersymmetric) thermal ensemble on $S^1\times S^3$ studied at weak coupling in \cite{Aharony:2003sx}. The key difference is that our surface defect probe, owing to the preserved supersymmetry, can be analyzed irrespective of the coupling strength and provides a non-perturbative diagnostic for this phase transition.

\begin{figure}[t!]
    \begin{center}
   \includegraphics[scale=.28]{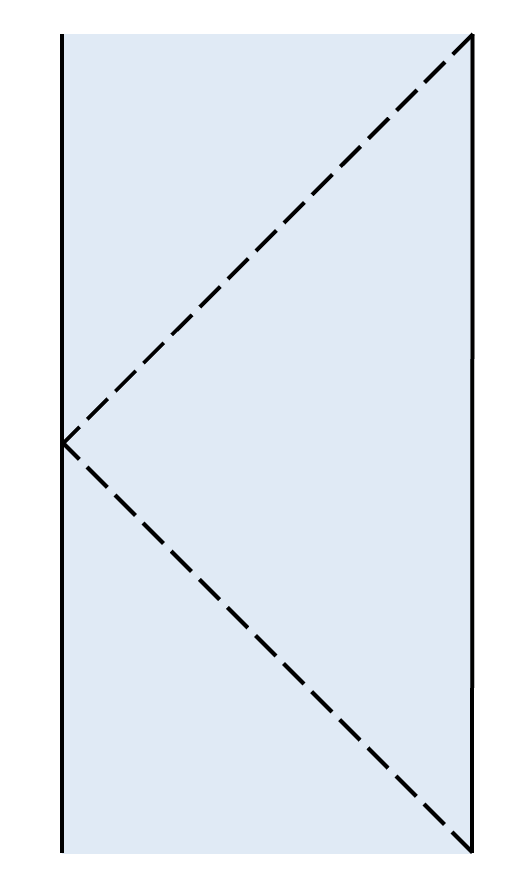}
    \end{center}
    \caption{A classical Penrose diagram representation of our set up in Lorentzian signature. In the diagram we focus on the near horizon AdS$_2$ part of the geometry. The black hole is at zero temperature and we have a D3-brane that fills the radial and time direction of the spacetime (represented by blue) plus some transverse directions. The brane extends into the interior in Lorentzian signature, though our calculation is in Euclidean signature and does not probe the interior in an obvious way. }
    \label{fig:penrose}
\end{figure}

In Lorentzian signature, the physical system that we consider is an extremal black hole coupled to a probe D3-brane in its environment, see Figure~\ref{fig:penrose}. It is natural to ask about the thermodynamics of the combined black hole + D-brane system. In the probe limit we are considering, the thermodynamic quantities such as the energy and the charges are simply given by the sum of those of black hole and those of the D-brane. In this paper, since we are computing the index of the combined system, we are considering a thermal ensemble where some of the chemical potentials are complex. Nonetheless we can still discuss the thermodynamic quantities of the D3-brane in the complex ensemble. We leave a detailed analysis of the properties of the Lorentzian system and the comparison with the complex ensemble to future study.

We also consider the subleading saddle points of $U_*$ in (\ref{ID}), some of which have been proposed in \cite{Choi:2021rxi}. 
The particular subleading saddles we study are associated with shifts of the complex chemical potentials by integers: they correspond to new bulk Euclidean solutions in which the various gauge fields obey the same boundary conditions as the original solution, but have nontrivial winding around the thermal circle. 
The semiclassical sum over the subleading saddles is important for ensuring the correct periodicity in the chemical potentials for the index, which is required for the microscopic interpretation as a trace over the CFT Hilbert space \cite{Aharony:2021zkr}. To our knowledge, the precise match between the subleading saddle points and the subdominant gravity solutions has not been fully understood, even for the original 4d index (\ref{4dUmm}). With some caveats in mind, we present evidence in Appendix \ref{sec:shifted} that the agreement we found for the dominant saddle persists for the subdominant ones.

The rest of the paper is organized as follows. In Section~\ref{sec:grav}, we describe the D3-brane configuration on the black hole background that is dual to the Gukov-Witten surface defect $\cD$ in SYM. We compute and regulate its on-shell action by subtracting the action of a brane on an empty AdS background. We also discuss the thermodynamics of the D3-brane in the ensemble that computes the index. In Section~\ref{sec:field theory}, we discuss the evaluation of the defect expectation value $\la \cD\ra$ in the field theory, first providing an exact expression for the defect index $\cI_\cD$ as a matrix integral and then carrying out the large $N$ analysis in order to reproduce the gravity result. We also explain how the same result can be derived in the Cardy limit of the defect index.
In Section~\ref{sec:conclusion}, we conclude and discuss a number of generalizations and future directions. Various technical details are collected in the Appendices.

\section{Gravity Analysis}\label{sec:grav}
\subsection{Review of the supergravity background and black holes}
\label{sec:10dgravity}

To describe the bulk dual of the $\mathcal{N}=4$ SYM GW surface defect, we first begin with a discussion of the gravity theory as well as the supersymmetric black hole solutions which appear in the semiclassical decoupling limit of the near horizon geometry of $N$ D3 branes, all placed at finite temperature. From this point of view, the defect configuration is ultimately a related brane system in which a single D3-brane is now placed transverse to the stack of $N$ and intersects along a 2-dimensional locus \cite{Constable:2002xt}. In the backreacted limit, the field theory description is that of ${\rm SU}(N)$ gauge theory with a surface defect, while the bulk is described by supergravity in the presence of a probe D3-brane. The probe approximation is justified at large $N$, which we review below. Additionally, depending on the chemical potentials chosen at the asymptotic AdS boundary, the dominant gravity saddle in the semiclassical limit might be empty AdS, or it might be a black hole.\footnote{It is logically possible that other gravitational saddles might contribute, and in Appendix. \ref{sec:shifted} we will indeed discuss a family of black hole solutions with subleading action as in \cite{Aharony:2021zkr}. With our desired amount of preserved supersymmetry as well as general semiclassical arguments about black hole entropy, we do not expect there to exist saddles with a larger action than the dominant black hole solution at sufficiently large $N$.}

Because the probe we consider wraps a profile in the full 10-dimensional geometry which is supported by $N$ units of Ramond-Ramond 5-form flux $F_{(5)} = d C_{(4)}$, the Dirac-Born-Infeld action of the brane must include this background as an input. This is simple for the case of empty (but thermally excited) AdS$_5 \times$ $S^5$, but finding general 10d black hole solutions with the same asymptotic boundary conditions is more difficult. A common strategy is to reduce to a 5-dimensional gauged supergravity which in general is maximally supersymmetric, but this theory has several consistent truncations \cite{Gunaydin:1984qu,Gunaydin:1985cu,Pernici:1985ju} which permit known five-dimensional black hole solutions in which we activate fewer fields. In terms of two conserved angular momenta $J_1$ and $J_2$ and three electric charges $Q_1, Q_2, Q_3$, the truncation to 5d $\mathcal{N}=2$ minimal gauged supergravity is achieved by setting $Q_1=Q_2=Q_3$; the 5d effective theory has a single gauge field corresponding to this symmetric rotation on the internal $S^5$ and no additional scalar fields. The electric charges are simply related to the R-charges by $Q_a = R_a$. Finally, the black hole solution may be uplifted to a full 10d background \cite{Cvetic:1999xp,Chamblin:1999tk} which serves as an input for the probe D3 action. 

For the gravity solutions we consider, the only nonvanishing fields are the 10d metric and 5-form, $(G_{MN}, F_{(5)})$, and the Einstein frame action is
\begin{equation}
    S_{\textrm{IIB}} = \frac{1}{16 \pi G_{10}} \int \! d^{10}x \, \sqrt{-G}\left (R_{10} -\frac{1}{4 \cdot 5!} F_{M_1\dots M_5}F^{M_1\dots M_5} \right )\,.
\end{equation}
Because the 5-form satisfies the self-duality condition $F_{(5)} = \ast F_{(5)}$, we must impose the covariant equations of motion by hand; these have the well known AdS$_5 \times$ $S^5$ flux background as the simplest nontrivial solution~\cite{Schwarz:1983qr}. The supersymmetry properties of a solution may be determined by studying the supersymmetry properties of the 10d gravitino around a given background, though we postpone a discussion of supersymmetry until Appendix \ref{sec:checkSUSY}.

In reducing on the S$^5$ to a 5d effective truncation, we may work in units in which the AdS radius is 1. In our conventions which match \cite{Aharony:2021zkr}, the action of this minimal theory is
\begin{equation}
\label{eq:5dsugraS}
    S_{5d} =  \frac{1}{16 \pi G_{5}} \int  \left[ (R_{5} + 12)\ast 1 -\frac23 F_{(2)} \wedge \ast F_{(2)} + \frac{8}{27} F_{(2)} \wedge F_{(2)} \wedge A_{(1)} \right ] + S_{bdy} \, ,
\end{equation}
where we have included possible Gibbons-Hawking-York boundary terms and counterterms in $S_{bdy}$ \cite{Gibbons:1976ue,Balasubramanian:1999re}.  Also, in this normalization we have $G_5 = \frac{\pi}{2N^2}$, so any bulk solution with nonvanishing renormalized action will contribute at order $N^2$.

In addition to the AdS vacuum solution of \eqref{eq:5dsugraS}, this theory also has charged black hole solutions in which the two AdS$_5$ angular momenta $J_1$ and $J_2$ are unequal. For reasons that will be explained in our field theory analysis, we choose to consider this general class of black hole rather than the more symmetrical case of equal spins. Thus the solutions we study are that of \cite{Chong:2005da}, but see also \cite{Gauntlett:1998fz,Gutowski:2004ez,Gutowski:2004yv,Kunduri:2006ek,Chong:2005hr,Cvetic:2005zi} for related solutions. 
Note however that for the consistency of our truncation, the electric charges satisfy $Q_1 = Q_2 = Q_3$; more general black hole solutions are possible~\cite{Wu:2011gq}, but we leave a discussion of this more general class for future work. Partially motivated by the successes of the superconformal index, there has been more recent work in better understanding the bulk black hole solutions, including their thermodynamics, higher derivative corrections, and novel solutions~\cite{Cassani:2019mms,Larsen:2019oll,Ezroura:2021vrt,Markeviciute:2018yal,Melo:2020amq,Dias:2022eyq,Bobev:2022bjm,Cassani:2022lrk}.

Using the 10d uplift formula \cite{Cvetic:1999xp}, the general black hole metric we consider is
\begin{equation}\label{appmetric10d}
	ds^2 = ds_5^2 + \sum_{a=1}^3 \left(d\mu_a^2 + \mu_a^2 \left(d\phi_a + \frac{2}{3}A_{(1)} \right)^2 \right)\,,
\end{equation}
with
\begin{equation}\label{ds5}
\begin{aligned}
	ds_5^2  = &- \frac{\Delta_\theta \left [ (1+r^2) \rho^2 dt + 2 q \nu \right ]dt }{\Xi_a \Xi_b \rho^2} + \frac{2q \nu w}{\rho^2}  + \frac{f_t}{ \rho^4} \left ( \frac{\Delta_\theta }{\Xi_a \Xi_b}dt -w \right)^2 + \frac{\rho^2}{\Delta_r}dr^2 \\
 &+\frac{\rho^2 }{\Delta_\theta}d\theta^2 +\frac{r^2 + a^2}{\Xi_a} \sin^2 \theta d\phi^2 + \frac{r^2 + b^2}{\Xi_b}\cos^2 \theta d\psi^2\,, 
\end{aligned}
\end{equation}
which includes the definitions
\begin{align}
		 \nu &= b \sin^2 \theta d\phi+ a \cos^2 \theta d\psi \, , \quad\quad \quad \quad   w = a \sin^2 \theta \frac{d\phi}{\Xi_a}+ b \cos^2 \theta \frac{d\psi}{\Xi_b}\,, \\
		 \Delta_r &= \frac{(r^2+a^2)(r^2+b^2)(1+r^2) + q^2 + 2abq}{r^2} - 2m, \quad\quad  \rho^2 = r^2 +b^2 \sin^2 \theta + a^2 \cos^2 \theta , \\
		 f_t &= 2(m+ a b q)\rho^2 - q^2\,, \quad\quad \quad \quad \quad  \Xi_a = 1-a^2\,, \, \quad \quad \Xi_b = 1-b^2 \, , \\ \Delta_\theta &= 1- b^2 \sin^2 \theta - a^2 \cos^2 \theta \,.
\end{align}
Here, the $\{\mu_a, \phi_a\}, \, \, a=1,2,3$ with $\mu_1^2 + \mu_2^2 + \mu_3^2  = 1$ coordinates parametrize the $S^5$ with an effective 5d gauge field:
\begin{equation}
\label{eq:5dgaugefield}
   A_{(1)} = \frac{3q}{2\rho^2 }\left ( \frac{\Delta_\theta dt}{\Xi_a \Xi_b} - w \right) - \alpha dt\,.
\end{equation}
We have included the parameter $\alpha$, which is the holonomy of the 5d gauge field which will be determined by smoothness of the Euclidean solution. 

Other than the metric $g_{\mu \nu}$ and gauge field $A_{(1)}$ (which uplift to a 10d rotating metric on ${\rm S}^5$), we also have a Ramond-Ramond four form gauge potential $C_{(4)}$, which couples to the D3-brane. Its full expression is complicated, and we provide it in Appendix~\ref{sec:checkSUSY} in a different coordinate system~\eqref{eq:C4form}.

The horizon of the black hole is located at $r_+$, the largest positive root of $\Delta_r$. 
This family of solutions is parameterized by four parameters $\{m,a,b,q\}$, where $m$ goes into the mass of the black hole, $a,b$ parameterize the two different angular momenta of black holes on $S^3$, while $q$ is proportional to the equal R-charges of the solution. We could equally parametrize the solutions using $\{r_+,a,b,q\}$, by expressing the mass parameter $m$ through
\begin{equation}\label{mass}
    m = \frac{ (r_+^2 +a^2)(r_+^2 + b^2) (1+r_+^2) + q^2 + 2ab q}{2r_+^2}\,.
\end{equation}

So far the discussion applies to general black holes in this class. There are special limits in which the black hole becomes extremal (in which the inverse temperature $\beta \rightarrow \infty$) or supersymmetric (in which the geometry permits a Killing spinor and the charges satisfy a BPS relation). In Euclidean signature the extremality and supersymmetry conditions are distinct, and we may consider genuine complex metrics which are not extremal but nevertheless supersymmetric.\footnote{The role of complex metrics in computing exact holographic observables is an interesting recent development, but the fundamental status of these geometries is unclear because we lack a definition of the bulk path integral (or an understanding of string theory on complex backgrounds). 
A systematic understanding of which complex saddles to include in the gravitational path integral is yet unknown and serves as one motivation for this work. One proposal for which such metrics are allowed is advanced in \cite{Witten:2021nzp} generalizing a criterion of Kontsevich and Segal \cite{Kontsevich:2021dmb}. 
} In terms of the elementary metric parameters, the supersymmetry condition is
\begin{equation}\label{susycon}
    q = \frac{m}{1+a+b} \, .
\end{equation} 
In Lorentzian signature, such a supersymmetric black hole with no closed timelike curves is automatically extremal~\cite{Chong:2005hr}, and in the end we may choose to take the extremal limit $r_+ \rightarrow r_* \equiv \sqrt{a+b+ab}$.\footnote{The charges of such black holes satisfy an extra constraint, which can be expressed as $m = (a+b)(1+a)(1+b)(1+a+b)$.} Without taking this limit, one in general obtains a solution with a complex action which can nevertheless be compared with a dual field theory partition function with complex fugacities. 

After imposing the supersymmetry constraint (but not the extremality constraint), the solutions are parameterized by three parameters $\{q,a,b\}$ and the charges of the black hole can be expressed as 
\begin{equation}\label{macro}
    \begin{aligned}
     &   E  = \frac{\pi  q (3+ab- (1+a)b^2 - (1+b)a^2)}{4 G_N (1-a)(1-a^2)(1-b)(1-b^2)}\,, \quad \quad & Q  = \frac{\pi q}{2  G_N (1-a^2)(1-b^2)}\, , \\
       & J_1   = \frac{\pi q (2a+b+ab)}{4G_N(1-a)(1-a^2)(1-b^2)}\,, \quad\quad & J_2 = \frac{\pi q (2b+a+ab)}{4G_N (1-b)(1-a^2)(1-b^2)}\,,
    \end{aligned}
\end{equation}
which satisfy the constraint $E = J_1 + J_2 + \frac{3}{2} Q$. 
Equation (\ref{susycon}) leads in general to a pair of complex values if we express the parameter $q$ in terms of $r_+$:
\begin{equation}
\label{eq:appelectriccharge}
    q = -ab + (1+a+b)r_+^2 \mp i r_+ (r_+^2 - r_*^2) \, .
\end{equation}
In the main part of this paper we will focus on the case of the upper sign which we refer to as the first branch of solutions, following the convention in \cite{Aharony:2021zkr}. 

On the first branch of solutions with the supersymmetry condition imposed, the various chemical potentials are found to be complex:
\begin{equation}\label{eq:appchempot}
\begin{aligned}
        \beta &= \frac{2 \pi (a-i r_+)(b-i r_+)(r_*^2+i r_+)}{(r_+^2-r_*^2)\left ( 2(1+a+b)r_+ + i(r_*^2-3 r_+^2)\right)} \, , \quad 
    \Phi = \frac{3 i r_+ (1- i r_+)}{2(r_*^2 + i r_+)} = \alpha  \, , \\
    \Omega_1 &= \frac{(r_*^2 + i a r_+)(1-i r_+)}{(r_*^2 + i r_+)(a- i r_+)} \, , \quad \quad \Omega_2 = \frac{(r_*^2 + i b r_+)(1-i r_+)}{(r_*^2 + i r_+)(b- i r_+)}\, .
\end{aligned}
\end{equation}
The second branch of solutions is obtained by sending $i \rightarrow - i$. Here, we also see the parameter $\alpha$ defined in \eqref{eq:5dgaugefield} is identified with the chemical potential for the electric charge. The chemical potentials in (\ref{eq:appchempot}) are defined with the normalization such that we have the statistical relation
\begin{equation}
\label{eq:quantumstatistical}
    I_{\textrm{Sugra}} = \beta E - S - \beta\Omega_1 J_1 - \beta \Omega_2 J_2 - \beta \Phi Q,
\end{equation}
where $I_{\textrm{Sugra}}$ is the on-shell action of the black hole. In terms of the potentials above, a supersymmetric black hole satisfies the important constraint:
\begin{equation}
\label{eq:susypotentialrelation}
    \beta (1 + \Omega_1 + \Omega_2 - 2 \Phi) = 2 \pi i \, .
\end{equation}

For many questions about the semiclassical properties of the black hole, the 5d analysis is sufficient. In preparation for the addition of probe D3-branes, this solution can be uplifted to 10d for a $C_{(4)}$ field given in \eqref{eq:C4form}. However, only some very specific terms in the expression will have a nonzero pull-back to the worldvolume of the brane we consider, which are the only terms relevant for the evaluation of the brane action. Such terms would involve a 2-form $\alpha_{(2)}$ for which the 5d gauge field equation of motion may be (locally) used to write:
\begin{align}
\label{eq:appalpha2}
\ast_5& F_{(2)} - \frac23 A_{(1)} \wedge F_{(2)} = d \alpha_{(2)} \, , \,  \\
   & \alpha_{(2)} \equiv \frac{3 q}{2 \rho^2}\left ( \frac{\Delta_\theta dt}{\Xi_a \Xi_b} -w\right )\wedge (\nu + \frac23 \alpha dt) - \frac{3 q \cos 2\theta }{4 \Xi_a \Xi_b}\left ( dt \wedge (b d\phi - a d \psi)+d\phi \wedge d \psi \right ) \, .
\end{align}

We now state the main modern result of the AdS$_5$ black hole microstate counting program. Viewing the superconformal index as a grand canonical partition function as in \eqref{eq:grandcanonicalindex}, the AdS/CFT correspondence implies that a suitable bulk calculation of $Z$ should reproduce the microscopic answer, at least at large $N$ where a semiclassical bulk calculation we have discussed is viable. Viewing the computation of $Z$ as a sum over Euclidean semiclassical saddles subject to the supersymmetric boundary condition \eqref{eq:susypotentialrelation}, the dominant contribution is given by the black hole solution. The on-shell action \eqref{eq:quantumstatistical} of the BPS black hole can be expressed into an elegant form that is a rational function of the chemical potentials \cite{Cabo-Bizet:2018ehj}, 
\begin{equation}
    \log Z = - I_{\textrm{Sugra}} = -i \pi N^2 \frac{\Delta^3}{\sigma \tau} \, 
\end{equation}
which matches precisely with the large $N$ limit of the index (\ref{eq:4dindexaction}) in the chosen ensemble with $\Delta_a = \Delta$. This matching was achieved in \cite{Cabo-Bizet:2018ehj,Choi:2018hmj,Benini:2018mlo} following a proposal in \cite{Hosseini:2017mds}, but also see the more comprehensive list of references in Section \ref{ssec:generalmotivation}. 
We will see in Section \ref{sec:branereg} that the action of the D3-brane probe has a similar form as a rational function of the potentials. 

Beyond the leading black hole saddle we have described so far, there are additional (complex Euclidean) solutions which should contribute to the sum over saddles in $Z$. In the full theory, all of these solutions should contribute to the partition function; and further one should include perturbative $1/N$ fluctuations from bulk fields as well as non-perturbative corrections from wrapped D3-branes and potentially other objects consistent with the supersymmetric boundary conditions. Exciting recent progress in this direction was made in \cite{Aharony:2021zkr}, which showed that various solutions of the Bethe Ansatz formulation of the index can be given a gravitational interpretation. In the context of supersymmetric defects, we study some of these other gravitational saddles in Appendix \ref{sec:shifted} and find agreement with the bulk analysis. \cite{Aharony:2021zkr} further identified the bulk dual of certain non-perturbative corrections to the index as D3-branes which wrap certain cycles of the black hole geometry. The brane configurations we study in what follows have many similar features, but the crucial difference is that our branes extend out to the boundary and thus correspond to (surface) operator insertions in the dual. Thus, rather than providing non-perturbative corrections to the black hole saddle which computes the index at large $N$, our branes have the dual description either as a supersymmetric surface operator insertion or as a quantum system which interacts with the black hole \eqref{ds5}.

\subsection{The probe D3-brane and its action}\label{sec:gravaction}

As explained in the introduction, AdS/CFT as derived from type IIB string theory predicts that a D3-brane system with $N$ coincident branes and a single transverse brane (intersecting along a two-dimensional surface, see Table \ref{table:D3}) has two dual descriptions in the decoupling limit; one in terms of a surface operator in ${\rm SU}(N)$ $\mathcal{N} = 4$ SYM and one in terms of a probe brane in a family of asymptotically AdS$_5$ $\times$ $S^5$ spacetimes. At finite temperature but with supersymmetric boundary conditions \eqref{eq:susypotentialrelation}, we are computing the superconformal index with a defect insertion provided the defect preserves the required background supersymmetry. At large $N$ in the bulk semiclassical limit, we expect to be able to approximate the gravitational path integral by saddle point in which we exponentiate the bulk + probe brane saddle with largest action. In the thermal AdS phase, which we will review later, it is already known \cite{Drukker:2008wr,Koh:2008kt} 
that the properly regulated brane action vanishes. Using the (complex) supersymmetric black hole metrics reviewed in Section~\ref{sec:10dgravity}, we compute the probe brane action around this new background. 
In the case of our surface defect insertion, we give an explicit expression for the bulk action for a brane that wraps the Euclidean black hole horizon, and find that it has order $N$ growth in the phase when the black hole saddle dominates.

As we discussed in the introduction, the field theory defect we are considering is supported on a great circle in $S^3$, with $\theta = \frac{\pi}{2}, \, 0 \leq \phi < 2\pi,\, \psi = \textrm{const}$.  In the holographic dual, these together with (\ref{GWconfig}) specify the boundary condition of the D3-brane at infinity. In other words, at the asymptotic boundary of the AdS$_5 \times S^5$, the brane sits at $\theta = \pi/2 ,\, (\psi, \, \phi_2, \, \phi_3) = \textrm{const}, \, \mu_1 = 1, \, \mu_2 =\mu_3=0,$ and fills the $t, \, \phi, \, \phi_1$ directions. As discussed in Section \ref{sec:GWrev}, we allow general holonomies of the worldvolume gauge field (and its dual) along the $S^1\subset S^5$, which correspond to the periodic parameters $\A,\eta$ for the surface defect (see around \eqref{contdefectparam}). They lead to constant gauge fields on the worldvolume which does not contribute to the bulk action, as expected for the defect index,
and thus will be implicit in the following discussion. 

Extending the brane into the bulk which now contains the black hole, we look for a configuration which has the same boundary conditions but a different bulk topology, as in Figure \ref{fig:topology}. It is easy to see from the symmetries of the metric \eqref{appmetric10d} and the gauge fields that the configuration which remains at $\theta = \pi/2, \, \psi, \, \phi_2, \, \phi_3 = \textrm{const}, \mu_1 = 1, \, \mu_2 =\mu_3=0$ and extends along the $t, \, r, \, \phi, \, \phi_1$ directions would be a solution to the equations of motion to the action of the D3-brane, which includes both the Dirac-Born-Infeld and the Wess-Zumino parts\footnote{As mentioned in Section \ref{sec:GWrev}, we should integrate over the holonomies of the worldvolume gauge field along the $t,\phi$ directions and in principle there could exist D3-brane profiles with non-zero worldvolume field strength. Here we assume that the solution with no worldvolume gauge field turned on dominates the integral. This is supported by the match with our field theory analysis where the holonomies are integrated over. See more comments in Section \ref{appC3}. We thank Ofer Aharony for questions about this point.}
\begin{equation}\label{actionbrane}
   S_{\textrm{D3}} = -  T_{\textrm{D3}} 
 \int \left( d^4 x \, \sqrt{- \textrm{det} (h_{\textrm{D3}})}  - P[C_{(4)}]
 \right)\,.
\end{equation}
In (\ref{actionbrane}), $h_{\textrm{D3}}$ is the induced metric on the worldvolume of the brane, and $P[C_{(4)}]$ is the pullback of the $C_{(4)}$ gauge field on the worldvolume. The coefficient of the action (\ref{actionbrane}) grows with $N$, this can be seen using our conventions for the 5d and 10d supergravity actions \eqref{eq:5dsugraS} upon restoring string units:
\begin{equation}
    T_{\textrm{D3}} = \frac{1}{g_s (2\pi)^3 (\alpha')^2} =\frac{N}{2\pi^2} \, , \quad \textrm{with} \quad N = \frac{1}{4 \pi g_s (\alpha')^2}\,.
\end{equation}

In the following we will simply use $\{ t,r,\phi,\phi_1\}$ as the worldvolume coordinates for the D-brane.  See Figure~\ref{fig:sketch} for a sketch of the brane configuration on the Euclidean black hole geometry.  We verify in Appendix \ref{sec:checkSUSY} that this configuration is supersymmetric, given that the black hole in the background is supersymmetric.

    \begin{figure}[t!]
    \begin{center}
   \includegraphics[scale=.33]{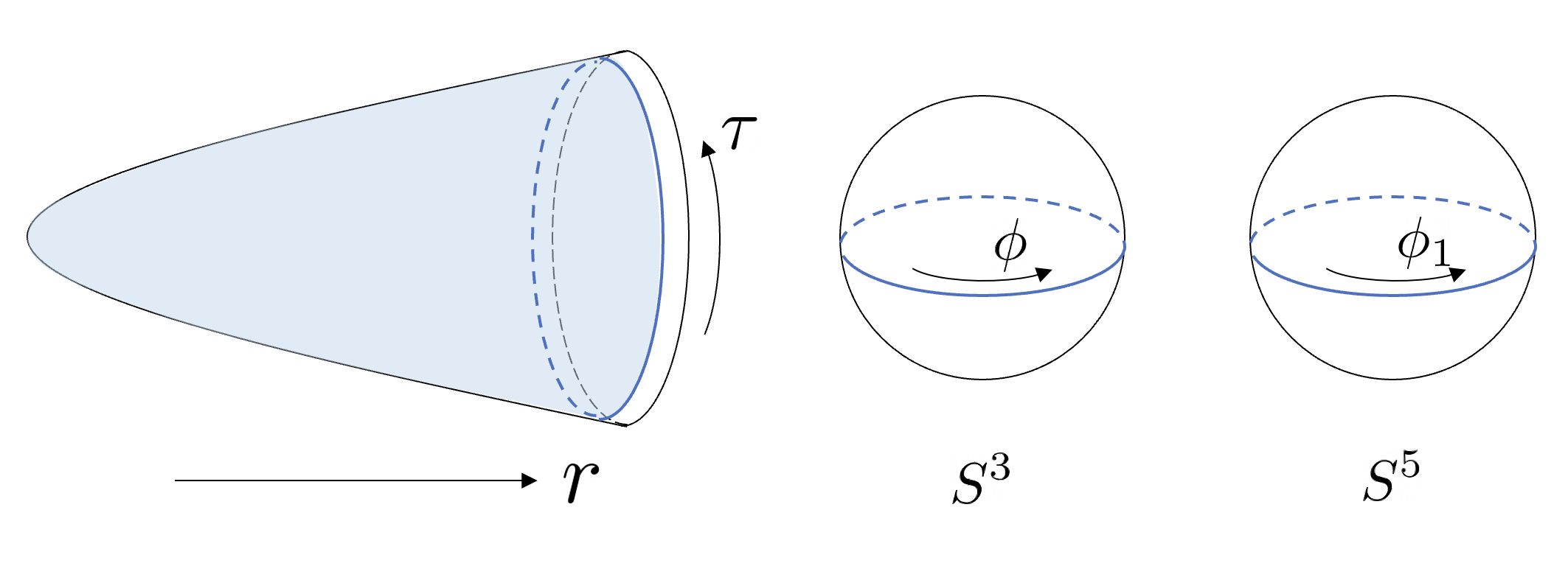}
    \end{center}
    \caption{Here we sketch the brane configuration (shown in blue) in the black hole geometry. It wraps the cigar part of the Euclidean black hole, while also wrapping two great circles, one in $S^3$ and the other in ${\rm S}^5$. }
    \label{fig:sketch}
\end{figure}

The evaluation of the DBI term before regulating follows from taking the determinant of the induced metric of \eqref{appmetric10d} on the worldvolume. This is obtained in \eqref{eq:induceddet} in the process of checking the supersymmetry condition; the action is:
\begin{equation}
\label{eq:appSDBI}
    S_{\textrm{DBI}} = -\frac{N}{2\pi^2} \int \! d^4x \left ( \frac{r}{1-a^2}\right) \, .
\end{equation}
The calculation of the WZ term requires a choice of $C_{(4)}$. The expression of $C_{(4)}$ is readily expressed in \cite{Aharony:2021zkr} using the so called ``orthotoric'' coordinates \cite{Cassani:2015upa}, which we reproduce in Appendix. \ref{sec:checkSUSY}. Specifically, we will use \eqref{eq:C4form} which solves the 10d equations of motion and is nonsingular along the brane worldvolume. 

To evaluate the pullback of this form to the brane worldvolume extending along $(t, r, \phi, \phi_1)$ in the original coordinates \eqref{ds5}, we make use of \eqref{eq:appalpha2} and the explicit expression \eqref{eq:C4form} for $C_{(4)}$ in our chosen gauge. The pullback then simplifies to
\begin{equation}
\label{eq:appPC4}
    P[C_{(4)}] =P\left[ \frac13 d\phi_1 \wedge d\alpha_{(2)} \right] = -\frac{qr (3b + 2 a \alpha)}{3(1-a^2)(r^2 + b^2)^2}dt\wedge dr \wedge d\phi \wedge d \phi_1 \, .
\end{equation}
Owing to the presence of $q$ and $\alpha$ in this expression, this form is typically complex, leading to a complex action for the D-brane. Only in the extremal limit do these quantities approach their real BPS values, but as in the calculation of the DBI term and supersymmetry projector, we need \emph{not} take the extremal limit ($r_+ \rightarrow r_* = \sqrt{a+b+ab}$) while still preserving supersymmetry in Euclidean signature. For reference, the BPS values are:
\begin{equation}
    q \rightarrow q_* = -ab + (1+a+b)r_*^2 \, , \quad \alpha \rightarrow \alpha_* = \frac32 \, .
\end{equation}
For a generic supersymmetric but not necessarily extremal black hole, the brane action \eqref{actionbrane} is obtained by combining \eqref{eq:appSDBI} and \eqref{eq:appPC4}:
\begin{equation}
    S_{D3} =  -\frac{N}{2\pi^2} \int dt dr d\phi d\phi_1 \left (\frac{r}{1-a^2} + \frac{qr (3b + 2 a \alpha)}{3(1-a^2)(r^2 + b^2)^2} \right ) \, .
\end{equation}
This is the action in the Lorentzian spacetime. To compare the action with the index, we would like to evaluate the action in the Euclidean signature, with $t =-i\tau,\, \tau \sim \tau + \beta$. 
Through Wick rotation, the Euclidean action reduces to
\begin{equation}\label{SD3E}
\begin{aligned}
    I_{D3,\textrm{E}}  = -i   S_{D3}&  = \frac{N \beta}{2\pi^2} \int dr d\phi  d\phi_1 \, \left[ \frac{r}{1-a^2} +   \frac{qr (3b + 2 a \alpha)}{3(1-a^2)(r^2 + b^2)^2}   \right]\\ & 
    = 2N\beta \int_{r_+}^{r_{\infty}} dr  \, \left[ \frac{r}{1-a^2} +  \frac{qr (3b + 2 a \alpha)}{3(1-a^2)(r^2 + b^2)^2} \right],
\end{aligned}
\end{equation}
where we used the fact that the angles $\phi,\phi_1$ have period $2\pi$.\footnote{More precisely, one has to evaluate the integrals of $\{\tau,\phi,\phi_1\}$ with twisted identifications $(\tau,\phi) \sim (\tau + \beta,\phi - i \beta \Omega),$ but as usual this twisting has no effect on the volume.} We use $I$ to denote the value of the Euclidean action, in order to distinguish it from $S$ for entropy.

\subsection{Subtracting divergences and imposing supersymmetry}\label{sec:branereg}

In (\ref{SD3E}), we introduced a radial cutoff $r_{\infty}$, which upon sending to infinity leads to a divergence due to the volume term. In principle, one should include suitable boundary terms and apply holographic renormalization to get a finite answer. In the following we will employ a different strategy; from the naively divergent answer we subtract the action of a brane in empty thermal AdS which satisfies the same asymptotic boundary conditions. More precisely, we consider a brane in a thermal AdS background, but with the coordinates identified at infinity in the same way as in the black hole geometry. The intuition is that at large radial coordinate, the black hole metric asymptotically approaches the thermal AdS metric, so the naive UV divergence should be common to both brane probes.

Taken at face value, what we are evaluating is the difference in the action of the brane in the black hole and thermal AdS background. However, the classical action of the brane in the empty AdS background has been studied in the literature \cite{Drukker:2008wr} and was found to be zero. We expect the brane action should continue to be zero even in the case that the coordinates are identified non-trivially at infinity, and given this holds, we can also interpret our final answer as the finite absolute value of the brane action in the black hole background, which one would presumably get by doing a careful holographic renormalization analysis. Both these perspectives will be consistent with our field theory analysis in Section \ref{sec:field theory}.

The empty AdS space has the metric
\begin{equation}\label{emptymet}
    ds_{\textrm{AdS}}^2 = (1 + \tilde{r}^2) d\tilde{\tau}^2 + \frac{d\tilde{r}^2}{1+\tilde{r}^2} + \tilde{r}^2 (d\tilde{\theta}^2 + \sin^2 \tilde{\theta} d\tilde{\phi}^2 + \cos^2 \tilde{\theta} d\tilde{\psi}^2) + d\Omega_5^2\,, \quad \tilde{\tau} \sim \tilde{\tau} + \beta\,.
\end{equation}
Note that we are adding tildes on the coordinates to emphasize that this is a new auxiliary spacetime that we introduced and is distinct from the black hole spacetime we are interested in. Note also that in terms of these coordinates, the metric is diagonal, but with non-trivial coordinate identifications according to the chemical potentials for the rotations and the electric charges. Such identifications will not affect the action of the brane so we don't spell them out in detail.

We consider the brane that shares the same boundary conditions as the black hole case. In the bulk, the brane simply extends in the $r$ direction, with other angles remaining the same (see Figure \ref{fig:topology} (a) for an illustration). The four form potential $C_{(4),\textrm{AdS}}$ does not contribute to the action in this case.
Then the Euclidean action of the brane is simply given by
\begin{equation}\label{Sempty}
       I_{D3,\textrm{E},\textrm{empty AdS}} = 2N \beta \int_{0}^{\tilde{r}_{\infty}} d\tilde{r}   \, \tilde{r} = N \beta \tilde{r}_\infty^2 \, .
\end{equation}
We also note that (\ref{Sempty}) is nothing but (\ref{SD3E}) after taking the formal limit in which the black hole has vanishing size and charges $r_+, \, a, \, b,\, q \rightarrow 0$. However, an important difference is that in (\ref{Sempty}) we are cutting the space at a different radial cutoff $\tilde{r}_{\infty}$. This has to be fixed such that when we do the background subtraction, we are comparing two spaces where the induced metric is the same on the two cutoff surfaces.

Passing to an asymptotically static coordinate system for the black hole,
\begin{equation}
    \hat{\phi_1} = \phi_1 - \frac23 \alpha t \, ,
\end{equation}
the induced metric on the cutoff surface to leading order in $r_\infty$ is
\begin{equation}
    ds^2_{\textrm{BH};\infty}\approx -\frac{(1+r_\infty^2)}{(1-a^2)} dt^2 + \frac{(a^2+r_\infty^2)}{(1-a^2)} d\phi^2 + d\hat{\phi}_1^2 + \mathcal{O}\left (\frac{1}{r_\infty^2} \right ) \, .
\end{equation}
If we now consider a D3-brane which wraps the same cycles but in empty AdS, the induced metric at the cutoff surface at $\tilde{r}_\infty$ is
\begin{equation}
    ds^2_{\textrm{AdS};\infty} = -(1+\tilde{r}_\infty^2)d\tilde{t}^2 + \tilde{r}_\infty^2 d\tilde{\phi}^2 + d\tilde{\phi}_1^2 \, .
\end{equation}
These can be made to agree for 
\begin{equation}\label{newrinf}
    \tilde{r}_\infty^2 = \frac{r_\infty^2+a^2}{1-a^2}.
\end{equation} 
Therefore, combining (\ref{newrinf}) with (\ref{SD3E}) and (\ref{Sempty}), we get
\begin{equation}\label{ID3nonSUSY}
\begin{aligned}
    I_{D3,\textrm{E, finite}} & = I_{D3,\textrm{E}} -  I_{D3,\textrm{E},\textrm{empty AdS}} \\
    &=  2 N \beta  \int_{r_+}^{r_\infty} dr  \left (\frac{r}{1-a^2} + \frac{qr (3b + 2 a \alpha)}{3(1-a^2)(r^2 + b^2)^2} \right )  -  N \beta \frac{r_\infty^2+a^2}{1-a^2} \, , \\
    &=  - \frac{N \beta}{1-a^2}  \left ( (a^2 + r_+^2) - \frac{q(3b + 2 a \alpha)}{3(r_+^2 + b^2)}\right)\, .
\end{aligned}
\end{equation}
We now insert the supersymmetry relations and expressions for the chemical potentials appropriate for the first branch of solutions via \eqref{eq:appelectriccharge}, \eqref{eq:appchempot}. This leads to a temperature independent (but complex) answer for the brane action:
\begin{equation}
   I_{D3,\textrm{E, finite}} = -2 \pi i N\frac{(a-i r_+)^2(b-i r_+)}{(1-a)(a+b+a b - 2i (1+a+b)r_+ - 3 r_+^2)}.
\end{equation}

We may now finally write the action of the brane in terms of the field theory variables. These are reduced chemical potentials which track the deviation from the BPS values, and are given as
\begin{equation}
    \sigma = \frac{\beta}{2\pi i}(\Omega_1-1) \, , \quad \tau = \frac{\beta}{2\pi i}(\Omega_2-1) \, , \quad \Delta = \frac{\beta}{2\pi i}\left(\frac{2}{3}\Phi-1 \right) \, ,
\end{equation}
which satisfy the supersymmetry constraint 
\begin{equation}
    \sigma + \tau - 3 \Delta =  1 \, .
\end{equation}
In terms of these on the first branch, the brane action may be written as
\begin{equation}\label{SD3final}
    I_{D3, \textrm{E, finite}} = -2\pi i N \frac{(\sigma + \tau - 1)^2}{9 \sigma} \, .
\end{equation}
This is our final result for the action of the D3-brane when the black hole has unequal $J_1 \neq J_2$ but all equal electric charges $Q$.

Finally, let us mention that we could repeat the calculation but with the second branch of solutions in (\ref{eq:appelectriccharge}) and we would get a similar answer with $\sigma + \tau - 1$ in (\ref{SD3final}) replaced by $\sigma + \tau + 1$.

\subsection{Explicit evaluation of the brane action}\label{sec:physical}

The final expression (\ref{SD3final}) that we derived applies to general complex parameters $\sigma,\tau$, but it is interesting to examine its behavior for parameters that correspond to physical black holes in the Lorentzian signature. 

\begin{figure}[t!]
    \begin{center}
   \includegraphics[scale=.35]{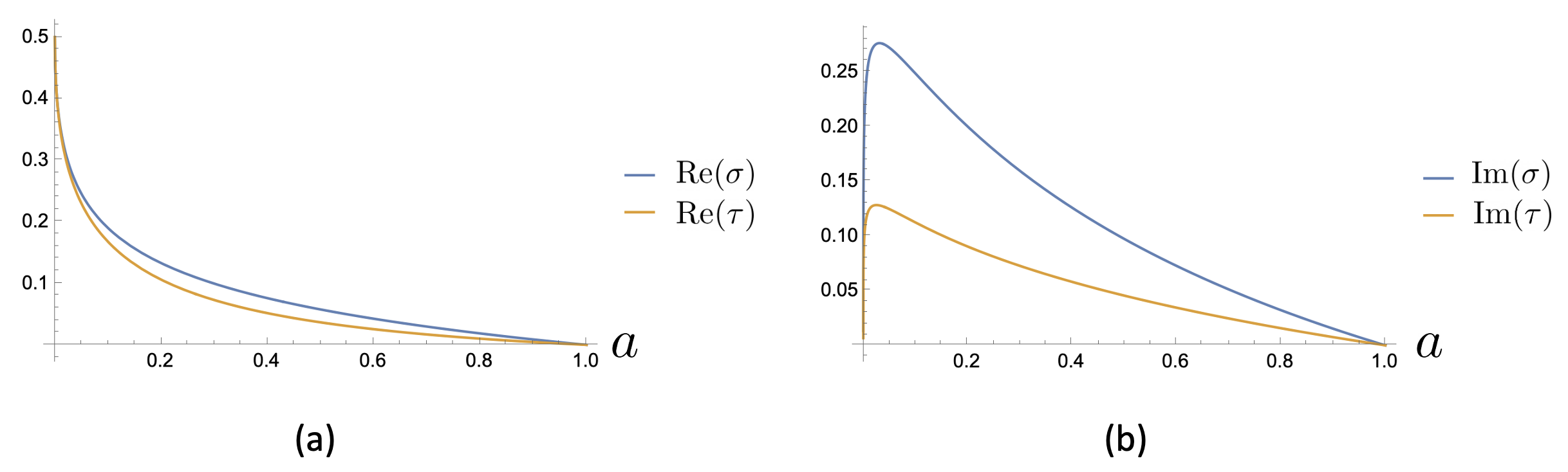}
    \end{center}
    \caption{We plot the (a) real and (b) imaginary parts of $\sigma, \tau$ for the specific family of supersymmetric black holes satisfying $J_2 = 2 J_1$.}
    \label{fig:example1}
\end{figure}

As an explicit example, we can look at black holes with $J_2 =2  J_1 $. From (\ref{macro}), this corresponds to black holes with parameters satisfying
\begin{equation}
    \frac{2b+a + ab}{1-b} = 2 \times \frac{2a+b+ab}{1-a} \, ,
\end{equation}
from which we get
\begin{equation}
    b = \frac{a(a-1) + \sqrt{a^4 + 6a^3 + 33 a^2 + 24 a}}{4 (1+a)}\, .
\end{equation}
We then have a one-parameter family of solutions labelled by $0<a<1$. In Figure~\ref{fig:example1}, we plot how the parameters $\sigma, \tau$ depend on $a$.

\begin{figure}[t!]
    \begin{center}
   \includegraphics[scale=.35]{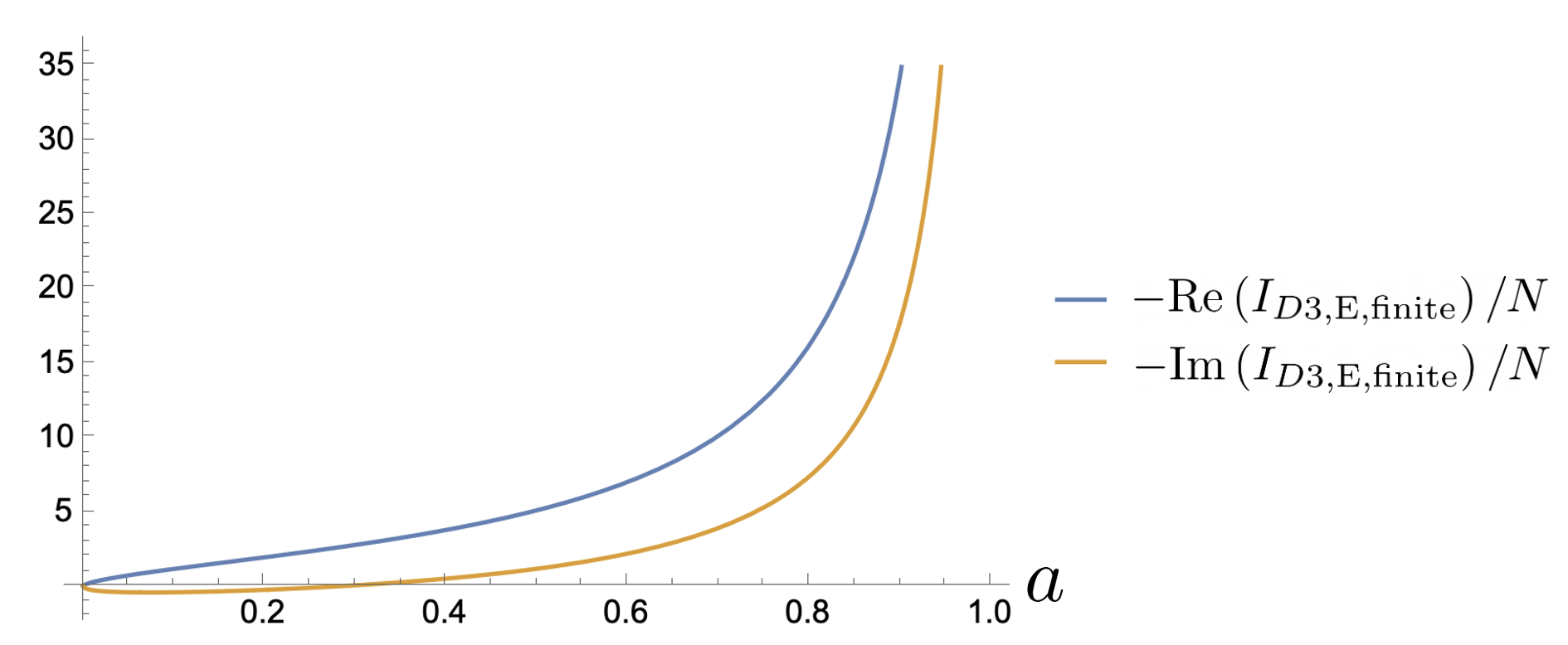}
    \end{center}
    \caption{We plot (minus) the action of the D3-brane as a function of $a$. We see that the action vanishes at $a=0$ while it diverges as $a\rightarrow 1$. }
    \label{fig:action}
\end{figure}

We can also plot the action of the D3-brane given by (\ref{SD3final}) as a function of $a$, see Figure~\ref{fig:action}. Note that $a=0$ is a special limit in which the black hole has vanishing charges and vanishing size. In this limit we have $\tau = \sigma = 1/2$, and the D3-brane action (\ref{SD3final}) becomes zero. This is a consistency check that we expect the action to be zero for the brane in the thermal AdS background, which can be also viewed as a formal limit where we take the black hole to have vanishing size.

 Notice that the action of the D3-brane action is complex even though the extremal black hole metric is real. Mathematically there is no contradiction since the direct calculation at extremality is singular since $\beta=\infty$. Instead, the way we evaluated the brane action was to consider a \emph{complex} geometry that is supersymmetric but is away from extremality.

\subsection{Thermodynamics of the D3-brane in the index ensemble}\label{thermo}

So far, we have been treating the D3-brane quite abstractly as a probe of the background black hole spacetime. However, in the Lorentzian spacetime of the extremal black hole (see Figure~\ref{fig:penrose}), we have a black hole in equilibrium with a D-brane in its environment, and therefore we can discuss the physical properties of this combined system, in particular its thermodynamics. 

One important caveat, however, is that in this paper we are studying the index, rather than the partition function, therefore at the level of computation, the thermal ensemble differs from the standard one in that some chemical potentials are \emph{complex}, a feature we emphasized in the introduction. This introduces subtleties in comparing directly with Lorentzian signature, and the various quantities calculated below are specific to the ``index ensemble'' but not obviously the Lorentzian system.

In the probe limit, the black hole degrees of freedom and the D3-brane separate cleanly, and therefore we can simply focus on the D3-brane when discussing its own thermodynamics. It is convenient to start with the expression (\ref{ID3nonSUSY}) before we impose the supersymmetric condition. The various conserved charges of the D3-brane can be derived by thinking of the action $ I \equiv I_{D3,\textrm{E, finite}}$ as a thermodynamic function of $\beta, \Omega_1 , \Omega_2 , \Phi$, and compute
\begin{equation}
\begin{aligned}
   & E_{\textrm{D3}} = \frac{\partial I}{\partial \beta} + \Omega_1 J_{1,\textrm{D3}} + \Omega_2 J_{2,\textrm{D3}} + \Phi Q_{\textrm{D3}}\, , \\
   & J_{1,\textrm{D3}} = - \frac{1}{\beta} \frac{\partial I}{\partial \Omega_1}\,, \quad J_{2,\textrm{D3}} = - \frac{1}{\beta} \frac{\partial I}{\partial \Omega_2}\,, \quad Q_{\textrm{D3}} = - \frac{1}{\beta} \frac{\partial I}{\partial \Phi}\, , \\
   & S_{\textrm{D3}} = - I + \beta E_{\textrm{D3}} - \beta \Omega_1  J_{1,\textrm{D3}} - \beta \Omega_2  J_{2,\textrm{D3}} -\beta \Phi Q_{\textrm{D3}}\, . 
\end{aligned}
\end{equation}
Applying this procedure, we arrive at some fairly complicated expressions which we won't record here. However, significant simplifications take place  once we enforce the supersymmetric conditions (\ref{eq:appelectriccharge}) (focusing on the first branch of solutions), and we obtain the energy and entropy in the generalized ensemble
\begin{equation}\label{ESD3}
    \begin{aligned}
         E_{\textrm{D3}} & =   J_{1,\textrm{D3}}  +   J_{2,\textrm{D3}}  + \frac{3}{2} Q_{\textrm{D3}}\,, \quad \quad
 S_{\textrm{D3}}  = \frac{4 \pi N (a-ir_+) (r_*^2 + i r_+) r_+ }{\Xi_a  (  r_+^2 - 2i (1+a+b)r_+ + r_*^2    ) }\,,
    \end{aligned}
\end{equation}
with the expressions for the D3-brane charges $J_{1,\textrm{D3}},J_{2,\textrm{D3}},Q_{\textrm{D3}}$ given by
\begin{equation}
\begin{aligned}
    \frac{J_{1,\textrm{D3}}}{N} & = \frac{r_+^2}{(1-a)^2} + \frac{2 (1+a^2) i r_+}{ \Xi_a (1-a)} \\ 
     \quad  - &\frac{ ((2-a)^2 (1+a) + 2(1-a)^2 b) r_+^2 + 2i (b - a (a(3+a+2b+a^2) + b -1)) r_+  - a^2 (1+a) r_*^2}{\Xi_a (1 - a) (r_+^2 - 2i (1+a+b)r_+ + r_*^2 )}\,,  \\
    \frac{J_{2,\textrm{D3}}}{N} & = \frac{2i (a-ir_+) (r_*^2 + ir_+) r_+}{\Xi_a (r_+^2 - 2i (1+a+b)r_+ + r_*^2 ) }\,,\\
     \frac{ Q_{\textrm{D3}}}{N} & = \frac{ 4i (a-2) r_+ }{3\Xi_a} \\
     \quad - &\frac{ 4 (-3(2+b) +a (-5+a+3b)) r_+^2 - 4i (a(5+a+2a^2) + (3-a) (1+a)b ) r_+ 
 + 4a (1+a) r_*^2 }{3\Xi_a (  r_+^2 - 2i (1+a+b)r_+ + r_*^2    )}\,.\\
\end{aligned}
\end{equation}
Note that the fact that the charges satisfy the BPS condition in (\ref{ESD3}) can be viewed as a consistency check of our analysis. It would be interesting to see whether the entropy $S_{\textrm{D3}}$ can be expressed in terms of some simple geometric quantities, which can be presumably derived via a Lewkowycz-Maldacena type argument \cite{Lewkowycz:2013nqa} generalized to brane action, similar to the case for a Wilson loop \cite{Lewkowycz:2013laa}. We note that the entropy $S_{\textrm{D3}}$ is an entropy at the classical level (of order $N$) and does not have an obvious state-counting interpretation in the bulk. In this sense, it is almost as mysterious as the black hole entropy itself.\footnote{We thank Juan Maldacena for comments on this point.}

We can further consider the extremal limit $r_+ \rightarrow r_* = \sqrt{a+b+ab}$. In contrast to the original black hole thermodynamic quantities which become real in the BPS limit, here the charges and the entropy remain complex even in this limit. This does not seem to be a problem \emph{a priori} since we are considering an ensemble with complex chemical potentials. However, it is an interesting question to determine the correct prescription to extract the thermodynamic quantities of the brane in Lorentzian signature. At the same time, one could perform an independent analysis of the physical properties of the brane in the Lorentzian background within the supergravity approximation. We hope to return to these problems soon in future work.

\section{Field Theory Analysis}\label{sec:field theory}

\subsection{Set-up of the calculation}

From the boundary point of view, the probe D3-brane is holographically dual to a certain half-BPS Gukov-Witten surface operator $\cD$ inserted in the 01-plane (see Section~\ref{sec:GWrev} for a review). This surface operator preserves a half-BPS subalgebra of the $\mathcal{N}=4$ superconformal symmetry, given by the 2d centrally-extended $\mathcal{N}=(4,4)$ superconformal algebra \eqref{defectalg}. In this section, we analyze the defect index $\cI_{\cD}$ that generalizes the 4d superconformal index with the insertion of this surface operator. As was shown in \cite{Gukov:2006jk,Gadde:2013dda}, the superconformal surface defect can be effectively described by the defect renormalization group (RG) fixed point of a supersymmetric 4d-2d coupled system. For the simplest GW surface defect in ${\rm SU}(N)$ SYM, labelled by Levi-type $L=[1,N-1]$ (see \eqref{Levi}), which corresponds to a single probe D3-brane in the bulk, the 4d-2d system is described by  a 2d $\cN=(4,4)$ ${\rm U}(1)$ gauge theory coupled to the 4d ${\rm SU}(N)$ gauge group by a 2d bifundamental hypermultiplet (namely an $\cN=(4,4)$ SQED on the defect worldvolume). 
Since the superconformal index is protected along the RG flow, we can calculate the index in the UV where the 2d fields become free. Consequently, at the level of the ${\rm SU}(N)$ matrix integral formula for the $\mathcal{N}=4$ superconformal index \eqref{4dSYMUmm}, 
the surface defect insertion amounts to 
simply inserting the index for the 2d fields as shown below,
\ie 
\label{eq:defectindexintegral}
\cI_{\cD}(p,q,y_i)={1\over N!}\int 
\left(\prod_{i=1}^{N-1}du_i\right)\mathcal{I}_{4d}(p,q,y_i;u_i)\mathcal{I}_{2d}(p,q,y_i;u_i)\,,
\fe 
where we have packaged the contributions from the 4d fields in $\mathcal{I}_{4d}$.

The field content of the 2d GLSM (here given by a SQED) that engineers the surface defect includes one $\cN=(4,4)$ hypermultiplet in the fundamental representation of ${\rm SU}(N)$, which is often written as a pair of $\cN=(2,2)$ chiral multiplets $(B,\Tilde{B})$ which transform in the fundamental and anti-fundamental representations of ${\rm SU}(N)$, respectively. In addition, the 2d GLSM involves a $\cN=(4,4)$ ${\rm U}(1)$ vector multiplet which, in $\cN=(2,2)$ notation, is given by $(\mathcal{S},\cV)$ for a chiral multiplet $\mathcal{S}$ and an $\cN=(2,2)$ vector multiplet $\cV$ (equivalent to a twisted chiral multiplet). The complete field content of these multiplets can be obtained by expanding the corresponding superfields in $\cN=(2,2)$ superspace $(\theta^{\pm},\bar{\theta}^{\pm})$ \cite{Constable:2002xt}: 
\begin{align}
    B=& b+\sqrt{2}\theta^+\psi^b_++\sqrt{2}\theta^-\psi_-^b-2\theta^+\theta^-F^b \,,\\
    \tilde{B}=& \tilde{b}+\sqrt{2}\theta^+\psi^{\tilde{b}}_++\sqrt{2}\theta^-\psi_-^{\tilde{b}}-2\theta^+\theta^-F^{\tilde{b}} \,,\\
    \mathcal{S} =& s+\sqrt{2}\theta^+\psi^{s}_{+}+\sqrt{2}\theta^-\psi_-^{s}-2\theta^+\theta^-F^{s} \,,\\
    \mathcal{V}=& \theta^- \bar{\theta}^-(\tilde{v}_0-\tilde{v}_1)+\theta^+\bar{\theta}^+(\tilde{v}_0+\tilde{v}_1)-\theta^-\bar{\theta}^+\omega-\theta^+\bar{\theta}^-\bar{\omega}\nonumber\\
    &+i\sqrt{2}\theta^-\theta^+(\bar{\theta}^-\bar{\lambda}_-+\bar{\theta}^+\bar{\lambda}_+)+i\sqrt{2}\bar{\theta}^+\bar{\theta}^-(\theta^-\lambda_-+\theta^+\lambda_+)\nonumber\\
    &+2\theta^-\theta^+\bar{\theta}^+\bar{\theta}^-(D-i(\partial \cdot \tilde v))\, .
\end{align}
The bosonic charges of these fields are summarized in Table~\ref{table:2dspectrum} and \ref{table:SVcharges} in Appendix \ref{sec:diffnotations}.

We expect the large $N$ limit of the defect index $\cI_\cD$ and the defect expectation value $\la \cD\ra$ to be dominated by contributions from the hypermultiplet fields $(B,\tilde{B})$ because their number grows with $N$. Nevertheless, 
including the contributions from the vector multiplet $(S,\mathcal{V})$ is crucial to ensure the correct periodicity property of the 2d contribution $\cI_{\rm 2d}$ to the defect index in \eqref{eq:defectindexintegral} and eventually to successfully match with the gravity answer.\footnote{See Appendix~\ref{app:nakayama} for related comments.}

To calculate the contributions from the GLSM fields localized on the temporal torus $T^2\subset S^1\times S^3$ to the defect index, we take advantage of existing results on the elliptic genus for 2d supersymmetric QFTs.
Here, with $\cN=(4,4)$ supersymmetry, the fully-refined elliptic genus for the GLSM fields is defined as the following trace in the RR sector (see Appendix~\ref{sec:diffnotations} for details),
\begin{equation}   \label{IRR} \mathcal{I}_{\text{RR}}=\text{Tr}_{\text{RR}}\, (-1)^F e^{2\pi i\tau_{\text{2d}}L_0}e^{-2\pi i\bar{\tau}_{\text{2d}}\bar{L}_0}e^{2\pi iz_{\text{R}}J_0}e^{2\pi i \chi J_A}e^{2\pi i u C}\,,
\end{equation}
where $\tau_{2d}$ is the complex structure of the temporal $T^2$. The bosonic charges in \eqref{IRR} are written in 2d notation: $L_0,\bar L_0$ are the left- and right- moving Hamiltonians, $J_0$ is the Cartan generator of the left-moving $\mf{su}(2)$ R-symmetry, $J_A$ generates a flavor symmetry of the GLSM, and $C$ is the central charge in \eqref{centralc}. Further details including the explicit relations to the Cartan generators of the $\cN=4$ superconformal algebra can be found in Appendix~\ref{sec:superalgebra}. Note that $\bar\tau_{\rm 2d}$ is the usual regulator and the elliptic genus does not depend on it.
 
The elliptic genus of 2d $\cN=(4,4)$ SQED has been computed in \cite{Gadde:2013dda,Benini:2013nda, Benini:2013xpa}. 
We write the answer in our conventions here and give a review of the derivation in Appendix \ref{sec:diffnotations},
\begin{equation}\label{EGfinal}
    \mathcal{I}_{\text{RR}} = \sum_{i=1}^N\prod_{i\neq j} \frac{\theta_1(\tau_{\text{2d}}|u_{ij}+z_{\text{R}}-2\chi)\theta_1(\tau_{\text{2d}}|u_{ij}-z_{\text{R}})}{\theta_1(\tau_{\text{2d}}|u_{ij})\theta_1(\tau_{\text{2d}}|u_{ij}-2\chi)}\,,
\end{equation}
where $u_{ij}\equiv u_i-u_j$ come from the $SU(N)$ holonomies and $\theta_1(\tau_{\rm 2d}|z)$ is the ordinary Jacobi Theta function (see Appendix \ref{sec:convention} for definitions and useful relations for Theta functions).

To obtain the 2d index that contributes to  \eqref{eq:defectindexintegral}, we work with an index in the NSNS sector of the 2d fields,
\begin{equation}\label{trace2d}
    \mathcal{I}_{\text{NSNS}} = \text{Tr}_{\text{NSNS}}\, (-1)^F e^{2\pi i \tau_{\text{2d}}L_0}e^{-2\pi i \bar{\tau}_{\text{2d}}(\bar{L}_0-\frac{1}{2}\bar{J}_0)}e^{2\pi i z_{\text{NS}}J_0}e^{2\pi i\chi J_A}e^{2\pi i u C}\,.
\end{equation}
The BPS condition translates into $\Bar{L}_0-\frac{1}{2}\bar{J}_0+\frac{C}{2}=0$ in the 2d notation and since all 2d fields are uncharged under $C$ (see Table~\ref{table:2dspectrum} and \ref{table:SVcharges}), the index is independent of $\bar{\tau}_{2d}$ as expected. Below we will obtain 
the 2d index $\cI_{\rm 2d}$ in \eqref{trace2d} from the elliptic genus \eqref{EGfinal} by spectral flow.

The $\mathcal{N}=2$ superconformal algebra has a nontrivial outer-automorphism generated by the spectral flow with parameter $\eta \in \frac{\mathbb{Z}}{2}$ \cite{Schwimmer:1986mf}, 
under which the bosonic generators transform as 
\begin{equation}
    L_n \rightarrow L_n +\eta J_n+\frac{c}{6}\eta^2 \delta_{n,0}\,,\quad J_n\rightarrow J_n+\frac{c}{3}\eta\delta_{n,0}\,,
\end{equation}
where $c$ is the conformal central charge. In particular, the spectral flows with $\eta = \pm \frac{1}{2}$ connect the $\text{NS}$ and $\text{R}$ sector algebras. Consequently, we have the following relation between the NSNS index and the elliptic genus,
\begin{equation}
    \mathcal{I}_{\text{NSNS}}(\tau_{\text{2d}},z_{\text{NS}},\chi,u) = e^{2\pi i \tau_{\text{2d}}(\frac{-c}{24})}e^{-\frac{\pi ic}{3}(z_{\text{NS}}-\frac{\tau_{\text{2d}}}{2})}\mathcal{I}_{\text{RR}}\left(\tau_{\text{2d}},z_{\text{NS}}-\frac{\tau_{\text{2d}}}{2},\chi,u \right)\,,
\end{equation}
where we identify $z_{\text{R}}$ with $z_{\text{NS}}-\frac{\tau_{\text{2d}}}{2}$ (i.e. $\eta = -\frac{1}{2}$). By implementing a further rescaling of $\mathcal{I}_{\text{NSNS}}$, such that the $q_{\text{2d}}\equiv e^{2\pi i\tau_{\text{2d}}}$-expansion starts with $q_{\text{2d}}^0$ (i.e. the ground state contributes $1$ in the $q_{\text{2d}}$-expansion), we obtain the 2d index $\mathcal{I}_{\text{2d}}$ needed in \eqref{eq:defectindexintegral},
\begin{equation}\label{I2dIRR}
    \mathcal{I}_{\text{2d}}(\tau_{\text{2d}},z_{\text{NS}},\chi,u) = e^{-\frac{\pi i c}{3}z_{\text{R}}}\mathcal{I}_{\text{RR}}(\tau_{\text{2d}},z_{\text{R}},\chi,u)\,.
\end{equation}
Here, as a consequence of the supersymmetry, the central charge of the 2d GLSM coincides with the defect central charge \eqref{c2d}, namely  $c = b_{\text{2d}} = 6(N-1)$ \cite{Chalabi:2020iie,Wang:2020xkc}.
Together with the relation between the 2d chemical potentials and the 4d ones from \eqref{4d2dcpr}
\begin{equation}
    \tau_{\text{2d}} = \sigma\,,\quad z_{\text{NS}} = \Delta_1+\Delta_2-\tau-\frac{\sigma}{2}\,,\quad u=\frac{\sigma-2\tau}{2}\,,\quad \chi = \frac{\Delta_1-\tau}{2}\,,
\end{equation}
we arrive at the following expression for the 2d index,
\begin{align}\label{I2dTachikawa}
    \mathcal{I}_{2d} 
    =&  \sum_{i=1}^N \prod_{j\neq i} \frac{\theta_0(\sigma|u_{ji} -\Delta_2 + \sigma)\theta_0(\sigma|u_{ij} +\sigma+\tau-\Delta_1-\Delta_2)}{\theta_0(\sigma|u_{ji}-\tau+\Delta_1) \theta_0(\sigma|u_{ij})}\,.
\end{align}

In the subsequent sections, we  will evaluate the matrix integral
\eqref{eq:defectindexintegral} together with \eqref{I2dTachikawa} in the large $N$ limit using a saddle point approximation. In other words, we use the large $N$ saddle points of the holonomies $u_i$ that are determined by the 4d index $\mathcal{I}_{4d}$. We only use the saddle point equations from the 4d fields because their contribution to the matrix potential is order $N^2$; the contribution from the 2d fields is order $N$ and thus suppressed.

Formally, (\ref{I2dTachikawa}) is defined for general chemical potentials $\tau,\sigma,\Delta_1 , \Delta_2 , \Delta_3$. However, we will restrict ourselves to choices of chemical potentials such that when all the holonomies $u_{i} \in \mathbb{R}$, i.e. on their defining integration contour, (\ref{I2dTachikawa}) has a convergent Plethystic exponential expression as\footnote{Strictly speaking, the term $\theta_0 (\sigma|u_{ij})$ lies on the boundary of having a convergent Plethystic exponential expression. In such cases, we could imagine adding an infinitesimal imaginary part so that $\theta_0 (\sigma |u_{ij} + i 0^+)$ has a valid Plethystic exponential. Such a small imaginary part is implicit in (\ref{I2dPl}). Note that the function $\theta_0 (\sigma|u_{ij})$ itself is perfectly regular when $u_{ij} \in \mathbb{R}$. }
\begin{equation}\label{I2dPl}
\begin{aligned}
       \mathcal{I}_{2d} &  = \sum_{i=1}^N \exp\left[ \sum_{n=1}^\infty \frac{1}{n} \left( \frac{ e^{2\pi i n (-\tau + \Delta_1)} + e^{2\pi i n (\sigma + \tau - \Delta_1)} -e^{2\pi i n (\sigma - \Delta_2) }  - e^{2\pi i n \Delta_2} }{1-e^{2\pi i n \sigma}} \right) \sum_{j\neq i} e^{2\pi i n u_{ji}} 
 \right. \\
 & \quad\quad\quad\quad\quad + \left. \sum_{n=1}^\infty \frac{1}{n} \left( \frac{ 2 -e^{2\pi i n (\sigma + \tau - \Delta_1 -  \Delta_2) }  - e^{2\pi i n (-\tau + \Delta_1 + \Delta_2)} }{1-e^{2\pi i n \sigma}} \right) \sum_{j\neq i} e^{2\pi i n u_{ij}} \right] \,.
 \end{aligned}
\end{equation}
This puts constraints on the chemical potentials:
\begin{equation}\label{range}
    0 < \textrm{Im}(\Delta_2) < \textrm{Im}(\sigma)\,, \quad 0< \textrm{Im}(\Delta_1 - \tau) < \textrm{Im}(\sigma- \Delta_2)\,. 
\end{equation}
One reason to motivate the choice (\ref{range}) is that within this range, the Jacobi $\theta_0$ function has the standard form of infinite product expansion (see (\ref{eqn:theta0})) which has the clear physical meaning of enumerating states, with the identity operator contributing as one.
For the case $\Delta_1 = \Delta_2 = \Delta_3 = (\tau + \sigma  - 1)/3$ that can be compared with the gravity analysis done in Section \ref{sec:grav}, the index becomes
\begin{equation}\label{I2dequal}
    \mathcal{I}_{2d} 
    =  \sum_{i=1}^N \prod_{j\neq i} \frac{\theta_0\left(\sigma|u_{ji} + \frac{2\sigma}{3} - \frac{\tau}{3} + \frac{1}{3}  \right)\theta_0(\sigma|u_{ij} + \frac{\sigma}{3} + \frac{\tau}{3} + \frac{2}{3} )}{\theta_0(\sigma|u_{ji} + \frac{\sigma}{3} - \frac{2\tau}{3} - \frac{1}{3}) \theta_0(\sigma|u_{ij})}
\end{equation}
and the range of chemical potentials we consider in (\ref{range}) becomes
\begin{equation}
\label{tausigmarelation}
    0< 2 \,\textrm{Im}(\tau) < \textrm{Im}(\sigma)\,. 
\end{equation}
We will restrict our field theory discussion to this range of chemical potentials. We emphasize this is not a very restrictive range and it includes a wide range of black hole solutions, including the explicit example we showed in Section \ref{sec:physical} (it can be verified explicitly by examining Figure~\ref{fig:example1} (b)). Of course, it is tempting to analytically continue (\ref{I2dTachikawa}) outside this range. However, naively applying the expression (\ref{I2dTachikawa}) outside the range seems to lead to answers that we haven't understood how to interpret. We will discuss more on this point in Section \ref{sec:puzzle}.

\subsection{Evaluating the defect expectation value at the AdS saddle}\label{sec:fieldAdS}

The closed form expression \eqref{I2dTachikawa} of the 2d index, similarly to the 4d index, is a periodic function in $u_i$ with period 1. This property implies that the ${\rm SU}(N)$-matrix integral admits a simple saddle point $u_j = j/N$, $j=0,...,N-1$. 
This saddle point where the holonomies $e^{2\pi i u_j}$ are distributed uniformly on the unit circle describes the confined phase of the unitary matrix integral, which corresponds to the empty AdS saddle in the bulk \cite{Sundborg:1999ue,Aharony:2003sx}.

In order to calculate the large-$N$ defect index for the pure AdS saddle, we simply set the gauge holonomies to be distributed as described above.
As we've constrained the chemical potentials to lie within the range \eqref{range}, the expression (\ref{I2dTachikawa}) has a valid Plethystic exponential expression as shown in (\ref{I2dPl}). In this case using the form of the sum, it is easy to see that
\begin{equation}\label{emptylargeN}
  - \log\, \langle \mathcal{D}\rangle  = 0
\end{equation}
in the large $N$ limit, simply because for the empty AdS saddle point, we have
\begin{equation}
   \begin{aligned}
       \sum_{j\neq i} e^{2\pi i n u_{ji}} &\stackrel{N\rightarrow\infty}{\approx}N\int_0^1 dx\, e^{2\pi i nx}=0\, ,\\
       \sum_{j\neq i} e^{2\pi i n u_{ij}} &\stackrel{N\rightarrow\infty}{\approx}N\int_0^1 dx\, e^{-2\pi i nx}=0
   \end{aligned}
\end{equation}
with $n>0$. This agrees with the gravity calculation where the classical action of the probe D3-brane was found to be zero in the empty AdS background \cite{Drukker:2008wr}. By zero in (\ref{emptylargeN}) we really only mean that the answer is zero at order $N$, but it could have subleading corrections such as $\mathcal{O}(\log N)$.

If we take the large $N$ saddle point action (\ref{emptylargeN}) and continue it beyond the regime (\ref{range}), we would conclude that it continues to be zero. However, a puzzling aspect is that if we start with the expression (\ref{I2dTachikawa}), analytically continued beyond (\ref{range}) and then do the large $N$ analysis, it appears that one can find a nonzero action. In particular, the naively simpler case $\sigma = \tau$, \, $\Delta_1 = \Delta_2 = \Delta_3 = (2\sigma - 1)/3$ lies in this region. We further discuss this issue in Appendix \ref{sec:puzzle}.  For this reason, in the main text we will focus on the range of chemical potentials (\ref{tausigmarelation}) where we do find a vanishing action, consistent with the gravity picture.

\subsection{Evaluating the defect expectation value at the black hole saddle}\label{sec:fieldBH}

\begin{figure}[t!]
    \begin{center}
   \includegraphics[scale=.25]{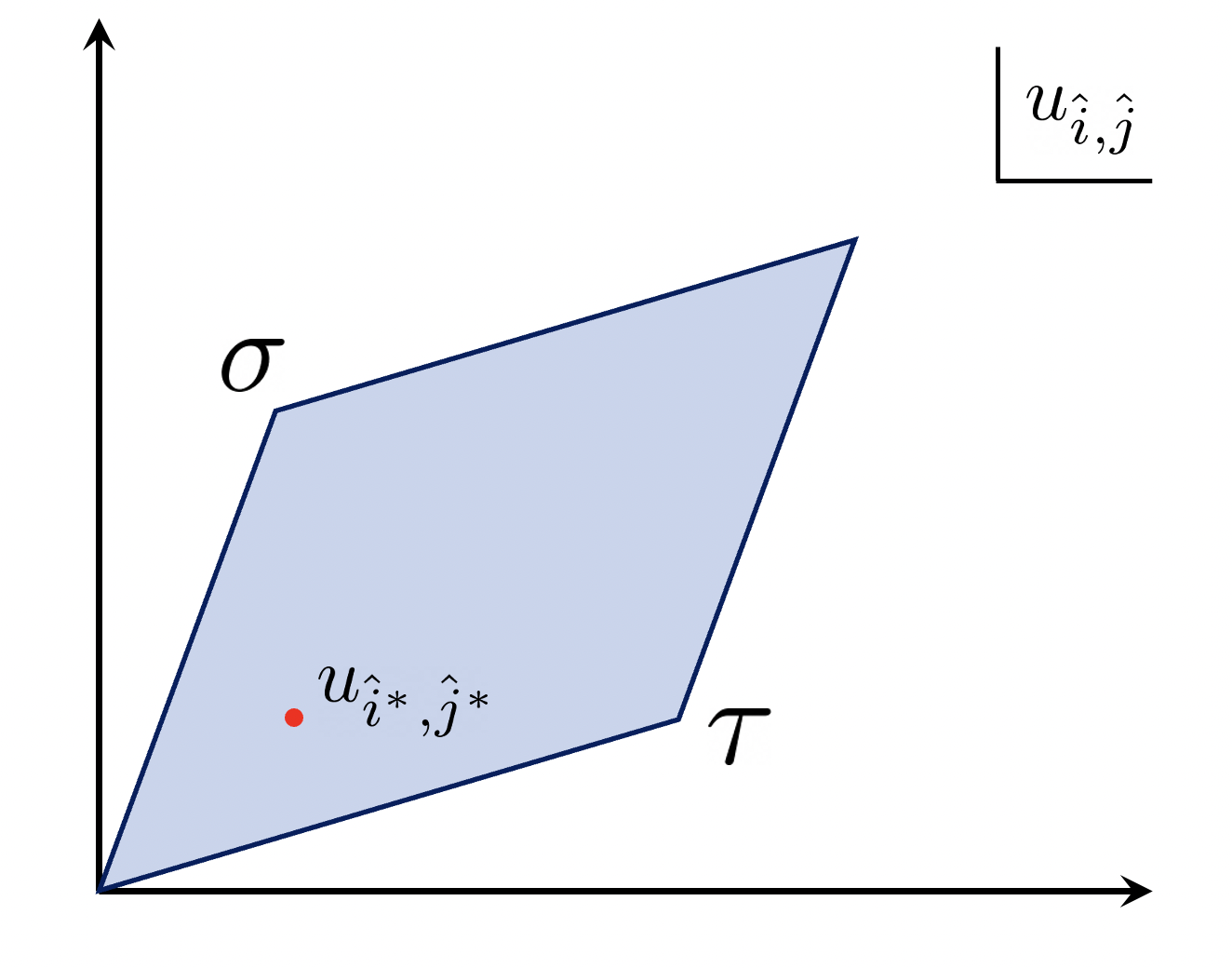}
    \end{center}
    \caption{Black hole saddle of the ${\rm SU}(N)$ matrix integral $(\sigma \neq \tau)$. The saddle-point values of the holonomies, labeled by $u_{\hat{i},\hat{j}}$ distribute  uniformly in the shaded parallelogram spanned by $\sigma$ and $\tau$ in the complex plane \cite{Choi:2021rxi}. The red point marked inside the parallelogram corresponds to $u_{\hat{i}^*,\hat{j}^*}$, which labels each individual term of the sum in \eqref{I2dequal}.}
    \label{fig:parallelmain}
\end{figure}

When $\tau \neq \sigma$, the saddle point of the ${\rm SU}(N)$ matrix integral corresponding to a bulk black hole is 
given by the uniform distribution of eigenvalues $u_{i}$ in the parallelogram spanned by $\sigma$ and $\tau$, see Figure~\ref{fig:parallelmain}.
This uniform parallelogram ansatz was recently proposed in \cite{Choi:2021rxi} as the solution to saddle-point problem of a generic ${\rm U} (N)$ matrix integral with the following form
\begin{equation}
\label{eq:parallelogramansatz}
  \mathcal{I}  = \int \left(\prod_{i=1}^{N} du_i\right)\,\prod_{i<j}\left(1-e^{\frac{2\pi iu_{ij}}{\tau}} \right)\, \exp{\left[-\sum_{i\neq j} (V_{\sigma}(u_{ij})-V_\tau(u_{ij}) )\right]},
\end{equation}
where $V_\sigma$ and $V_{\tau}$ are periodic functions in $u_{ij}$ with $\sigma$ and $\tau$ as periods respectively. It is then shown that the ${\rm SU}(N)$ matrix integral for the index of 4d $\mathcal{N}=4$ SYM  can be recast into the same form \eqref{eq:parallelogramansatz}, and the large-$N$ contribution from this parallelogram saddle is calculated and matched with the gravity action of the BPS black hole background \cite{Choi:2021rxi}.\footnote{In the case with the defect, this distribution of $u_i$ is not guaranteed to be the saddle point of the full 4d-2d index $\mathcal{I}_{\mathcal{D}}$; here it satisfies only the saddle point equation of the 4d index, an approximation that is valid at large $N$. In this limit we can still evaluate the 2d index using the same distribution of $u_i$ without modification. As we have emphasized, this is dual to the statement that in gravity we are not taking the backreaction from probe D3-brane into account.} In order to prove that the provided $u_i$ distribution solves the saddle-point equations, $V_\sigma$ and $V_\tau$ are required to be holomorphic when $u_i$ takes values inside the parallelogram. Strictly speaking, due to this additional requirement, we need to impose further restrictions on the chemical potentials on top of (\ref{tausigmarelation}). Such conditions are detailed in \cite{Choi:2021rxi}. However, these conditions do not play an important role in the evaluation of $\mathcal{I}_{2d}$, so we will not keep track of them in the following discussion.

We denote the saddle point value of the holonomies as:
\begin{equation}
    u_j \longrightarrow u_{\hat{i},\hat{j}} = \frac{\hat{i}}{\sqrt{N}}\tau+ \frac{\hat{j}}{\sqrt{N}}\sigma \, ,
\end{equation}
where $\hat{i},\,\hat{j}$ are integers running between 0 and $\sqrt{N}-1$, where we assume $N$ is a perfect square before passing to the continuum. In (\ref{I2dequal}), each term contains a reference holonomy labelled by $i$ which we subtract off from every other holonomy. Here we denote the reference holonomy as $u_{\hat{i}^*,\hat{j}^*}$. In terms of this notation, (\ref{I2dequal}) can be recast into
\begin{equation}\label{I2dnew}
    \mathcal{I}_{2d} = \sum_{\hat{i}^*,\hat{j}^* } \, \, \prod_{(\hat{i},\hat{j} ) \neq (\hat{i}^*,\hat{j}^* )}\frac{\theta_0\left(\sigma| u_{\hat{i},\hat{j} } - u_{\hat{i}^*,\hat{j}^* } + \frac{2\sigma}{3} - \frac{\tau}{3} + \frac{1}{3}  \right)\theta_0(\sigma| u_{\hat{i}^*,\hat{j}^* } - u_{\hat{i},\hat{j} }  + \frac{\sigma}{3} + \frac{\tau}{3} + \frac{2}{3} )}{\theta_0(\sigma|u_{\hat{i},\hat{j} } - u_{\hat{i}^*,\hat{j}^* } + \frac{\sigma}{3} - \frac{2\tau}{3} - \frac{1}{3}) \theta_0(\sigma| u_{\hat{i}^*,\hat{j}^* } - u_{\hat{i},\hat{j} }  )} \, .
\end{equation}

We could in principle plug in the saddle point value of $u_{\hat{i},\hat{j}}$ into (\ref{I2dnew}) and attempt to evaluate its large $N$ limit. We perform such a direct calculation in Appendix \ref{sec:direct}. However, there is in fact a simpler way to do the calculation, which involves first performing some manipulations on $\mathcal{I}_{\text{2d}}$. We will make use of a well known strategy for extracting the asymptotic behavior of 2d partition functions; several combined modular transformations on $\mathcal{I}_{\text{2d}}$ lead to a Cardy-like growth multiplying theta functions evaluated in the dual channel. This dual expression (which we present in (\ref{eq:I2dmodular})) has two advantages -- first, it will allow us to take the Cardy limit of the surface operator index (as we have reviewed in the introduction, the Cardy limit in this context is usually applied to higher dimensional superconformal indices and refers to taking $\sigma$, $\tau$ $\rightarrow 0$ with the ratio $\tau / \sigma$ fixed and finite). This limit on the potentials results in a significant simplification in extracting the result. Moreover, we will also find that using the modular transformed expression for $\mathcal{I}_{\text{2d}}$ makes it significantly easier to evaluate in the ordinary large $N$ limit, with $\sigma,\tau$ kept finite.

To begin with, 
using formulae from Appendix \ref{sec:convention}, we can first make a modular transformation on each individual theta function in (\ref{I2dnew}) for each value of the argument $u$,
\begin{equation}
    \theta_0(\sigma | u) = e^{i \pi B_{2,2}(u|\sigma, -1)}\theta_0 \left(-\frac{1}{\sigma} \left| \frac{u}{\sigma}  \right. \right) \, ,
\end{equation}
where the overall factor depends on the multiple Bernoulli polynomial
\begin{equation}
    B_{2,2}(u|\sigma, -1) = -\frac{((u+1)^2-(u+1)+\frac16)}{\sigma} + \left(u+\frac12 \right) - \frac{\sigma}{6} \, .
\end{equation}
Doing this to each factor gives
\begin{equation}\label{I2dP}
\begin{aligned}
    \mathcal{I}_{2d} 
    =  &  \sum_{\hat{i}^*,\hat{j}^* } \prod_{(\hat{i},\hat{j} )  \neq (\hat{i}^*,\hat{j}^*  )} \exp{(i\pi P) } \times \\
   &  \frac{\theta_0\left(-\frac{1}{\sigma} \left|\frac{u_{\hat{i},\hat{j} } - u_{\hat{i}^*,\hat{j}^* }}{\sigma} + \frac{2}{3} - \frac{\tau}{3\sigma} + \frac{1}{3\sigma}  \right. \right)\theta_0\left(-\frac{1}{\sigma}\left|\frac{u_{\hat{i}^*,\hat{j}^* } - u_{\hat{i},\hat{j} }}{\sigma} + \frac{1}{3} + \frac{\tau}{3\sigma} + \frac{2}{3\sigma} \right.  \right)}{\theta_0\left(-\frac{1}{\sigma} \left|\frac{u_{u_{\hat{i},\hat{j} } - u_{\hat{i}^*,\hat{j}^* } }}{\sigma} + \frac{1}{3} - \frac{2\tau}{3\sigma} - \frac{1}{3\sigma} \right. \right) \theta_0 \left(-\frac{1}{\sigma} \left|\frac{u_{\hat{i}^*,\hat{j}^* } - u_{\hat{i},\hat{j} } }{\sigma} \right. \right)} \, ,
\end{aligned}
\end{equation}
where we have collected the phases in $P$, writing the polynomials in shorthand $B(u)\equiv B_{2,2}(u|\sigma,-1)$,
\begin{equation}
\begin{aligned}
    P & =  -B(u_{\hat{i}^*,\hat{j}^* }- u_{\hat{i},\hat{j}}) - B\left( u_{\hat{i},\hat{j}}- u_{\hat{i}^*,\hat{j}^* } + \frac{\sigma}{3} - \frac{2\tau}{3} - \frac{1}{3} \right) \\
   & \quad  + B\left( u_{\hat{i},\hat{j}}- u_{\hat{i}^*,\hat{j}^* } + \frac{2\sigma}{3} - \frac{\tau}{3} + \frac{1}{3} \right) + B\left(u_{\hat{i}^*,\hat{j}^* }- u_{\hat{i},\hat{j}} + \frac{\sigma}{3} + \frac{\tau}{3} + \frac{2}{3} \right) \\
   & = \frac{ 2(\sigma + \tau + 2) (\sigma + \tau -4) }{9\sigma}\, .
\end{aligned}
\end{equation}
A further simplification occurs when we use the periodicity of the $\theta_0$ function to shift the upper right theta function in (\ref{I2dP}) down by one,
\begin{equation}
    \theta_0\left( -\frac{1}{\sigma}\left| \frac{ u_{\hat{i}^*,\hat{j}^* }- u_{\hat{i},\hat{j}} }{\sigma} + \frac{1}{3} + \frac{\tau}{3\sigma} + \frac{2}{3\sigma} \right.\right) = \theta_0\left( -\frac{1}{\sigma} \left| \frac{  u_{\hat{i}^*,\hat{j}^* }- u_{\hat{i},\hat{j}}  }{\sigma} - \frac{2}{3} + \frac{\tau}{3\sigma} + \frac{2}{3\sigma} \right.  \right).
\end{equation}
After this manipulation, we now shift the two theta functions in the numerator of (\ref{I2dP}) using the quasiperiodicity of the $\theta_0$ function (see Appendix \ref{sec:convention})
\begin{equation}\label{quasi}
    \theta_0\left( -\frac{1}{\sigma} | u  \right) = -e^{2\pi i u}\theta_0 \left( -\frac{1}{\sigma}\left| u - \frac{1}{\sigma} \right.\right) \, ,
\end{equation}
which leads to an extra phase $\exp(i\pi P')$, where
\begin{equation}
     P' =  2\left (\frac{u_{\hat{i},\hat{j}}- u_{\hat{i}^*,\hat{j}^* } }{\sigma}  +   \frac{u_{\hat{i}^*,\hat{j}^* }- u_{\hat{i},\hat{j}} }{\sigma}   + \frac{1}{\sigma}\right ) = \frac{2}{\sigma} \, .
\end{equation}
Combining $P$ and $P'$ we find
\begin{equation}
  \pi i(  P + P' )= 2 \pi i \frac{(\sigma + \tau - 1)^2}{9 \sigma} \, .
\end{equation}
We get the same factor like this for each term of the product in (\ref{I2dP}), and the resulting expression is
\begin{equation}
\label{eq:I2dmodular}
\begin{aligned}
    \mathcal{I}_{2d} 
    = &  \exp\left(2 \pi i (N-1) \frac{(\sigma + \tau - 1)^2}{9 \sigma}\right) \times \\
 &  \sum_{\hat{i}^*,\hat{j}^* } \, \, \prod_{(\hat{i},\hat{j} )  \neq (\hat{i}^*,\hat{j}^*  )} \! \frac{\theta_0\left(-\frac{1}{\sigma} \left|\frac{u_{\hat{i},\hat{j} } - u_{\hat{i}^*,\hat{j}^* }}{\sigma} + \frac{2}{3} - \frac{\tau}{3\sigma} - \frac{2}{3\sigma}  \right. \right)
 \theta_0\left(-\frac{1}{\sigma}\left|\frac{u_{\hat{i}^*,\hat{j}^* } - u_{\hat{i},\hat{j} }}{\sigma} -  \frac{2}{3} + \frac{\tau}{3\sigma} - \frac{1 }{3\sigma} \right.  \right)}{\theta_0\left(-\frac{1}{\sigma} \left|\frac{u_{\hat{i},\hat{j} } - u_{\hat{i}^*,\hat{j}^*  }}{\sigma} + \frac{1}{3} - \frac{2\tau}{3\sigma} - \frac{1}{3\sigma} \right. \right) \theta_0 \left(-\frac{1}{\sigma} \left|\frac{u_{\hat{i}^*,\hat{j}^* } - u_{\hat{i},\hat{j} } }{\sigma} \right. \right)} \,.
\end{aligned}
\end{equation}
At large $N$, we recognize the first line as exactly the expected result from gravity, \eqref{SD3final}. However, we still need to evaluate the large $N$ limit of the remaining combination of theta functions on the second line on the support of the parallelogram black hole saddle, see Figure~\ref{fig:parallelmain}. 
We will do this shortly, but instead we will first focus on the Cardy limit of the surface operator index, which leads to a significant simplification.

\subsubsection{Defect expectation value in the Cardy limit}
\label{sec:dindCardy}

In the so-called Cardy limit, we would like to take $N$ finite and fixed, while taking $\sigma,\tau\rightarrow 0$ and at the same time keep the ratio $\tau/\sigma$ finite. It is possible to expand out the theta functions in \eqref{eq:I2dmodular} in this limit. In the argument of each $\theta_0$-function other than the bottom right one in (\ref{eq:I2dmodular}), we drop all terms which remain finite in the Cardy limit, which includes the terms depending on the holonomies, since
\begin{equation}
 \frac{u_{\hat{i},\hat{j}} - u_{\hat{i}^*,\hat{j}^* }}{\sigma} =  \frac{\hat{i}  - \hat{i}^* }{\sqrt{N}} \frac{\tau}{\sigma} + \frac{\hat{j}  - \hat{j}^* }{\sqrt{N}} 
\end{equation}
which remains order one as $\sigma,\tau\rightarrow 0$. These terms can be ignored when there is a term in the argument of the $\theta_0$ function that is $\propto 1/\sigma$ which will dominate in the Cardy limit. Using the infinite product expression for the $\theta_0$ function in (\ref{eqn:theta0}),  for the first $\theta_0$ in (\ref{eq:I2dmodular}), we have
\begin{equation}
    \theta_0\left(-\frac{1}{\sigma} \left|\frac{u_{\hat{i},\hat{j} } - u_{\hat{i}^*,\hat{j}^* }}{\sigma} + \frac{2}{3} - \frac{\tau}{3\sigma} - \frac{2}{3\sigma}  \right. \right) \sim 1 -  \tilde{q}^{\frac{1}{3}} - \tilde{q}^{\frac{2}{3}} + \mathcal{O} \left(\tilde{q}  \right)\, , 
\end{equation}
where $\tilde{q} \equiv e^{-\frac{2 \pi i}{\sigma}} \rightarrow 0$ in the Cardy limit (note that $\textrm{Im}(\sigma )>0$). Similarly, the upper-right and bottom-left $\theta_0$ in (\ref{eq:I2dmodular}) also go to one in the Cardy limit. Notice that in the Cardy limit we keep $N$ finite while sending $\sigma,\tau$ to zero, therefore the product of $N-1$ terms in (\ref{eq:I2dmodular}) will remain one if each individual term goes to one. 

However, the bottom right $\theta_0$-function in (\ref{eq:I2dmodular}) requires some extra care since there is no $1/\sigma$ term present, so we cannot simply ignore the dependence on the holonomy. 
Fortunately, this is not a major issue, since we will have
\begin{equation}\label{cardyextra}
    \prod_{(\hat{i},\hat{j} )  \neq (\hat{i}^*,\hat{j}^*  )} \frac{1}{\theta_0 \left(-\frac{1}{\sigma} \left|\frac{u_{\hat{i}^*,\hat{j}^* } - u_{\hat{i},\hat{j} } }{\sigma} \right. \right)} =   \prod_{(\hat{i},\hat{j} )  \neq (\hat{i}^*,\hat{j}^*  )} \frac{1}{ 1 - \exp\left[2\pi i \left(\frac{\hat{i}  - \hat{i}^* }{\sqrt{N}} \frac{\tau}{\sigma} + \frac{\hat{j}  - \hat{j}^* }{\sqrt{N}} \right)\right] + \mathcal{O}(\tilde{q})} \, . 
\end{equation}
Even though this does not go to one in the Cardy limit, the main observation is that this expression only depends on $\tau/\sigma$, which remains finite in the Cardy limit. On the other hand, the phase factor on the first line of (\ref{eq:I2dmodular}) leads to a large contribution which diverges in the $\sigma \rightarrow 0$ limit, namely
\begin{equation}
    \exp\left(2 \pi i (N-1)\frac{(\sigma + \tau - 1)^2}{9\sigma}\right) \sim  \exp\left((N-1) \frac{2 \pi i }{9\sigma}\right) ,\quad \sigma \rightarrow 0\,.
\end{equation}
Therefore, the conclusion is that contributions from the second line in (\ref{eq:I2dmodular}) can be safely ignored compared to the phase factor. In summary, we get 
\begin{equation}\label{I2dCardy}
    \langle \mathcal{D}\rangle  \sim \exp\left((N-1) \frac{2 \pi i }{9\sigma}\right) \,, \quad {\rm as~}\sigma\rightarrow 0, \,\, \frac{\tau}{\sigma} \,\, \textrm{finite}\,.
\end{equation}

Notice that the Cardy limit result (\ref{I2dCardy})  of the field theory calculation, where we keep $N$ finite while sending $\sigma, \tau$ to zero, agrees with the $\sigma, \tau \rightarrow 0$ limit of the gravity answer (\ref{SD3final}), where we had taken the large $N$ limit first.  
This Cardy limit calculation demonstrates compatibility of our answer with gravity in the regime $N\rightarrow \infty,  \sigma, \tau \rightarrow 0$.  The factor $(N-1)$ in (\ref{I2dCardy}) reflects the surface defect central charge (\ref{c2d}). In principle, by carefully including (\ref{cardyextra}), we can obtain the full Cardy limit result, up to non-perturbatively small corrections of order $\tilde{q}$. Alternatively, it can also be derived from an EFT argument generalizing that of \cite{Cassani:2021fyv} by including the defect (see Section~\ref{sec:CardyGen} for further comments). We leave a detailed analysis for the Cardy limit of (general) surface defects to future. Instead
we will now turn our attention to the evaluation of the large $N$ limit \eqref{eq:I2dmodular} on the black hole saddle. The following derivation is valid for general finite chemical potentials for which the parallelogram ansatz is valid and we can compare it with the full large $N$ answer from gravity in (\ref{SD3final}).

\subsubsection{Defect expectation value in the large $N$ limit}
\label{sec:dindlargeN}

In the large-$N$ limit, we could replace $u_{\hat{i},\hat{j}}$ by a continuous distribution $u(x,y) = x\tau+y\sigma$ with $x,y \in [0,1]$. For each term in the sum of (\ref{eq:I2dmodular}), we can study its large $N$ limit by first computing its logarithm, where the product $\prod_{(\hat{i},\hat{j} ) \neq (\hat{i}^*,\hat{j}^* )}$ turns into a sum $\sum_{(\hat{i},\hat{j} ) \neq (\hat{i}^*,\hat{j}^* )}$, which can be in turn approximated by an integral $N\int dx\,dy$. Generally we do not expect the integral to vanish, and therefore we get an contribution of order $e^{\mathcal{O}(N)}$ after exponentiating it back. At the same time, we can write $u_{\hat{i}^*,\hat{j}^*} = x^*\tau+y^*\sigma$; then we have that in the large $N$ limit, (\ref{eq:I2dmodular}) is given by an integral over $x^*,y^*$, namely the coordinates of the red dot along the $\tau,\sigma$ directions in Figure~\ref{fig:parallelmain}, with the integrand being the product inside the sum. Since the integrand is generally of order $e^{\mathcal{O}(N)}$, we expect that the integral over $x^*,y^*$ will be dominated by a saddle point coming from the following extremization 
\begin{equation}
\label{I2dlargeN}
\begin{aligned}
       - \frac{1}{N} \log  \langle \mathcal{D}\rangle 
    =& -2 \pi i  \frac{(\sigma + \tau - 1)^2}{9 \sigma} \\
    & - \textrm{ext}_{x^*,y^*} \left\{  \int_0^1 dx\, \int_0^1 dy\, \left[ \log \theta_0\left(-\frac{1}{\sigma} \left| (x-x^*) \frac{\tau}{\sigma} + (y-y^*) + \frac{2}{3} - \frac{\tau}{3\sigma} - \frac{2}{3\sigma}  \right. \right)  \right. \right. \\
    & \quad\quad\quad \quad\quad + \log \theta_0\left(-\frac{1}{\sigma}\left|  (x^* -x) \frac{\tau}{\sigma} + (y^* -y)   -  \frac{2}{3} + \frac{\tau}{3\sigma} - \frac{1 }{3\sigma} \right.  \right)  \\
    & \quad\quad\quad \quad\quad - \log \theta_0\left(-\frac{1}{\sigma} \left| (x-x^*) \frac{\tau}{\sigma} + (y-y^*)  + \frac{1}{3} - \frac{2\tau}{3\sigma} - \frac{1}{3\sigma} \right. \right) \\
    & \quad\quad\quad \quad\quad \left.\left. - \log  \theta_0 \left(-\frac{1}{\sigma} \left|(x^* -x) \frac{\tau}{\sigma} + (y^* -y)   \right. \right)  \right] \right\} \,.
\end{aligned}
\end{equation}
While somewhat formidable looking, the integrals in (\ref{I2dlargeN}) are not hard to evaluate. Similar to what we did for the empty AdS saddle in Section \ref{sec:fieldAdS}, one way to proceed is to turn the $\theta_0$ function into a Plethystic exponential expression, given that such an expression is convergent. To simplify the discussion, here we assume $\tau \in \mathbb{R} + i0^+$ and $0<\tau < \frac{1}{4}$. The benefit is that the first three lines of the $\theta_0$ functions in (\ref{I2dlargeN}) will all have convergent Plethystic exponential expressions, even though the fourth one doesn't and still has to be treated with extra care. 
This assumption on $\tau$ makes the discussion simpler but is not necessary. The final expression we derive is an analytic function of $\tau$ so can be simply continued to general values of $\tau$, given that the parallelogram saddle point still applies.\footnote{We have also checked that at least in the range of $0<\tau< 1$ we will arrive at the same answer. } 

For the first three $\theta_0$ functions in (\ref{I2dlargeN}), it is evident that the integrals are simply zero once we write them in the form of Plethystic exponentials. This can be seen as follows. These three $\theta_0$-functions in (\ref{I2dlargeN}) take the form of $\theta_0 (- \frac{1}{\sigma} | y + c)$, where $c$ is independent of $y$. Using formulae in Appendix \ref{sec:convention}, we have
\begin{equation}\label{intzero}
\begin{aligned}
  \log \theta_0 \left( \left. - \frac{1}{\sigma}  \right| y + c  \right)   &  =  \log \frac{1}{\textrm{PE}\left( \frac{e^{2\pi i (y + c)}}{ 1 - \tilde{q}} \right)\textrm{PE}\left(  \frac{\tilde{q} e^{-2\pi i (y + c)}}{ 1 - \tilde{q}} \right) }  \\
  & = - \sum_{n=1}^{\infty} \frac{1}{n } \left[ \frac{ e^{2\pi i n (y + c)}  }{1 - \tilde{q}^n } + \frac{\tilde{q}^n  e^{-2\pi i n (y + c)}  }{1 - \tilde{q}^n } \right],
\end{aligned}
\end{equation}
therefore the integral of $y$ from $0$ to $1$ on each term vanishes identically. 
Of course, the same conclusion can be also reached by considering the infinite product formula for $\theta_0$. This logic would be correct for the first three $\log\theta_0$ integrals in (\ref{I2dlargeN}) but would be wrong for the last one. The reason is that the last $\theta_0$ does not always have a Plethystic exponential representation but rather depends on the value of $x$. Note that in (\ref{tausigmarelation}) we introduced criteria on the chemical potentials by demanding the $\theta_0$-functions have convergent Plethystic or infinite product representations, assuming that the holonomies are real. Here we are seeing that extra non-convergence issues could arise once we consider the black hole saddle point where the holonomies are complex. We don't view this new effect as something that has significant physical meaning, but rather as an extra hurdle in the calculation one needs to overcome. 

The criterion for the $\theta_0$ function on the last line of (\ref{I2dlargeN}) to have a valid Plethystic exponential representation is
\begin{equation}
  0< \textrm{Im} \left( (x^* - x) \frac{\tau}{\sigma} \right) < \textrm{Im} \left( - \frac{1}{\sigma} \right)\,,
\end{equation}
and with our simplifying assumption on $\tau$, this translates to $x^*<x<1$. When $0<x<x^*$, we need to take an extra step by shifting the argument using the quasiperiodicity of the $\theta_0$ function as
\begin{equation}
     \theta_0 \left(-\frac{1}{\sigma} \left|(x^* -x ) \frac{\tau}{\sigma} + (y^* -y)   \right. \right) =  e^{2\pi i  \left[ (x^* -x ) \frac{\tau}{\sigma} + (y^* -y)   - \frac{1}{2}\right]} \theta_0 \left(-\frac{1}{\sigma} \left|(x^* -x ) \frac{\tau}{\sigma} - \frac{1}{\sigma} + (y^* -y)   \right. \right) \,.
\end{equation}
After this step, the shifted $\theta_0$ on the right hand side now can be expressed in terms of a convergent Plethystic exponential. After this manipulation, we have
\begin{equation}
\begin{aligned}
&   \int_0^1 dx\, \int_0^1 dy\,  \log  \theta_0 \left(-\frac{1}{\sigma} \left|(x^* -x) \frac{\tau}{\sigma} + (y^* -y)   \right. \right) 
\\
=&   \int_0^{x^*} dx\, \int_0^1 dy\, 2\pi i  \left[ (x^* -x ) \frac{\tau}{\sigma} + (y^* -y)   - \frac{1}{2}\right] \\
  & + \int_0^{x^*} dx\, \int_0^1 dy\, \log \theta_0 \left(-\frac{1}{\sigma} \left|(x^* -x ) \frac{\tau}{\sigma} - \frac{1}{\sigma} + (y^* -y)   \right. \right) \\
  & + \int_{x^*}^1 dx\, \int_0^1 dy\, \log \theta_0 \left(-\frac{1}{\sigma} \left|(x^* -x ) \frac{\tau}{\sigma} + (y^* -y)   \right. \right) \\
   = & \pi   i \frac{\tau}{\sigma} x^{*2} - 2\pi i  x^* (1-y^*)\,.
\end{aligned}
\end{equation}
The integral on the third line and the fourth line vanishes due to the $y$ integral for the same reason as in (\ref{intzero}), while the answer on the last line entirely comes from the elementary integral on the second line. 

In summary, we've performed all the integrals in (\ref{I2dlargeN}) and we get
\begin{equation}
          - \frac{1}{N} \log  \langle \mathcal{D}\rangle  =  -2 \pi i  \frac{(\sigma + \tau - 1)^2}{9 \sigma} +  \textrm{ext}_{x^*,y^*} \left[ \pi   i \frac{\tau}{\sigma} x^{*2} - 2\pi i  x^* (1-y^*) \right] \, .
\end{equation}
The extremum singles out the point at $x^* = 0, \, y^* = 1$. We could check that this extremum is a local maximum and thus leads to a dominating answer. Therefore, in the large $N$ limit, we conclude
\begin{equation}\label{I2dfinalfield}
 -\log  \,\langle \mathcal{D}\rangle    = -2\pi i N\frac{ (\sigma + \tau - 1)^2 }{9\sigma} \,,
\end{equation}
which is in perfect agreement with the answer \eqref{SD3final} from the gravity analysis.

\section{Conclusion and Future Directions}
\label{sec:conclusion}

In this paper, we proposed and studied a new type of supersymmetric observable in the $\mathcal{N} = 4$ ${\rm SU}(N)$ SYM theory that probes the bulk black hole geometry while being accessible on both sides of the duality. 
It is an extended supersymmetric surface defect analogous to the familiar Polyakov loop which detects the confinement/deconfinement transition in the non-abelian gauge theory and equivalently the Hawking-Page transition for the dual black holes.
In the gravity description, the surface defect is identified with a calibrated D3-brane that wraps the Euclidean cigar of the (complex) Euclidean black hole and extends to asymptotic infinity. We then evaluated the renormalized on-shell action of the brane in the probe limit, which is a reliable calculation in the semiclassical $G_N \rightarrow 0$ regime, and derived a simple elegant expression for this defect observable as an exact function of the boundary chemical potentials. On the field theory side, we have a Gukov-Witten surface defect extending along the Euclidean time direction and an $S^1$ inside the spatial $S^3$. This defines the defect index that generalizes the usual superconformal index. Using the gauged linear sigma model description of this defect, we computed the defect index in the large $N$ limit as well as in the Cardy limit and found a precise match with the gravity analysis. This opens up a new window to study the properties of black holes interacting with external degrees of freedom, in this case a D3-brane. It also demonstrates the role of the surface defect as a supersymmetric order parameter for the deconfinement phase transition in large $N$ SYM that is interesting in it's own right.

They are many open questions that are worth further exploration. We will discuss some of them below.

\subsection{Lorentzian geometry and the AdS$_2$ limit}

We evaluated the surface 
defect expectation value $\la \cD\ra$ on a family of complex Euclidean gravity solutions that are supersymmetric but non-extremal. Even though the final answer is independent of $\beta$ (see around \eqref{N4SCI} and \eqref{eq:grandcanonicalindex}), it is important that we set up the calculation at finite $\beta$, where the classical action can be evaluated unambiguously. The defect expectation value reflects features of this family of complex black hole saddles. It would be interesting to understand precisely how our answer translates into features of the geometry of the \emph{Lorentzian} extremal black hole. Relatedly, we hope to investigate how the thermodynamics of the coupled setup discussed in Section \ref{thermo} translates properly into the properties of the Lorentzian system.

A related direction worth exploring is whether our answer can be reproduced by a calculation purely in the near horizon AdS$_2$ limit. 
We have already discussed in Section \ref{ssec:generalmotivation} how this low temperature limit must include quantum fluctuations of the AdS$_2$ metric and its superpartners, and these fluctuations are captured by $\mathcal{N}=2$ JT supergravity and the associated super-Schwarzian boundary mode\cite{Heydeman:2020hhw,Boruch:2022tno,Iliesiu:2022kny}.\footnote{Here we strictly speaking mean the leading topology in super JT gravity which is that of the AdS$_2$ disk. If one only uses the 2d gravitational degrees of freedom, higher topologies like wormholes are not expected to contribute to the index\cite{Iliesiu:2021are}.} In the case with the probe D3-brane insertion, it is natural to first ask whether the defect action and entropy is reproduced using only rigid AdS$_2$ without the near BPS backreacting theory. We believe the answer is no, owing to the fact that the rigid AdS$_2$ geometry has a real metric and chemical potentials, so it seems difficult 
to reproduce the complex 
($\beta$ independent) brane action we found. Note that our set-up is different from the situation studied in \cite{Giombi:2020amn,Giombi:2022pas}, in which a defect wraps an AdS$_2$ sub-manifold of the empty AdS$_5$. In our case, the AdS$_2$ limit is approached from a black hole solution with generically complex parameters and a physical Euclidean horizon, so the brane dynamics in this background can be quite different. An independent field theory check of the low temperature limit of $\mathcal{N}=4$ SYM with the surface operator would naively involve a one-dimensional quantum mechanics with the insertion of a line operator (from reduction on the spatial $S^3$). However, it is difficult to directly tackle this problem because the gauge theory is strongly coupled and we do not know how to set up the defect path integral away from the BPS limit. There are quantum mechanical systems that are known to contain supersymmetric near-AdS$_2$\cite{Fu:2016vas,Heydeman:2022lse}, but these systems are disordered and do not have an obvious connection to higher dimensional gauge theory.  At least in the bulk, in future work we plan to take the near AdS$_2$ limit more carefully and study the effective theory of the defect in this limit.

\subsection{Finite $N$ and the entropy formula} 

One main benefit of our set up is that on the field theory side we have an expression for the index (\ref{defectmm}) that is exact and can be in principle computed even at finite $N$. This provides an opportunity for doing precision holography for a black hole coupled to an external system, in our case a D3-brane. 
As an intriguing possible application of this, we can wonder about the entropy formula for our system. To simplify the discussion, here we ignore the difference between the degeneracy and the (logarithm of the) index of BPS states.
In the large $N$ limit (the limit we have focused on in this paper), the total entropy of the system is given by
\begin{equation}\label{entfor}
    S_{\textrm{total}} \approx S_{\textrm{BH}}  + S_{\textrm{D3}}\,,
\end{equation}
where the first term is of order $N^2$ while the second term is of order $N$. This is simply a special case of the entropy of a black hole coupled to some external system. However, from the field theory point of view, there should not be a clear separation between the two terms in (\ref{entfor}) at finite $N$, so it would be interesting to understand how (\ref{entfor}) extends away from the large $N$ limit. We note that in the black hole phase the D3-brane acquires a classical entropy that does not obviously have a state-counting interpretation in the bulk. This is similar to the situation of the black hole entropy, but the D3-brane might provide a simpler laboratory to approach this problem.

\subsection{Other field theory observables in SYM}
\label{sec:PLcomment}
The Gukov-Witten defect is not the simplest extended operator one can imagine utilizing to probe the black hole geometry in the bulk gravity and the deconfinement phase transition in the gauge theory. A natural candidate one could consider is the supersymmetric fundamental Wilson loop, which is dual to an extended fundamental string in the bulk \cite{Rey:1998ik,Maldacena:1998im}. It is well-known that the Polyakov loop \cite{Polyakov:1975rs}, namely the expectation value of Wilson loop extending along the thermal circle, serves as an order parameter for confinement/deconfinement transition. In particular, in the black hole phase, the string worldsheet wraps the black hole cigar geometry similar to our D3-brane in Figure~\ref{fig:sketch}, leading to a nonzero expectation value for the Polyakov loop in the large $N$ limit \cite{Witten:1998zw}. More generally, the Polyakov loop and its generalizations can be used to classify various subdominant saddles showing up in the type of unitary matrix model (\ref{4dUmm}) and diagnose the fine-grained features of the spectrum \cite{Chen:2022hbi,Choi:2022asl}.

However, an immediate difficulty is to identify a Wilson loop configuration that preserves the ${1\over 16}$ supersymmetries of the black hole.\footnote{Note that there are certainly interesting superconformal indices refined by line defects in 4d $\cN=2$ SCFTs (see for example \cite{Gang:2012yr,Drukker:2015spa,Cordova:2016uwk,Neitzke:2017cxz,Hatsuda:2023iwi}). However in all such examples, a different supercharge is used to define the index which turns out to count BPS states preserving more supersymmetry instead of the minimal amount preserved by the black hole states.} A straight line defect that sits at a particular point on $S^3$ while extending along the time direction would not work since there is no isolated fixed point on $S^3$ under the vector field $J_1 + J_2$ that appears in (\ref{QSac}).\footnote{We thank Juan Maldacena for discussions on this point.}  Nonetheless, one can potentially consider a smeared version of the Wilson loops that would be compatible with the desired supercharges $\cQ,\cQ^\dagger$ (see \cite{Nishioka:2014zpa} for such an example). It would be interesting to investigate this possibility and its relation to the surface defect probe considered here.

\subsection{Giant graviton expansion}

The superconformal index for $\cN=4$ SYM is known to admit a mysterious expansion \cite{Arai:2020qaj,Imamura:2021ytr,Gaiotto:2021xce,Murthy:2022ien,Lee:2022vig,Imamura:2022aua,Holguin:2022drf,Choi:2022ovw,Liu:2022olj,Lin:2022gbu,Eniceicu:2023uvd} that echoes the intuition that the black holes can be built from the bound states of  an $\cO(N)$ number of intersecting D3-branes in the bulk of the ${\rm AdS}_5\times S^5$ vacuum. However
its precise physical origin is yet to be understood. 
This is known as the giant graviton expansion because the constituent D3-branes wrap $S^3\subset S^5$ and behave as particles in AdS$_5$ \cite{
McGreevy:2000cw,Grisaru:2000zn}. While this bound state picture is only valid at strong coupling, the superconformal index, being coupling independent, allows one to extrapolate calculations involving D3-branes from weak to strong coupling. In particular, the index of the BPS states on the intersecting D3-branes (along a two dimensional locus fixed by supersymmetry) are crucial ingredients on top of the ``classical contributions'' from the D3-brane actions that ensures consistency of the giant graviton expansion (i.e. avoiding the naive overcounting) \cite{Arai:2020qaj,Imamura:2021ytr,Gaiotto:2021xce,Lee:2022vig}.

The physical interpretation of these giant graviton D3-branes differ from the defect D3-branes considered here because they reside in the bulk of AdS$_5$ as opposed to extending to the asymptotic boundary. Consequently they give rise to normalizable states in the usual Hilbert space (dual to local operators in the CFT) in contrast to defining a different Hilbert space due to the defect insertion.
Nonetheless, the kinematic configurations of these two intersecting D3-brane setups are similar\footnote{For the surface defect, this is the setup studied in detail in \cite{Constable:2002xt} where the defect D3-brane intersects with the stack of $N$ D3-branes along two common directions.} and the one-loop determinants for the defect index studied here are reminiscent of 
the BPS indices that counts supersymmetric states at the two-dimensional intersections between giant graviton D3-branes in \cite{Imamura:2021ytr} (up to a redefinition of the chemical potentials and analytic continuation). This is all but a snapshot of the multiple roles played by D3-branes in the generalized grand canonical ensemble of ${1\over 16}$-BPS states in $\cN=4$ SYM and the dual type IIB string theory.
 
It would be interesting to study 
an extension of the graviton expansion of the superconformal index by incorporating the surface defect. At each order in the giant graviton expansion, which corresponds to the total wrapping number of the giant graviton D3-branes, including the defect D3-brane enriches the BPS state counting problem. By comparing the result with the explicit matrix integral for the surface defect expectation value \eqref{defectmm}, this will provide a nontrivial test of the giant graviton expansion.

\subsection{Cardy limit for general surface defects} 
\label{sec:CardyGen}

Our analysis of the surface defect expectation value is implemented in the large $N$ limit where we have kept exact dependence on the chemical potentials. A complementary approach would be a generalization of the Cardy limit studied in the context of the ordinary superconformal index applied to the defect index. Here we take a high-temperature-like limit on the chemical potentials where $\sigma,\tau \to 0$ (from the upper half-plane) with fixed $\sigma\over \tau$. A technical merit of the Cardy limit is the drastic simplification in the unitary matrix model expression for the (defect) index irrespective of whether $N$ is taken to be large,
which enables a quick derivation of the defect expectation value in this limit
 as we have seen in Section~\ref{sec:dindCardy}. Conceptually, the Cardy limit of the (defect) index highlights universal features of the (defect) CFT (e.g. the asymptotic density of states) that can be inferred from an effective field theory (EFT) from circle reduction that emerges in this high-temperature-like limit \cite{DiPietro:2014bca,DiPietro:2016ond}. In particular, as explained in \cite{Cassani:2021fyv} (see also \cite{ArabiArdehali:2021nsx,Lezcano:2021qbj,Amariti:2021ubd,Ohmori:2020wpk}), in the absence of defect insertions, the EFT consideration determines the Cardy limit of the superconformal index in terms of the anomalies of the SCFT. Importantly this analysis produces even stronger universal results when the EFT is gapped, in which case the subleading terms in $\sigma,\tau$  are completely fixed by the anomaly coefficients alone, up to exponentially small contributions \cite{Cassani:2021fyv}.  
 
With defect insertions along the Euclidean time direction (such as the surface defects considered here), the Cardy limit of the defect expectation value  is naturally produced by studying the circle reduction of the defect in the above-mentioned high-temperature EFT.  By a similar reasoning, we expect the Cardy limit of the defect expectation value to be determined by the defect conformal anomalies, and a similar truncation in the perturbative $\sigma,\tau$ expansion to occur when the EFT is gapped. We leave the  general analysis of the Cardy limit of defect expectation value to future. Below we comment on our findings in Section~\ref{sec:dindCardy} from this point of view.\footnote{We thank Zohar Komargodski for discussions on this perspective.}

In the special case of $\Delta_1=\Delta_2={\sigma+\tau-1\over 3}$ (which we have also assumed in the main text to evaluate the surface defect index), the $\cN=4$ superconformal index \eqref{N4SCI} reduces to the so-called universal superconformal index $\cI_{\rm univ}(p,q)$, which is universally defined for all $\cN=1$ SCFTs.
As explained in \cite{Cassani:2021fyv}, due to the fractional charges under the $\cN=1$ $U(1)_r$ symmetry (generated by $r={R_1+R_2+R_3\over 3}$ in $\cN=4$ SYM), the universal superconformal index $\cI_{\rm univ}(p,q)$  is not single valued in the angular chemical potentials $(\sigma,\tau)$. For $\cN=4$ SYM, there are three Riemann sheets related by $\sigma\to \sigma + n_0$ keeping $\tau$ fixed with ${\rm Re}(\tau)\in [0,1)$ and $n_0=0,1,2$, commonly referred to as the first, second and third sheet respectively. Furthermore, the index on the second and the third sheet are related by complex conjugation, and thus the index only behaves distinctly between the first and the second (or third) sheet. In particular, the Cardy limit is highly sensitive to this choice of sheet: while the SYM index is 1 in the Cardy limit on the first sheet, it behaves as 
\ie 
\log {\cI}_{\rm SCI} = { -\pi i (N^2-1) (\sigma+\tau-1)^3\over  27\sigma\tau  } 
+ \log N + {\rm exponentially~small}\,,
\fe 
on the second sheet (similarly on the third sheet) \cite{Cassani:2021fyv}. This drastic difference comes from the fact that the ``black hole saddles'' only dominate the generalized thermal ensemble for the index on the second (and third) sheet. Further taking the large $N$ limit, this is explained by the exchange in dominance between the thermal AdS and supersymmetric black hole solutions reviewed in Section~\ref{sec:10dgravity}. In particular, in our convention, the first branch and the second branch of the black hole solutions in \eqref{eq:appelectriccharge} 
are responsible for the exponential divergence in the Cardy limit on the second and on the third sheets respectively. 
The high-temperature EFT on the second sheet is an $\cN=2$ supersymmetric ${\rm SU}(N)_N$ Chern-Simons theory which turns out to be a trivially gapped theory (trivial TQFT) due to non-abelian confinement in 3d \cite{Cassani:2021fyv}. This enables the powerful prediction in \cite{Cassani:2021fyv} for the Cardy limit of the $\cN=4$ SYM index that is exact in $N$ and up to exponentially small corrections in the angular chemical potentials $\sigma,\tau$.

In the Cardy limit of the surface defect index $\cI_{\cal D}$ (and the normalized expectation value $\la \cD\ra$), the surface defect insertion reduces to a line defect in the high-temperature EFT. Here, since the EFT is a trivial TQFT (whose partition function in the presence of background gauge fields remembers the anomalies of the 4d theory), the resulting line defect is in general a 1d supersymmetric quantum mechanics coupled to  the background gauge fields corresponding to the chemical potentials. Now, from the Lagrangian description of the GW surface defect as a 2d $\cN=(4,4)$ ${\rm U} (1)$ gauge theory coupled to the 4d ${\rm SU}(N)$ gauge group through one 2d bifundamental hypermultiplet, it is easy to see that there are no zero modes from the 2d bifundamental hypermultiplet upon circle reduction on the second sheet, though there are zero modes from the 2d gaugino.\footnote{This can be deduced from the quantum numbers for the 2d hypermultiplet in Table~\ref{table:2dspectrum} and for the 2d vector multiplet in Table~\ref{table:SVcharges}, together with the fact that the fields are subject to the periodicity condition $\chi(t_E+2\pi r_1)=e^{\pi i(r+F)} \chi(t_E)$ along the Euclidean circle of radius $r_1$ and coordinate $t_E$ in order to contribute to the index on the second sheet \cite{Cassani:2021fyv}.} Consequently, this line defect is also trivial to leading order in $N$ and the Cardy limit of the surface defect expectation value $\la \cD\ra$ is completely determined by integrating out the massive modes from KK reduction of the 2d bifundamental hypermultiplets, thus producing a formula in terms of the defect conformal anomalies, up to exponentially small corrections in $\sigma,\tau$ as we have found in \eqref{I2dCardy}.

\subsection{Generalizations to other defect theories and other dimensions} 
\label{sec:genOtherDCFT}

Our study of superconformal surface defects as probes of supersymmetric black holes and the confinement/deconfinement phase transition has immediate generalizations in multiple directions. 

First of all, we have focused on the simplest Gukov-Witten surface defect described by a single probe D3-brane in the AdS dual.\footnote{A further generalization involves the $\cN=(0,8)$ surface defect in \cite{Harvey:2008zz} which corresponds to a probe D7-brane (and more general $(p,q)$ 7-branes) wrapping the entire ${\rm S}^5$ and extending in ${\rm AdS}_5$.} As reviewed in Section~\ref{sec:GWrev}, there is a large family of half-BPS Gukov-Witten surface defects labelled by integer partitions $[k_1,\dots,k_n]$ of $N$. As long as the defect central charge satisfies $b_{2d}\ll N^2$, the probe approximation remains valid.\footnote{It would also be interesting to study the subleading effects by taking into account the back-reaction of these probe branes, as well as the case when $b_{2d}$ is of order $\cO(N^2)$ where the probe approximation is no longer valid.} This is the case for $\sum_{i=1}^{n-1}k_i=M$ of order $\cO(N^0) $ in the large $N$ limit and then $b_{2d}=6N M +\cO(N^{0})$\,. In the holographic dual, this surface defect probe is described by $M$ mutually-BPS D3-branes. Our bulk analysis immediately implies that the surface defect expectation value in the large $N$ limit is given by 
\ie 
\log \, \la {\cD} \ra= 2\pi iMN {  (\sigma+\tau-1)^2\over 9\sigma}\,.
\fe
It would be interesting to verify this directly from the field theory which involves a more complicated linear quiver description for the surface defect (see Section~\ref{sec:GWrev} for a review). 

Secondly, as we explain in detail in Appendix~\ref{sec:superalgebra},
since the defect index only relies on the supercharges $\cQ,\cQ^\dagger$ 
which are contained in an $\cN=(0,2)$ subalgebra of the universal $\cN=1$ superconformal subalgebra $\mf{su}(2,2|1)$,
\ie 
 \mf{sl}(2,\mR) \times  \mf{su}(1,1|1)  \times \mf{u}(1)_Z\subset \mf{su}(2,2|1) \,,
 \label{2dminimalalg}
\fe
where $\mf{u}(1)_Z$ 
is generated by $Z=-J_2-{r\over 2}$ \cite{Bianchi:2019sxz} (see also Appendix~\ref{sec:superalgebra}),
 our setup generalizes straightforwardly
to more general $\cN=1$ SCFTs in $d=4$ with type IIB string theory duals on ${\rm AdS}_5\times {\rm SE}_5$ where ${\rm SE}_5$ denotes a general Sasaki-Einstein manifold that replaces the $S^5$ in the $\cN=4$ case.
In particular,
the supersymmetric black hole solutions reviewed in Section~\ref{sec:10dgravity} only rely on the universal 5d $\cN=1$ supergravity sector and thus remains valid for these more general SCFTs. A detailed analysis of the matching of the bare superconformal index of these more general theories to gravity can be found in \cite{Lanir:2019abx,Amariti:2019mgp,Benini:2020gjh}. 
The relevant surface defects that refines the $\cN=1$ (universal) superconformal index here preserve the $\cN=(0,2)$ superconformal symmetry together with the $\mf{u}(1)_Z$ commutant in \eqref{2dminimalalg}. On the gravity side, these surface defect probes are described by D3-branes wrapping an orbit of the Reeb vector field on ${\rm SE}_5$ and extend in ${\rm AdS}_5$ (see \cite{Koh:2009cj} for an example). On the field theory side, they are generalizations of the Gukov-Witten surface defects, whose explicit descriptions are more complicated in general. 
Nonetheless, we expect our bulk analysis to go through, leading to the following conjecture for the surface defect expectation value in general 4d $\cN=1$ SCFTs, 
in the large $N$ limit,
\ie 
\log\,  \la \cD\ra  = \pi i{  (\sigma+\tau-1)^2\over 27\sigma}b_{\rm 2d}(\cD) \,,
\fe
where $b_{\rm 2d}(\cD)$ is the defect central charge (and $b_{\rm 2d}\ll N^2$) for the $\cN=(0,2)$ supersymmetric surface defect $\cD$. Similarly we expect the same to follow from the generalized Cardy limit for the defect index as described in Section~\ref{sec:CardyGen}. It would be interesting to derive this directly from the field theory. 

Finally, our setup generalizes to defect indices in other spacetime dimensions, where the probes are provided by other types of branes.  
 For example, the 6d $\cN=(2,0)$ SCFTs of the $A_{N-1}$ type contain two types of half-BPS surface defects that correspond either to M2-branes in the bulk M-theory on ${\rm AdS}_7\times S^4$ that sit at a point on the $S^4$ \cite{Berenstein:1998ij,Lunin:2007ab,Berman:2007bv} or to M5-branes wrapping $S^3\subset S^4$\cite{Lunin:2007ab,Chen:2007ir,Chen:2008ds,Berman:2007bv}. Similarly, in the 3d $\cN=8$ ABJM SCFTs which have M-theory duals on  ${\rm AdS}_4\times S^7$, a class of half-BPS surface defects are described by M5 branes wrapping $S^3\subset S^7$ \cite{Lunin:2007ab,Chen:2007tt,Berman:2007bv}. It would be interesting to investigate such brane probes in relation to the confinement/deconfinement phase transitions in these spacetime dimensions and the possibility of probing black holes in the dual AdS spaces.

\section*{Acknowledgments}

We would like to thank Ofer Aharony, Francesco Benini, Simone Giombi, Seok Kim, Zohar Komargodski, Shota Komatsu, Henry Lin, Juan Maldacena, Ohad Mamroud,  Mark Mezei, Sameer Murthy, Leopoldo Pando Zayas, Sahand Seifnashri, Joaquin Turiaci, and Edward Witten for helpful discussions and comments on the draft. Y.C. is supported by a Procter Fellowship from Princeton University. M.H. is supported by the U.S. Department of Energy Grant DE-SC0009988, the Sivian Fund, and the Institute for Advanced Study. The work of YW was supported in part by NSF grant PHY-2210420 and by the Simons Junior Faculty Fellows program.

\appendix

\section{Calibrated D3-brane in the black hole geometry}\label{sec:checkSUSY}

The AdS$_5$ black holes we are considering (viewed as 10d solutions of type IIB supergravity) preserve ${1\over 16}$-th of the supersymmetry. This is also the minimum required supersymmetry for a BPS state in the dual $\mathcal{N}=4$ SYM theory. In the presence of a probe D3-brane, it is a non-trivial requirement that the full bulk + defect system preserve this required amount of supersymmetry due to the presence of the brane. At the level of the probe worldvolume action \eqref{actionbrane}, we have only given the bosonic terms, but this should always be supplemented by terms involving worldvolume fermions determined by supersymmetry. Naively, a purely bosonic embedding into the 10d target space can never be supersymmetric because a supersymmetry variation parametrized by the bulk killing spinor turns on these fermions. Supersymmetry is only possible when the local fermionic kappa symmetry on the worldvolume may be used to gauge away this variation.

Rather than checking that the total action $S_{\textrm{bulk}} + S_{\textrm{brane}}$ in invariant under the combined supergravity and worldvolume supersymmetries, we will instead check the necessary kappa symmetry projection condition \cite{Cederwall:1996pv,Cederwall:1996ri,Bergshoeff:1996tu,Bergshoeff:1997kr}. 
Similar supersymmetric brane configurations were considered in \cite{Aharony:2021zkr}, and we closely follow these conventions in what follows. The primary difference is that the branes considered here extend out to the AdS boundary, and it is nontrivial to show that the supersymmetries preserved by this profile and the supersymmetries preseved by the black hole are compatible.\footnote{See \cite{Drukker:2008wr,Koh:2008kt} for similar studies for D3-branes in empty AdS.}

If $\varepsilon$ is a bulk Killing spinor satisfying
\begin{equation}
\label{eq:Killingspinoreq}
    \hat{D}_M \varepsilon = \left ( \partial_M + \frac14 \omega^{AB}_M \Gamma_{AB} + \frac{i}{16 \cdot 5!}F_{N_1 N_2 \dots N_5}\Gamma^{N_1 N_2 \dots N_5}\Gamma_M \right )\varepsilon = 0 \, ,
\end{equation}
then the kappa symmetry projection condition is
\begin{equation}\label{kappa}
    \Theta \varepsilon = \mp i \varepsilon \, ,
\end{equation}
where
\begin{equation}
\label{eq:Kappamatrix}
    \Theta = \frac{1}{4!}\frac{\epsilon^{\alpha \beta \gamma \delta}}{\sqrt{-h}}\frac{\partial X^M}{\partial \sigma^\alpha}\frac{\partial X^N}{\partial \sigma^\beta}\frac{\partial X^P}{\partial \sigma^\gamma}\frac{\partial X^Q}{\partial \sigma^\delta}\Gamma_{MNPQ} \, ,
\end{equation}
where $\sigma^\alpha$ are the worldvolume coordinates of the brane and $\sqrt{-h}$ is the volume of the pullback metric in Lorentzian signature. We have chosen the Lorentzian problem to avoid the complications of supersymmetry in Euclidean signature. We assume that a kappa symmetric brane configuration may safely be continued to Euclidean where the metric will in general be complex and non-extremal. We will see that the existence of supersymmetric (but finite temperature) black hole solutions implies the existence of kappa symmetric brane configurations on the non-extremal background.

In the main text, we chose the coordinates in (\ref{appmetric10d}) to evaluate the action of the D3-brane. However, as in \cite{Aharony:2021zkr}, for the purpose of studying supersymmetric embedding, it is convenient to introduce orthotoric coordinates \cite{Cassani:2015upa}  $(t, \xi, \eta, \Phi, \Psi)$ on the AdS black hole part of the spacetime, as well as new coordinates $(\rho_s, \theta_s, \varphi_s, \zeta_s, \psi_s)$ on the $S^5$. These coordinates are related to the original ones \eqref{appmetric10d} via:
\begin{equation}
    \begin{aligned}
        r^2 & = r_*^2 + (a+b) \, \widetilde{m} + \frac{(a+b)^2}2 \, \widetilde{m} + \frac{a^2 - b^2}2 \, \widetilde{m}\, \xi \,,  \quad \quad \quad  \theta  = \frac12 \arccos (\eta) \, , \\
\phi &= t + \frac{4 (1-a^2)}{(a^2-b^2) \widetilde{m}} \, (\Psi - \Phi ) \, ,
 \quad\quad\quad  \psi  = t - \frac{4 (1-b^2)}{(a^2 - b^2) \widetilde{m}} \, (\Phi + \Psi) \, , \\ 
\mu_1 &= \sin\rho_s \cos\frac{\theta_s}{2} \, ,  \quad \quad \quad \quad \phi_1 = \frac16 (2 \psi_s + \zeta_s - 3 \varphi_s)\, , \\
\mu_2 &= \sin \rho_s \sin \frac{\theta_s}{2}   \, , \quad \quad \quad \quad \phi_2 = \frac16 (2 \psi_s + \zeta_s + 3 \varphi_s)\, , \\
\mu_3 &= \cos\rho_s \, , \quad \quad \quad \quad \quad \quad \, \, \, \phi_3 = \frac13 (\psi_s - \zeta_s)\, , 
    \end{aligned}
\end{equation}
where the new angular variables take the ranges $0 \leq \rho_s \leq \frac\pi2 $, $0 \leq \theta_s \leq \pi$, $0 \leq \varphi_s < 2\pi$, $0 \leq \zeta_s < 4\pi$, and $0 \leq \psi_s < 6\pi$. We also use a new mass parameter
\begin{equation}
\widetilde{m} = \frac{m}{(a+b) (1+a)(1+b) (1+a + b)} - 1 \;.
\end{equation}
In the extremal limit, $\widetilde{m} = 0$ and we could consider taking this limit before evaluating the supersymmetry condition of the brane. However, our coordinate system has naive singularities for $\widetilde{m}=0$ so this limit would need to be taken with care. Instead, we will find supersymmetric brane configurations for any value of $\widetilde{m}$; this corresponds to the fact that in Euclidean signature the supersymmetric and $T=0$ conditions of the gravity background are decoupled \cite{Cabo-Bizet:2018ehj}. Additionally note that the equal rotation limit of $a=b$ is also singular in these coordinates. We've independently verified the kappa symmetry condition in the metric suitable for the $a=b$ limit, even though we do not focus on this case in our paper. 

The advantage of using orthotoric coordinates is that the vielbeins and gauge potential are mostly diagonal and may be written in terms of a few simple functions of $\cG,\cF, f, \omega$ of $\xi, \eta$:
\begin{equation}\label{eq:tet012}
   \begin{aligned}
        e^0 = f (dt - \omega) \, , \quad 
e^1 =  \, \sqrt{ \frac{\eta - \xi}{ \cF(\xi) \, f}} \, d\xi \, , \quad 
e^2 =  \, \sqrt{ \frac{\cF(\xi) }{(\eta- \xi) \, f}} \, \bigl( d\Phi + \eta\, d\Psi \bigr) \, , \quad  e^3 = -  \, \sqrt{ \frac{\eta - \xi}{\cG(\eta) \, f}} \, d\eta \, ,
    \end{aligned}
\end{equation}
\begin{equation}\label{eq:tet345}
    \begin{aligned}
e^4 =  \, \sqrt{ \frac{\cG(\eta)}{(\eta- \xi) \, f}} \, \bigl( d\Phi + \xi \, d\Psi \bigr) \, , \quad\quad
e^5 = d\rho_s \, , \quad\quad  e^6 = \frac14 \sin(2\rho_s) \, \bigl( d\zeta_s - \cos(\theta_s)\, d\varphi_s \bigr) \, , 
    \end{aligned}
\end{equation}
\begin{equation}\label{eq:tet9}
    \begin{aligned}
 & e^7 = \frac12 \sin(\rho_s) \, d\theta_s \, , \quad
e^8 = \frac12 \sin(\rho_s) \, \sin(\theta_s) \, d\varphi_s \, ,\quad 
e^9 = \frac13 \bigl( d\psi_s + \mathcal{A}  + 2A_{(1)} \bigr) \, , \\
& \mathcal{A} = 3 \tan(\rho_s) \, e^6 - d\zeta_s \, , \\
& A_{(1)} = - \frac{6(1+a)(1+b)(1+\widetilde{m}) f}{ (a^2-b^2) (a-b) \widetilde{m}^2 (\eta - \xi) } \, \biggl[ \left( a+b-(a-b)\eta \right) d\Phi + \left( (a+b) \eta - (a-b) \right) d\Psi \biggr] \,  \\
& + \frac{3(1-f)}2 dt  - \alpha\, dt  \, ,
    \end{aligned}
\end{equation}
where we have used the functions,
\begin{equation}
    \begin{aligned}
       & \cG(\eta) = - \frac{4 \, (1-\eta^2) }{(a^2 - b^2) \, \widetilde{m}} \, \Bigl[ (1-a^2 ) (1+\eta) + (1-b^2)(1-\eta) \Bigr]\,, \\
      &  \cF(\xi) = - \cG(\xi) - 4 \, \frac{1+\widetilde{m}}{\widetilde{m}} \left( \frac{2+a+b + (a-b) \xi}{a-b} \right)^3 \, , \\
      & f =  - \frac{\widetilde{m} \, (a-b) \, (\eta - \xi) }{ (2+a+b)(1+\widetilde{m}) + (a-b)(\eta + \widetilde{m} \xi) } \, ,  \\
     & \omega =  \frac{ \cF''' \, \cG'''}{288} \Bigl[ (\eta + \xi) \, d\Phi + \eta    \, \xi\, d\Psi \Bigr] + \frac2{\widetilde{m}} \, d\Psi  \\
& \quad - \frac{\cF''' + \cG'''}{48 \, (\eta - \xi)^2} \left \{ \left[ \cF + (\eta - \xi) \left( \frac{\cF'}2 - \frac14 \left( \frac{2+a+b}{a-b} + \xi \right)^2 \cF''' \right) \right] (d\Phi + \eta\, d\Psi)  + \cG \, (d\Phi + \xi \, d\Psi) \right \} .
    \end{aligned}
\end{equation}

In addition to the 10d metric in the new coordinates, the equations of motion demand a specific choice of the self-dual RR 5-form $F_{(5)}$ which we have not yet written. This form includes the volume form of AdS plus its dual, but also additional terms due to the rotation of the black hole on the $S^5$. Up to gauge transformation, one may then find a gauge field $C_{(4)}$ such that $F_{(5)} = d C_{(4)}$. The one form $\mathcal{A}$ enters the expression for $C_{(4)}$. The natural choice giving a valid solution for $F_{(5)}$ in our coordinates \eqref{eq:tet9} is singular on the brane worldvolume where $\rho_s = \frac{\pi}{2}$ and $\theta_s = 0$; the circles corresponding to $\mu_2 = \mu_3 = 0$ are shrinking, so similar to \cite{Aharony:2021zkr} we perform a gauge transformation to remove the $\phi_2$ and $\phi_3$ components from $\mathcal{A}$:
\begin{equation}
    \mathcal{A} \rightarrow \tilde{\mathcal{A}} = \mathcal{A} + d \psi_s  \, .
\end{equation}
When evaluated on the worldvolume, $\rho_s = \frac{\pi}{2}$, $\theta_s = 0$, one obtains $\tilde{\mathcal{A}} = 3 d\phi_1$, leading to the quantization condition $\int \tilde{\mathcal{A}} = 6 \pi$ and a nonsingular choice of $C_{(4)}$:
\begin{equation}
\label{eq:C4form}
    C_{(4)} = -4 \beta_{(4)} + \frac19 \tilde{\mathcal{A}} \wedge \left [ \ast_{5}F_{(2)} + \left(\frac12 d\tilde{\mathcal{A}} - F_{(2)} \right ) \wedge e^9  \right] \, ,
\end{equation}
where $\beta_{(4)}$ is any 4-form such that $d\beta_{(4)} = \epsilon_{(5)}$ is the volume form of the 5d asymptotically AdS metric and $F_{(2)} = d A_{(1)}$ is the 5d gauge field strength. The $F_{(5)} = d C_{(4)}$ derived from the above expression solves the 10d equations of motion, but we do not need its explicit form. 
This is the expression we utilized in Section \ref{sec:gravaction} to evaluate the brane action. Only the $\frac19 \tilde{\mathcal{A}} \wedge \ast_{5}F_{(2)} $ term in the expression contains components that cover all the brane worldvolume coordinates.

One may explicitly check that this metric is equivalent to the original 10d one used in the main text. Further, the Killing spinor solving \eqref{eq:Killingspinoreq} associated with this solution in these coordinates was reported in \cite{Aharony:2021zkr}. This led to the discovery of a class of wrapped supersymmetric Euclidean D3-branes in AdS$_5 \times S^5$ which contribute to the dual superconformal index. The bulk calculation here is almost identical, and we only need to check that \eqref{eq:Kappamatrix} preserves the kappa symmetry condition for our D3-branes which extend out to the AdS boundary. In the new coordinates, the static gauge embedding of the brane which wraps the $(\sigma^0, \sigma^1, \sigma^2, \sigma^3)= (t, r, \phi, \phi_1)$ coordinates is given by
\begin{align}
    t &= \sigma^0 \, , \quad \xi = \frac{2(a+b+ab) + (a^2 +b^2)\widetilde{m}  +2(a + b + ab) \widetilde{m} - 2 (\sigma^1)^2}{(a^2-b^2)\widetilde{m}} \, , \quad \eta = -1 \, , \\
    \Phi &= - \frac{(a^2-b^2)\widetilde{m}}{8(1-a^2)(1-b^2)}\left ( (1-a^2)\psi_0 + (-2 +a^2 +b^2)\sigma^0 + (1-b^2) \sigma^2 \right ) \, , \\
    \Psi &= - \frac{(a^2-b^2)\widetilde{m}}{8(1-a^2)(1-b^2)}\left ( (a^2 - b^2)\sigma^0 + (1-a^2)\psi_0 - (1-b^2) \sigma^2 \right ) \, , \\
    \rho_s &= \frac{\pi}{2} \, , \quad \theta_s = 0\, , \quad \varphi_s = \textrm{const} - \sigma^3 \, , \quad \zeta_s = \textrm{const} + \sigma^3 \, , \quad \psi_s = \textrm{const} + \sigma^3 \, .
\end{align}
Here, $\psi_0$ was a constant angle where we placed the brane in the original coordinates, and the further constants are similar and do not affect the analysis.

Using the above vielbeins evaluated on this embedding, some computation shows the determinant of the induced metric, $\sqrt{-\textrm{det}(g_{D3})} = \sqrt{-\textrm{det}(\partial_\alpha X_M \partial_\beta X^M)}$, is
\begin{equation}
\label{eq:induceddet}
    \sqrt{-\textrm{det}(g_{D3})} = \frac{r}{1-a^2} \, .
\end{equation}
We've also done the sanity check that we get the same answer using the coordinate system in (\ref{appmetric10d}). 
Further, the combination of pullbacks in \eqref{eq:Kappamatrix} is
\begin{align}
        \frac{1}{4!}\epsilon^{\alpha \beta \gamma \delta}\partial_\alpha X^M \partial_\beta X^N \partial_\gamma  X^P \partial_\delta X^Q \Gamma_{MNPQ} = \left (\frac{r}{1-a^2} \right )\Gamma_{0129} \, ,
\end{align}
where $\Gamma_{0129}$ is a particular rank-four Gamma matrix in the frame basis. This allows us to conclude
\begin{equation}\label{Thetatwosigns}
    \Theta = \pm \Gamma_{0129} \, .
\end{equation}
The reason why a $\pm$ ambiguity appears is because there are square roots found in obtaining the determinant and pullbacks. The sign also relies on the ordering of the worldvolume coordinates. 
To fully resolve this one would need to more carefully define the orientation of the brane as opposed to an anti-brane. Of course, we know from \cite{Constable:2002xt} that it is D3-brane that is supersymmetric on the background instead of anti D3-brane. Up to these choices which affect the overall sign, we see that there exists a projection matrix with this simple form. As discussed in Appendix. C of \cite{Aharony:2021zkr}, there exists Killing spinor $\varepsilon$ on the black hole background satisfying
\begin{equation}
    \Gamma_{0129}\, \varepsilon = i \varepsilon,
\end{equation}
therefore the kappa symmetry projection condition (\ref{kappa}) holds given the lower sign in (\ref{Thetatwosigns}). 

Further note that we did not have to make any assumptions about the extremality. Note that no divergences appear in the extremal limit $T=0$ and thus $\widetilde{m} = 0$ in the background. 
We did impose the supersymmetric condition (\ref{eq:appelectriccharge}) on the charge $q$, which enters implicitly through the definition of the gauge potential $A$.

\section{Superconformal algebra}
\label{sec:superalgebra}

In this paper, we consider the insertion of a half-BPS GW surface operator in 4d $\mathcal{N}=4$ ${\rm SU}(N)$ SYM, breaking the 4d superconformal algebra to a half-BPS subalgebra,
\ie 
\mf{u}(1)_A \ltimes (\mf{psu}(1,1|2)\times \mf{psu}(1,1|2) 
 ) \ltimes \mf{u}(1)_C \subset \mf{psu}(2,2|4)\,,
 \label{2dsubalgebra}
 \fe  which is a central extension of the usual 2d $\mathcal{N}=(4,4)$ superconformal algebra by $\mf{u}(1)_C$ and we have included the $\mf{u}(1)_A$ that acts as an outer-automorphism on the $\cN=(4,4)$ algebra (and in general is not a symmetry of the surface defect\footnote{In the case of the half-BPS Gukov-Witten surface defect \cite{Gukov:2006jk}, only when $\B=\C=0$ in \eqref{GWconfig} does $\mf{u}(1)_A$ become a symmetry of the defect.}). 

The Cartan generators for the subalgebra on the LHS of \eqref{2dsubalgebra} can be identified with those in 4d as follows,
\begin{equation}\label{4d2dcharge}
\begin{aligned}
   &  L_0 = \frac{E+J_1}{2} , \quad \bar L_0 = \frac{E-J_1}{2}, \quad J_0 = \frac{R_2 -  R_3 }{2}    , \quad \bar{J}_0 = \frac{R_2 + R_3}{2}\,,  \\
   &  J_A = R_1 - R_2 , \quad C = - J_2 - \frac{R_1}{2}\,,
\end{aligned}
\end{equation}
where $L_0$ and $\bar L_0$ are the left- and the right-moving Hamiltonians on the 2d worldvolume, $J_0$ and $\bar J_0$
 are the Cartan generators for the left- and the right-moving $\mf{su}(2)$ R-symmetries, $J_A$ generates the outer-automorphism $\mf{u}(1)_A$ and $C$ generates the central extension $\mf{u}(1)_C$.

We denote the Poincar\'e supercharges of the $\cN=4$ superconformal algebra by $Q^{\pm,i},\, \bar{Q}_{\pm, i}$ and their hermitian conjugate conformal supercharges by $S_{\pm, i}, \bar{S}^{\pm,i}$ respectively \cite{Kinney:2005ej}, where the index $i=1,2,3,4$ labels the fundamental (upper index) and anti-fundamental (lower index) representations of the $\mf{su}
(4)_R$ R-symmetry. We list the the charges of $Q^{\pm,i}$ and $\bar S^{\pm,i}$ under the Cartan generators $R_{1,2,3}$ below,
\begin{equation}
\begin{aligned}
    i=1\,&:\,R_1=1,\,R_2=1,\,R_3=1\,,\\
    i=2\,&:\, R_1 = 1,\,R_2=-1,\,R_3 = -1\,,\\
    i=3\,&:\,R_1 = -1,\,R_2 = 1,\,R_3=-1\,,\\
    i=4\,&:\,R_1 = -1,\,R_2=-1,\,R_3 = 1\,,
\end{aligned}
\end{equation}
and the charges of $\bar Q_{\pm,i}$ and $S_{\pm,i}$ come from flipping the sign of every entry above. The bars on the supercharges keep track of their chiralities and the $\pm$ are the corresponding spinor indices (charge under $\frac{J_1 + J_2}{2}$ for the chiral spinor and $\frac{J_1 - J_2}{2}$ for the anti-chiral spinor). 

We denote the left- and the right-moving Poincar\'e supercharges of the 2d $\cN=(4,4)$ algebra by $G^{\pm a}$ and $\bar G^{\pm \dot a}$ respectively, where $a,\dot a=1,2$ are the doublet indices for the left- and the right-moving $\mf{su}(2)$ R-symmetries and $\pm$ keeps track of the charges under $R_1$ (which together with \eqref{4d2dcharge} determines the corresponding charges under the outer-automorphism $\mf{u}(1)_A$). 
 The explicit identification between the fermionic generators of the 2d $\cN=(4,4)$ algebra and those in the 4d $\cN=4$ algebra is recorded in Table~\ref{tb:embedding}, along with their anticommutation relations.
 \begin{table}[!htb]
\centering
\renewcommand{\arraystretch}{1.28}
\begin{tabular}{||c c||} 
 \hline
 4d $\mathcal{N}=4$ & 2d $\mathcal{N}=(4,4)$  \\ [0.5ex] 
 \hline\hline
 $\{Q^{-,1},S_{-,1}\} = (E-J_{1})-\frac{R_2+R_3}{2}-(J_{2}+\frac{R_1}{2})$  & $\{\bar{G}^{+,1},(\bar{G}^{+,1})^\dagger\} = 2\bar{L}_0-\bar{J}_0+C$  \\ 
 \hline
 $\{Q^{-,2},S_{-,2}\} = (E-J_{1})-\frac{-R_2-R_3}{2}-(J_{2}+\frac{R_1}{2})$ & $\{\bar{G}^{+,2},(\bar{G}^{+,2})^\dagger\} = 2\bar{L}_0+\bar{J}_0+C$  \\
 \hline
 $\{\Bar{Q}_{-,1},\Bar{S}^{-,1}\} = (E-J_{1})+\frac{R_2+R_3}{2}+(J_{2}+\frac{R_1}{2})$ & $\{\bar{G}^{-,2},(\bar{G}^{-,2})^\dagger\} = 2\bar{L}_0+\bar{J}_0-C$ \\
 \hline
  $\{\Bar{Q}_{-,2},\Bar{S}^{-,2}\} = (E-J_{1})+\frac{-R_2-R_3}{2}+(J_{2}+\frac{R_1}{2})$ & $\{\bar{G}^{-,1},(\bar{G}^{-,1})^\dagger\} = 2\bar{L}_0-\bar{J}_0-C$  \\
 \hline
 $\{Q^{+,3},S_{+,3}\} = (E+J_{1})+\frac{-R_2+R_3}{2}+(J_{2}+\frac{R_1}{2})$ & $\{G^{-,1},(G^{-,1})^\dagger\} = 2L_0-J_0-C$   \\
 \hline
$\{Q^{+,4},S_{+,4}\} = (E+J_{1})+\frac{R_2-R_3}{2}+(J_{2}+\frac{R_1}{2})$ & $\{G^{-,2},(G^{-,2})^\dagger\} = 2L_0+J_0-C$\\\hline $\{\Bar{Q}_{+,3},\Bar{S}^{+,3}\} = (E+J_{1})-\frac{-R_2+R_3}{2}-(J_{2}+\frac{R_1}{2})$ & $\{G^{+,2},(G^{+,2})^\dagger\} = 2L_0+J_0+C$ \\\hline $\{\Bar{Q}_{+,4},\Bar{S}^{+,4}\} = (E+J_{1})-\frac{R_2-R_3}{2}-(J_{2}+\frac{R_1}{2})$ & $\{G^{+,1},(G^{+,1})^\dagger\} = 2L_0-J_0+C$ \\
 \hline
\end{tabular}
\caption{Identification between the supercharges in the 2d $\mathcal{N}=(4,4)$ superalgebra and those in the 4d $\mathcal{N}=4$ superconformal algebra together with their anti-commutation relations \cite{Constable:2002xt,Nakayama:2011pa}.}
\label{tb:embedding}
\end{table}

In particular, the ${1\over 16}$-BPS superconformal index  $\cI_{\rm SCI}$ \eqref{N4SCI} is defined by the following complex supercharge and its hermitian conjugate
\ie 
\cQ = Q^{-,1} = \bar{G}^{+,1}\,,\quad \cQ^\dagger= S_{-,1}= (\bar{G}^{+,1})^\dagger\,,
\label{indexSUSY}
\fe 
which are obviously contained the 2d superalgebra \eqref{2dsubalgebra}. The bosonic quantum numbers of $\cQ$ are $E={1\over 2},J_{1,2}=-{1\over 2},R_{1,2,3}=1$ (equivalently $L_0=J_0=0,  2 \bar L_0=\bar J_0=1,C=J_A=0$).

In this paper, we consider the GW surface defect $\cD$ with the continuous parameters restricted to $\B=\C=0$ (see Section~\ref{sec:GWrev} for a review). In this case, the entire subalgebra in \eqref{2dsubalgebra} including the $\mf{u}(1)_A$ outer-automorphism is preserved by $\cD$.
Consequently, $\cD$ can be inserted in the ${1\over 16}$-BPS superconformal index  $\cI_{\rm SCI}$, giving rise to the defect index $\cI_{\cD}$ \eqref{dSCI} that count states in the defect Hilbert space  $\mathcal{H}_{\mathcal{D}}(S^3)$ with the same set of fugacities as for the bare index $\cI_{\rm SCI}$. 

Our consideration above generalizes straightforwardly to general superconformal surface defects in general 4d $\cN=1$ SCFTs (see \cite{Gukov:2014gja} for a review on superconformal surface defects). 
Under the most general setup, they are half-BPS surface defects that preserve a
2d $\mathcal{N}=(0,2)$ half-BPS subalgebra in the $\cN=1$ superconformal algebra,
\ie 
\mf{sl}(2,\mathbb{R})\times\mf{su}(1,1|1)\times  \mf{u}(1)_Z\subset \mf{su}(2,2|1)\,,
\label{2d02subalgebra}
\fe
where $\mf{u}(1)_Z$ is the commutant of the 2d superconformal subalgebra. We denote the supercharges for the $\cN=1$ superconformal algebra by $({\bm Q}^{\pm},{\bm S}_{\pm},\bar{\bm Q}_{\pm},\bar{\bm S}^{\pm})$. The GW surface operators in the $\cN=4$ SYM theory are special cases where we identify the 4d $\cN=1$ superconformal algebra as the subalgebra $\mf{su}(2,2|1)\subset \mf{psu}(2,2|4)$ generated by
\ie
({\bm Q}^{\pm},{\bm S}_{\pm},\bar{\bm Q}_{\pm},\bar{\bm S}^{\pm})=(Q^{\pm,1},S_{\pm,1},\bar{Q}_{\pm,1},\bar{S}^{\pm,1})\,.
\fe 
The 4d supercharges $\cQ,\cQ^\dagger$ in \eqref{indexSUSY} then translate to
\ie 
\cQ= {\bm Q}^-\,,\quad \cQ^\dagger={\bm S}_-\,.
\label{N1indexSUSY}
\fe
in the $\cN=1$ language, which are the familiar supercharges that define the $\cN=1$ superconformal index $\cI_{\rm univ}(p,q)$\cite{Romelsberger:2005eg}, also referred to as the universal superconformal index in SCFTs with enhanced supersymmetry \cite{Aharony:2021zkr}. In a similar way, as we explain below, incorporating the $\cN=(0,2)$ supersymmetric surface defect defines a \textit{universal defect index} for general $\cN=1$ SCFTs.
 
The explicit relation between the Cartan generators in the 2d $\cN=(0,2)$ subalgebra in \eqref{2d02subalgebra} and those in the 4d $\cN=1$ algebra are,\footnote{For a $\cN=(4,4)$ supersymmetric surface defect in $\cN=4$ SYM (e.g. the GW surface defects), the bosonic generators of the $\cN=(0,2)$ subalgebra are given by $\bar {\bm J}_0=\bar J_0-C$ and $Z=C+{J_A+J_0\over 3}$.}
\ie 
 {L}_0 = \frac{E+J_1}{2}\,,\quad \bar{L}_0 = \frac{E-J_1}{2}\,,\quad \bar{\bm J}_0 = J_2+\frac{3}{2}r\,,\quad  Z = -J_2-\frac{r}{2}\,,
\fe 
where $L_0,\bar L_0$ are 2d conformal generators as defined before, $\bar {\bm J}_0$ is the generator for the right-moving $\mf{u}(1)$ R-symmetry, $Z$ generates the $\mf{u}(1)_Z$ factor, and
\ie 
r={R_1+R_2+R_3\over 3}\,
\fe
is the generator for the $\mf{u}(1)_r$ symmetry in the 4d $\cN=1$ superconformal algebra. The 2d $\cN=(0,2)$ supercharges are identified with the four supercharges $({\bm Q}^{-},{\bm S}_{-},\bar{\bm Q}_{-},\bar{\bm S}^{-})$ in the $\cN=1$ superconformal algebra, which include \eqref{N1indexSUSY} as desired.
Therefore, the $\cN=(0,2)$ supersymmetric surface defect defines a refinement of the $\cN=1$ (universal) superconformal index $\cI_{\rm univ}(p,q)$ with the same set of fugacities $p,q$ as long as $\mf{u}(1)_Z$ in \eqref{2d02subalgebra} is preserved.

\section{Details on the 4d-2d index}\label{sec:diffnotations}
\subsection{Generalities on the defect superconformal index}\label{sec:index2d}

The 2d surface defect considered in this paper is half-BPS, which realizes a centrally extended 2d $\cN=(4,4)$ superalgebra embedded in the 4d $\mathcal{N}=4$ superconformal algebra (see Appendix \ref{sec:superalgebra}). 
The superconformal index for the coupled system, namely the defect index \eqref{dSCI} counting states in the defect Hilbert space $\cH_{\cD}(S^3)$, is a supersymmetric partition function on $S^1\times S^3$ with the surface defect wrapping a temporal $T^2$. We will work with the GLSM description of the GW surface defect below.

The contributions from the degrees of freedom localized on the surface defect to $\cI_{\cD}$ can be written in term of a 2d superconformal index defined as
\begin{equation}\label{index2d}
    \mathcal{I}_{\rm 2d} = \textrm{Tr}_{\text{NSNS}} \left[ (-1)^F e^{-\beta \{\cQ,\cQ^\dagger\}} e^{2\pi i \tau_{2d} L_0 } e^{2\pi i z J_0 } e^{2\pi i  \chi J_A } e^{2\pi i  u C } \right],
\end{equation}
where $\tau_{2d}$ is the complex structure of the worldvolume $T^2$ and the trace is taken over the 2d states in the NSNS sector (i.e. satisfying anti-periodic boundary condition for both left- and right-moving modes along the equator $S^1\subset S^3$). The BPS condition for the 2d index  $\{\cQ,\cQ^\dagger\}=2\bar{L}_0-\bar{J}_0+C=0$ coincides with that for the 4d index \eqref{N4SCI} by construction. 
 
In \eqref{index2d}, the 2d bosonic charges that commute with $\cQ,\cQ^\dagger$ are related to the 4d charges by \eqref{4d2dcharge}. Accordingly, by comparing \eqref{index2d} with \eqref{N4SCI}, we deduce the following dictionary between the 4d and 2d chemical potentials (fugacities)
\begin{equation}\label{4d2dcpr}
\begin{aligned}
   & \sigma = \tau_{2d}\,,\quad  \tau = \frac{\tau_{2d}}{2} - u \,, \quad  \Delta_1 = \frac{\tau_{2d}}{2} +2\chi - u\,, \quad \Delta_2 = \frac{\tau_{2d}}{2} + z - 2\chi\,.
\end{aligned}
\end{equation}

The GLSMs that engineer the GW surface defects in ${\rm SU}(N)$ SYM are $\cN=(4,4)$ linear quiver gauge theories with bifundamental hypermultiplets between consecutive quiver nodes. For the surface defect that corresponds to a probe D3 brane in the bulk (which is our focus in the main text), this GLSM is simply $\cN=(4,4)$ SQED with $N$ hypermultiplets. In $\cN=(2,2)$ notation, we denote the $\cN=(4,4)$ hypermultiplet as $(B,\tilde B)$ which packages two $\cN=(2,2)$ chiral multiplets in conjugate representations of the gauge (flavor) symmetry, and write the $\cN=(4,4)$ vector multiplet as $(\mathcal{S},\cV)$ where $\mathcal{S}$ is another chiral multiplet and $\cV$ is a $\cN=(2,2)$ vector multiplet (equivalent to a twisted chiral multiplet).

The charges of the GLSM fields under the Cartan generators of \eqref{2dsubalgebra} can be fixed from the explicit coupling between the 2d GLSM and the 4d SYM fields (see e.g. \cite{Constable:2002xt}). In our conventions, these charges of the 2d hypermultiplets $(B,\tilde B)$ are listed in Table~\ref{table:2dspectrum}, where for later convenience we have included entries in both the 4d and the 2d notations (which are equivalent by \eqref{4d2dcharge}).
\begin{table}[!htb]
\centering
    \renewcommand{\arraystretch}{1.28}
\begin{tabular}{||c | c c c c c c | c c c c c c||} 
 \hline
 Field & $E$ & $J_1$ & $J_2$ & $R_1$ & $R_2$ & $R_3$  & $L_0$ & $\bar{L}_0$ & $J_0$ & $\bar{J}_0$ & $J_A$ & $C$\\ [0.5ex] 
 \hline\hline
 $b$ & 0 & 0 & $\frac{1}{2}$  & $-1$ &  0  &  0 & 0 & 0 & 0 & 0 &  $-1$  & 0\\ 
 \hline
 $b^*$ & 0 & 0 & $-\frac{1}{2}$  & 1 &  0  & 0  & 0 & 0 & 0 & 0 &  1  & 0\\
 \hline
  $\psi^b_{+}$ & $\frac{1}{2}$ & $\frac{1}{2}$ & 0 & 0 &  1 & $-1$
  & $\frac{1}{2}$ & 0 & 1 & 0 &  $-1$ & 0\\
 \hline
  $\bar{\psi}^{b}_{+}$ & $\frac{1}{2}$ & $\frac{1}{2}$ & 0 & 0 &  $-1$ & 1 
  & $\frac{1}{2}$  &  0  & $-1$ &0  &  $1$ & 0  
  \\
 \hline
$\psi^{b}_-$ & $\frac{1}{2}$ & $-\frac{1}{2}$ & 0 & 0 &  $1$ & 1
&  0 & $\frac{1}{2}$ & 0 & 1 &  $-1$ & 0
\\
\hline
$\bar{\psi}^{b}_{-}$ & $\frac{1}{2}$ & $-\frac{1}{2}$ & 0 & 0 &  $-1$ & $-1$  
& 0  & $\frac{1}{2}$  & 0 & $-1$ &  $1$ & 0 \\
 [1ex] 
 \hline
\end{tabular}
\caption{Bosonic charges of fields in the 2d chiral multiplet $B$ (the same for $\tilde{B}$).
}
\label{table:2dspectrum}
\end{table}
Similarly, the charges of the 2d $\cN=(4,4)$ vector multiplet $(\mathcal{S},\cV)$ are listed in Table~\ref{table:SVcharges}, which can be deduced from the superpotential couplings in the GLSM from the charges of the hypermultiplets, together with the $\mf{su}(2)_1\times \mf{su}(2)_2$ R-symmetry and $\mf{su}(2)_F$ flavor symmetry whose Cartan generator is $R_1=J_0+\bar J_0+J_A$, which are all symmetries of the $\cN=(4,4)$ GLSM (see for example \cite{Harvey:2014nha}). In particular, under $\mf{su}(2)_1\times \mf{su}(2)_2\times \mf{su}(2)_F$, the four real scalars in $(s,s^*,\omega,\omega^*)$ transform as $({\bm 2},{\bm 2},{\bm 1})$ and the fermions $(\psi,\bar\psi,\lambda,\bar \lambda)$ transform as $({\bm 2},{\bm 1},{\bm 2})_+\oplus ({\bm 1},{\bm 2},{\bm 2})_-$ depending on the chirality \cite{Harvey:2014nha}.

\begin{table}[!htb]
\begin{center}
\renewcommand{\arraystretch}{1.28}
\begin{tabular}{||c | c c c c c c||} 
 \hline
 Field & $L_0$ & $\bar{L}_0$ & $J_0$ & $\Bar{J}_0$ & $J_A$ & $C$ \\ [0.5ex] 
 \hline\hline
 $s$ & ${1\over 2}$ & ${1\over 2}$ & $-1$ & $-1$ & $2$ &$0$\\ 
 \hline
 $s^*$ & ${1\over 2}$ & ${1\over 2}$ & $1$ & $1$ & $-2$ &$0$\\ 
 \hline
 $\psi^{s}_{+}$ & $1$ & ${1\over 2}$ & $0$ & $-1$ &$2$ &$0$\\
 \hline
 $\bar{\psi}^{s}_{+}$ & $1$ & ${1\over 2}$ & $0$ & $1$ &$-2$ &$0$\\
 \hline
  $\psi^{s}_{-}$ & ${1\over 2}$ & $1$ &$-1$ & $0
$ & $2$ &$0$\\
 \hline
 $\bar{\psi}^{s}_{-}$ & ${1\over 2}$ & $1$ &$1$ & $0$ & $-2$ &$0$\\
 \hline
 \hline
  $\omega$ &${1\over 2}$ & ${1\over 2}$ & $-1$ & $1$ & $0$& $0$ \\
  \hline
  $\omega^*$ &${1\over 2}$ & ${1\over 2}$ & $1$ & $-1$ & $0$& $0$ \\
 \hline
$\lambda_{+}$ & $1$ & ${1\over 2}$ & $0$ & $-1$& $0$ & $0$ \\
\hline
$\lambda_{-}$ & ${1\over 2}$ & $1$ &$-1$ & $0$& $0$ & $0$ \\
\hline
$\bar{\lambda}_{+}$ & $1$ & ${1\over 2}$ & $0$ & $1$& $0$ & $0$ \\
\hline
$\bar{\lambda}_{-}$ & ${1\over 2}$ & $1$ & $1$ & $0$& $0$ & $0$ \\
 [1ex] 
 \hline
\end{tabular}
\caption{Bosonic charges of the fields in the 2d chiral multiplet $\mathcal{S}$ and the 2d vector multiplet $\mathcal{V}$ (equivalent to the twisted chiral multiplet).
}
\label{table:SVcharges}
\end{center}
\end{table}

\subsection{Comparison to previous results}\label{app:nakayama}
The defect superconformal index for a different type of surface defect preserving the same symmetry \eqref{2dsubalgebra} was studied in \cite{Nakayama:2011pa} for ${\rm U}(N)$ SYM. Here the surface defect is described directly by coupling 2d $\cN=(4,4)$ hypermultiplets in the fundamental representation of ${\rm U}(N)$ to the bulk 4d ${\rm U}(N)$ SYM, we will refer to it as the \textit{free-hyper surface defect}. This is physically distinct from the Gukov-Witten surface defects we study in this paper which have well-defined holographic duals as D3-branes in the type IIB string theory. Nonetheless, at the level of the 4d-2d description of the surface defect, the free-hyper surface defect only differs from the GW surface defect of the Levi-type $L=[1,N-1]$ by the 2d $\cN=(4,4)$ $\rm{U}(1)$ gauge interaction on the defect worldvolume. Therefore we naively expect to obtain the same answer in the leading large $N$ limit. Below we connect the computation in \cite{Nakayama:2011pa} to our calculation here and comment on the differences and relations.

In \cite{Nakayama:2011pa}, a different set of linearly independent bosonic charges was chosen  to express the defect superconformal index \eqref{dSCI},
\begin{equation}\label{indexNY}
    \mathcal{I}_{\cD} = \textrm{Tr}_{\mathcal{H}_{\cD}(S^3)} \left[ (-1)^F e^{-\beta \{\cQ,\cQ^\dagger\}} t^{2(\tilde{E} +\tilde{j}_1)} y^{2\tilde{j}_2} v^{ \tilde{R}_2 }  w^{ \tilde{R}_3 } \right]\,.
\end{equation}
These charges in \cite{Nakayama:2011pa} (which we denote with an extra tilde) are related to the ones here by,
\begin{equation}
\begin{aligned}
	& \tilde{E} = E\,, \quad \tilde{j}_1 =  \frac{J_1 + J_2}{2}\,, \quad \tilde{j}_2 = \frac{J_1 - J_2}{2}\,, \quad \tilde{R}_1 = \frac{R_2 + R_1}{2} \,, \quad \tilde{R}_2 =  \frac{R_1 - R_2}{2} \,,\\
	&  \tilde{R}_3 = \frac{R_2 - R_3}{2} \,, \quad \tilde{C} = -J_2 - \frac{R_1}{2}\,.
\end{aligned}
\end{equation}

The relation between the chemical potentials in \cite{Nakayama:2011pa} and our convention follows by comparing \eqref{dSCI} and \eqref{indexNY}, 
\begin{equation}
    e^{2\pi i \sigma} = t^3 y\, , \quad e^{2\pi i \tau} = \frac{t^3}{y}\,, \quad e^{2\pi i \Delta_1} = t^2 v\, ,\quad e^{2\pi i \Delta_2} = \frac{t^2 w}{v} \,.
\end{equation}
By explicitly enumerating the single-particle letters, the contribution to the defect index from the hypermultiplets $(B,\tilde{B})$ is obtained in the form of a Plethystic exponential, which then produces an explicit form of the modified unitary matrix model for the defect index \cite{Nakayama:2011pa}.  

Let us point out a few differences between the 2d index $\cI_{\text{2d}}$ expression \eqref{I2dPl} considered in this paper for the GW surface defect and the Plethystic exponential expression for the free-hyper defect insertion in \cite{Nakayama:2011pa}. Firstly, 
 the bosonic 
 zero-modes of the hypermultiplet fields $b,\tilde{b}$ were manually excluded from the 2d index calculation in \cite{Nakayama:2011pa}, while these fields have been taken into account in \eqref{I2dPl}. 
 At a technical level, including these contributions makes it possible to write 
 the Plethystic exponential in terms of the $\theta_0$-functions that produce the desired modular and elliptic properties (in particular  periodicity upon certain integer shifts of the chemical potentials, see e.g. \eqref{shiftedgrav}). Furthermore we find that including these modes is crucial to reproduce the gravity answer in the holographic dual.\footnote{As mentioned in the Introduction, the GW surface defect is dual to the probe D3 brane with Neumann boundary condition for its AdS$_3$ gauge field. On the other hand, we expect the analog of the free-hyper surface defect in the SU$(N)$ SYM to be dual to the case with Dirichlet boundary condition for the AdS$_3$ gauge field. In particular, there is a residual U$(1)$ global symmetry and decoupled 2d degrees of freedom on this free-hyper defect, unlike the GW surface defect. We thank Ofer Aharony for discussions on this point.} Secondly, the contribution from the 2d $\cN=(4,4)$ vector multiplet $(\mathcal{V},\mathcal{S})$ is missing for the free-hyper defect in \cite{Nakayama:2011pa} by construction. Relatedly, unlike the GW surface defect, the ${\rm U}(1)$ subgroup of the ${\rm U}(N)$ symmetry of the free hypermultiplets is gauged not 
 by the 2d vector multiplet but instead by the 4d vector multiplet in the bulk SYM \cite{Nakayama:2011pa}, and consequently the corresponding 2d contribution $\cI_{\rm 2d}$ to the defect index (compared to \eqref{I2dPl}) has not been projected to the ${\rm U}(1)$-singlet sector.

\subsection{2d contribution to the defect index from the elliptic genus}\label{appC3}
 
Here we briefly review the calculation of the elliptic genus of 2d $\cN=(4,4)$ SQED with a pair of hypermultiplets $(B,\tilde{B})$  with charge $\pm 1$. Upon spectral flow, this will produce the 2d contribution $\cI_{\rm 2d}$ to the defect index for the Gukov-Witten surface defect that corresponds to a probe D3-brane in the bulk.

The elliptic genus is an important invariant defined for general 2d $\cN=(0,1)$ supersymmetric QFTs (SQFTs) that is preserved under supersymmetry preserving deformations such as supersymmetric RG flows. It is typically defined in the RR sector of the 2d SQFT where the left- and right-moving fermions have periodic boundary conditions along the spatial ${\rm S}^1$. Here, for the $\cN=(4,4)$ GLSM that describes the surface defect, the fully refined elliptic genus takes the following form,
\begin{equation}
    \mathcal{I}_{\text{RR}} \equiv  \text{Tr}_{\text{RR}}\,(-1)^{F}e^{2\pi i \tau_{\text{2d}}L_0}e^{-2\pi i \bar{\tau}_{\text{2d}}\bar{L}_0}e^{2\pi i z_{\text{R}}J_0}e^{2\pi i\chi J_A}e^{2\pi i uC}\,,
\end{equation} 
where the bosonic charges and chemical potentials are as defined in Appendix~\ref{sec:index2d}.

General discussions on how to compute the elliptic genus for GLSMs can be found in \cite{Gadde:2013dda,Benini:2013xpa,Benini:2013nda}. The main point is that $\cI_{\rm RR}$ can be written as an integral over the holonomies of the gauge fields on the spacetime torus $T^2$ with complex structure $\tau_{\rm 2d}=\sigma$ where the integrand is given by a product of one-loop determinants that capture the fluctuations. This general formula was derived in \cite{Benini:2013xpa,Benini:2013nda} by supersymmetric localization (see also the review \cite{Benini:2016qnm}). Here, specializing to the case of $\cN=(4,4)$ SQED with $N$ unit-charge hypermultiplets, we have
\begin{align}
\label{eq:ungaugedindex}
    \mathcal{I}_{\text{RR}} = &\int d {\bm u}\, \frac{2\pi(\eta(\tau_{\text{2d}}))^3}{\theta_1(\tau_{\text{2d}}|-z_{\text{R}})}\frac{\theta_1(\tau_{\text{2d}}|-2\chi)}{\theta_1(\tau_{\text{2d}}|z_{\text{R}}-2\chi)}\nonumber\\
    &\prod_{i=1}^N \frac{\theta_1 \left(\tau_{\text{2d}}|-{\bm u}+u_i+ \chi-z_{\text{R}}\right) \theta_1 \left(\tau_{\text{2d}}|{\bm u} - u_i+ \chi-z_{\text{R}} \right)}{ \theta_1 \left(\tau_{\text{2d}}|-{\bm u}+u_i +\chi  \right) \theta_1 \left(\tau_{\text{2d}}|{\bm u}-u_i +\chi  \right) }\,,
\end{align}
where $\eta(\tau_{\text{2d}})$ denotes the Dedekind eta function and $\theta_1(\tau_{\rm 2d}|z)$ is a Jacobi theta function. 
The two terms on the first line of \eqref{eq:ungaugedindex} come from $S$ and $\mathcal{V}$ respectively in the abelian $\cN=(4,4)$ vector multiplet. The terms in the second line account for the contribution from the hypermuliplets $(B,\tilde B)$.
The integral is over the complex variable ${\bm u}$ which encodes the independent holonomies of the abelian gauge field on $T^2$,
\begin{equation}
    {\bm u} = \oint_{S^1} dx^0\, A_0 -\sigma \oint_{S^1}dx^1\,A_1\,,\quad {\bm u}\in {\mC/(\mZ+\sigma\mZ)}\,,
\end{equation}
where the periodic identifications on $\bm u$ come from the large gauge transformations.

The complex holonomy integral (e.g. \eqref{eq:ungaugedindex}) can be further reduced to a contour integral on the torus of abelian holonomies and the latter can be evaluated by the Jeffery-Kirwan residue prescription \cite{Benini:2013nda,Benini:2013xpa,Benini:2016qnm}.
In the case of $\cN=(4,4)$ SQED, this amounts to collecting the residues at the $N$ simple poles  in \eqref{eq:ungaugedindex} associated to the positive-charged field at ${\bm u}= u_i-\chi$. The final answer is
\begin{equation}
    \mathcal{I}_{\text{RR}} = \sum_{i=1}^N\prod_{i\neq j} \frac{\theta_1(\tau_{\text{2d}}|u_{ij}+z_{\text{R}}-2\chi)\theta_1(\tau_{\text{2d}}|u_{ij}-z_{\text{R}})}{\theta_1(\tau_{\text{2d}}|u_{ij})\theta_1(\tau_{\text{2d}}|u_{ij}-2\chi)}\,,
\end{equation}
where $u_{ij}\equiv u_i - u_j$. This expression already appeared in \cite{Benini:2013xpa} and is here written in terms of the chemical potentials in our conventions.

Instead of doing the integral over holonomy $\bm u$ exactly using the Jeffery-Kirwan residue prescription, we can also fix the value of $\bm u$, perform the large $N$ analysis of the integrand in  (\ref{eq:ungaugedindex}) and eventually integrate over $\bm u$. We found that both in the empty AdS phase and the black hole phase, $\bm u = 0$ dominates the integral. This justifies our choice in the gravity analysis of Section \ref{sec:grav} to focus on the simplest D3-brane configuration with no worldvolume gauge field turned on.

\section{Direct evaluation of the defect  at the black hole saddle}\label{sec:direct}

In this appendix, we present a different route to derive the large $N$ limit answer (\ref{I2dfinalfield}) of the 2d defect expectation value at the black hole saddle. The difference from the main text is that here we don't perform modular transformations on the following expression
\begin{equation}\label{I2ddirect}
    \mathcal{I}_{2d} = \sum_{\hat{i}^*,\hat{j}^* } \, \, \prod_{(\hat{i},\hat{j} ) \neq (\hat{i}^*,\hat{j}^* )}\frac{\theta_0\left(\sigma| u_{\hat{i},\hat{j} } - u_{\hat{i}^*,\hat{j}^* } + \frac{2\sigma}{3} - \frac{\tau}{3} + \frac{1}{3}  \right)\theta_0(\sigma| u_{\hat{i}^*,\hat{j}^* } - u_{\hat{i},\hat{j} }  + \frac{\sigma}{3} + \frac{\tau}{3} + \frac{2}{3} )}{\theta_0(\sigma|u_{\hat{i},\hat{j} } - u_{\hat{i}^*,\hat{j}^* } + \frac{\sigma}{3} - \frac{2\tau}{3} - \frac{1}{3}) \theta_0(\sigma| u_{\hat{i}^*,\hat{j}^* } - u_{\hat{i},\hat{j} }  )}\,,
\end{equation}
but rather attempt to evaluate it directly. Of course, in the end we will land on the same final answer. The goal of this appendix is to provide a sanity check, while also providing a different method that could potentially be useful for other purposes.

\begin{figure}[t!]
    \begin{center}
   \includegraphics[scale=.3]{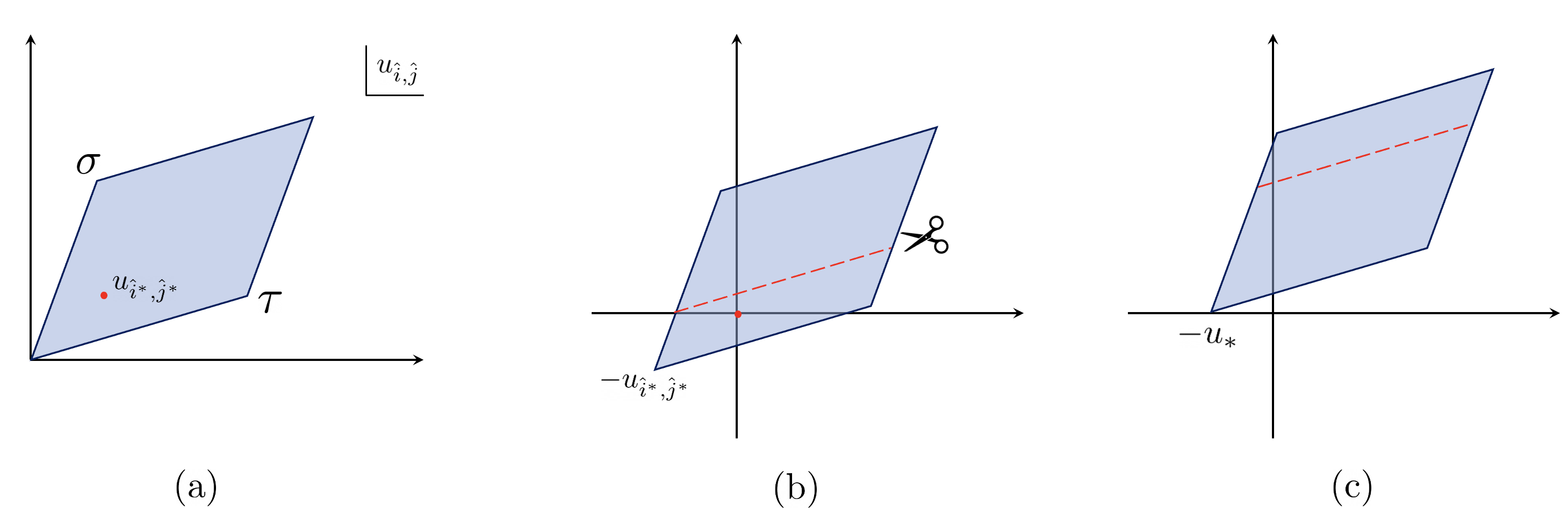}
    \end{center}
    \caption{Black hole saddle of the ${\rm SU}(N)$ matrix integral $(\sigma \neq \tau)$. In (a), we reproduce Figure~\ref{fig:parallelmain} of the parallelogram saddle. The red point marked inside the parallelogram corresponds to $u_{\hat{i}^*,\hat{j}^*}$, which labels each individual term of the sum in \eqref{I2ddirect}. In (b), we draw a new shaded parallelogram, which represents the difference $u_{\hat{i},\hat{j}}-u_{\hat{i}^*,\hat{j}^*}$. Note that the red point now sits at the origin. In order to absorb the $\tilde{u}_*$ parameter in \eqref{uijdiff} into $y$, we cut the parallelogram along the dashed red line in $\tau$ direction, the left end of which lies on the real axis. In (c), we shift the parallelogram below the dashed red line by $\sigma$ and glue two parallelograms together again. Since the expression \eqref{I2ddirect} is invariant under $u_{\hat{i},\hat{j}}\rightarrow u_{\hat{i},\hat{j}}+\sigma$, the move from (b) to (c) will not change the final result. What's remaining is an off-set on the real axis, labelled by $u_*>0$, which cannot be simply removed.}
    \label{fig:parallelogram}
\end{figure}

As in Section \ref{sec:fieldBH}, we parametrize the saddle point of the holonomies in the continuum limit as $u_{\hat{i},\hat{j}} = x\tau+y\sigma$. For fixed $u_{\hat{i}^*,\hat{j}^*} = x^*\tau+y^*\sigma \in \mathbb{C}$, we can always rewrite $u_{\hat{i}^*,\hat{j}^*} = \tilde{u}_*\sigma+u_*$, such that $\tilde{u}_*,u_* \in \mathbb{R}$. Explicitly, we have 
\begin{equation}
   \tilde{u}_* = x^*\frac{\text{Im}(\tau)}{\text{Im}(\sigma)}+y^*,\quad u_* = x^* \left( \tau-\sigma \frac{\text{Im}(\tau)}{\text{Im}(\sigma)} \right).
\end{equation}
The difference between $u_{\hat{i},\hat{j}}$ and $u_{\hat{i}^*,\hat{j}^*}$ in the large-$N$ limit can be expressed in terms of $x, y, \tilde{u}_* , u_*$
\begin{equation}\label{uijdiff}
    u_{\hat{i},\hat{j}}-u_{\hat{i}^*,\hat{j}^*} = x\tau + (y- \tilde{u}_* )\sigma -u_*.
\end{equation}
Note that the summand in \eqref{I2ddirect} is invariant under each individual shift $u_{\hat{i},\hat{j}} \rightarrow u_{\hat{i},\hat{j}} +\sigma$. Therefore, the constant $\tilde{u}_*$ can always be absorbed into $y$ via this kind of shift. We demonstrate this shift explicitly in Figure~\ref{fig:parallelogram} by cutting the parallelogram along the $\tau$ direction into two parts, shifting the lower part by $\sigma$ and gluing them back together.

On the other hand, the dependence on $u_*$ in (\ref{uijdiff}) cannot be simply removed. The situation here is different from the 4D index $\mathcal{I}_{4d}$, where the argument of (\ref{4dSYMUmm}) is both periodic in the $\tau$ and $\sigma$ direction. Here we only have periodicity in the $\sigma$ direction. Therefore we will carry $u_*$ through the calculation. Since in the index we need to sum over all possible choices of $u_{\hat{i}^*,\hat{j}^*}$, the large $N$ value of the 2d index should be given by extremizing over all possible values of $u_*$.
 With all this in mind, the explicit expression for the defect expectation value in the large $N$ limit reads
\begin{equation}\label{I2dBHapp}
\begin{aligned}
    \frac{- \log   \,\langle \mathcal{D}\rangle  }{N} 
   = &  - \underset{u_*}{\textrm{ext}} \int_0^1 dx\int_0^1 dy\, \left[ \log \theta_0\left(\sigma|\left(x- \frac{1}{3} \right)\tau + \left(y+ \frac{2}{3} \right) \sigma  -u_* + \frac{1}{3}  \right) \right. \\
   & \left. + \log \theta_0 \left(\sigma|\left(\frac{1}{3}-x \right) \tau + \left(\frac{1}{3}- y \right) \sigma + u_*   + \frac{2}{3} \right)- \log \theta_0 \left(\sigma|-x\tau - y \sigma + u_* \right)  \right. \\
   & \left.  - \log \theta_0 \left(\sigma|\left( x -\frac{2}{3} \right) \tau + \left(y + \frac{1}{3}\right) \sigma - u_*   - \frac{1}{3} \right) \right].
\end{aligned}
\end{equation}

One way to evaluate the integral is to rewrite the $\theta_0$-functions into Plethystic exponential form, similar to what we did in the main text. As discussed in Section \ref{sec:fieldBH}, we need to make sure that the Plethystic exponential expression is convergent. 
Similar to what we did there, we need to use the periodicity of the $\theta_0$ function listed in Appendix \ref{sec:convention} to shift the arguments of each $\theta_0$ function into the convergent region. To simplify the discussion, here we again assume $\tau \in \mathbb{R} + i 0^+$. This assumption makes the calculation simpler but is not necessary. The final expression we derive is an analytic function of $\tau$ so it can be simply continued to general values of $\tau$. The relevant transformations for the four $\theta_0$'s in (\ref{I2dBHapp}) are performed as follows 
\begin{equation}\label{firsttheta}
    \begin{aligned}
 & \quad \quad \theta_0\left(\sigma \left|\left(x-\frac{1}{3} \right)\tau+ \left(y+\frac{2}{3} \right)\sigma+\frac{1}{3}\right.-u_*\right) \\
 & = \begin{cases}
         \theta_0\left(\sigma \left|(x-\frac{1}{3})\tau+(y+\frac{2}{3})\sigma+\frac{1}{3}\right.-u_*\right), & y \in [0,\frac{1}{3}) ,\\
             e^{-2\pi i\left[\left(x-\frac{1}{3}\right)\tau+\left(y-\frac{1}{3}\right)\sigma-\frac{1}{6}-u_*\right]} \times\theta_0\left(\sigma \left|(x-\frac{1}{3})\tau+(y-\frac{1}{3})\sigma+\frac{1}{3}\right.-u_*\right) , & y \in [\frac{1}{3},1],
    \end{cases}
    \end{aligned}
\end{equation}
\begin{equation}
    \begin{aligned}
 &  \quad\quad \theta_0\left(\sigma\left| \left(-x+\frac{1}{3} \right)\tau + \left(-y+\frac{1}{3} \right)\sigma+\frac{2}{3}\right.+u_*\right) \\
 & = \begin{cases}
         \theta_0\left(\sigma\left| (-x+\frac{1}{3})\tau +(-y+\frac{1}{3})\sigma+\frac{2}{3}\right.+u_*\right) , & y\in [0,\frac{1}{3}) , \\
             e^{2\pi i \left[\left(-x+\frac{1}{3}\right)\tau+(-y+\frac{1}{3})\sigma+\frac{1}{6}+u_*\right]}
          \times\theta_0\left(\sigma\left| (-x+\frac{1}{3})\tau +(-y+\frac{4}{3})\sigma+\frac{2}{3}\right.+u_*\right), & y\in[\frac{1}{3},1] ,
    \end{cases}
    \end{aligned}
\end{equation}
\begin{equation}
     \theta_0(\sigma|-x\tau-y\sigma+u_*) = e^{2\pi i(-x\tau-y\sigma+ \frac{1}{2}+u_*)}\theta_0(\sigma|-x\tau+(1-y)\sigma+u_*),
\end{equation}
\begin{equation}
    \begin{aligned}
 &   \quad\quad  \theta_0\left(\sigma\left|\left(x-\frac{2}{3}\right)\tau+ \left(y+\frac{1}{3} \right)\sigma-\frac{1}{3}\right.-u_*\right) \\
 & = \begin{cases}
        \theta_0\left(\sigma\left|(x-\frac{2}{3})\tau+(y+\frac{1}{3})\sigma-\frac{1}{3}\right.-u_*\right), & y\in [0,\frac{2}{3}), \\
            e^{-2\pi i\left[\left(x-\frac{2}{3}\right)\tau+\left(y-\frac{2}{3}\right)\sigma +\frac{1}{6}-u_*\right]}
            \times\theta_0\left(\sigma\left|(x-\frac{2}{3})\tau+(y-\frac{2}{3})\sigma-\frac{1}{3}\right.-u_*\right) , & y\in[\frac{2}{3},1] .
    \end{cases}
    \end{aligned}
\end{equation}
Let's now focus on the first piece (\ref{firsttheta}), which appears as the first term in the integral (\ref{I2dBHapp}), and explain how to do the integral.  Using (\ref{firsttheta}), we have
\begin{equation}
\begin{aligned}
  &    - \int_0^1 dx\,\int_0^1 dy \, \log \theta_0\left(\sigma \left|\left(x-\frac{1}{3} \right)\tau+ \left(y+\frac{2}{3} \right)\sigma+\frac{1}{3}\right.-u_*\right) \\
  = &  - \int_0^1 dx\,\int_0^{\frac{1}{3}} dy \, \log \theta_0\left(\sigma \left|\left(x-\frac{1}{3} \right)\tau+ \left(y+\frac{2}{3} \right)\sigma+\frac{1}{3}-u_* \right. \right) \\
  & + \int_0^1 dx\int_{\frac{1}{3}}^{1} dy  \left[ 2\pi i  \left( \left(x-\frac{1}{3}\right)\tau+\left(y-\frac{1}{3}\right)\sigma-\frac{1}{6}-u_* \right) \right.\\&\left.- \log \theta_0\left(\sigma \left|\left(x-\frac{1}{3} \right)\tau+ \left(y- \frac{1}{3} \right)\sigma+\frac{1}{3}-u_* \right. \right) \right] \\
 = & \frac{2\pi i ( 2\sigma + \tau - 6u_* - 1)}{9} - \int_0^1 dx\,\int_0^{1} dy \, \log \theta_0\left(\sigma \left|\left(x-\frac{1}{3} \right)\tau+ y\sigma+\frac{1}{3}-u_* \right. \right).
\end{aligned}
\end{equation}
In the last line, we've done the elementary integral, and combined the two integrals involving $\theta_0$ into one. After this procedure, we can now express the $\theta_0$ into Plethystic exponential form and perform the integral. Using (\ref{eqn:plethystic}) - (\ref{eqn:theta0}), we have
\begin{equation}
    \begin{aligned}
        &  -\int_0^1 dx\, \int_0^1 dy\, \log \theta_0 \left(\sigma \left|\left(x-\frac{1}{3}\right)\tau+y\sigma+\frac{1}{3}\right.-u_*\right)\\
        & = \int_0^1 dx\, \int_0^1 dy\,  \sum_{n=1}^\infty \frac{1}{n (1-e^{2\pi i n \sigma}) } \left[ e^{2\pi i n \left(\left(x-\frac{1}{3}\right)\tau +  y\sigma+\frac{1}{3}-u_* \right) } +  e^{2\pi i n \left(-\left(x-\frac{1}{3}\right)\tau + ( 1- y)\sigma-\frac{1}{3}+u_* \right)}  \right]\\
        &= \sum_{n=1}^{\infty}\frac{1}{(2\pi i n\sigma)(2\pi i n\tau)n(1-p^n)}\left[ (p^n-1)(q^n-1)e^{2\pi i n\left(-\frac{1}{3}\tau+\frac{1}{3}-u_*\right)}+(p^n-1)(1-q^{-n})e^{2\pi i n\left(\frac{1}{3}\tau-\frac{1}{3}+u_*\right)}\right] \\
        &= -\frac{1}{4\pi^2\sigma \tau}\left[-\text{Li}_3(e^{2\pi i\left(\frac{2}{3}\tau+\frac{1}{3}-u_*\right)})+\text{Li}_3(e^{2\pi i\left(-\frac{1}{3}\tau+\frac{1}{3}-u_*\right)})-
        \text{Li}_3(e^{2\pi i\left(\frac{1}{3}\tau-\frac{1}{3}+u_*\right)})+\text{Li}_3(e^{2\pi i\left(-\frac{2}{3}\tau-\frac{1}{3}+u_*\right)})\right] \\
        & = -\frac{\pi i }{3\sigma \tau}\left[   - B_3 \left( \frac{2}{3} \tau + \frac{1}{3} - u_*\right) +B_3 \left(- \frac{1}{3} \tau + \frac{1}{3} - u_*\right)  \right].
    \end{aligned}
\end{equation}
where we used the definition of the $\textrm{Li}_3$ function
\begin{equation}
    \textrm{Li}_3 (z) = \sum_{n=1}^\infty \frac{z^n}{n^3} 
 \end{equation}
 and its identity\footnote{Here we assume that the real parts of $\tau,u_*$ are small in order to apply the identity (\ref{Liidentity}). When they are not, one should apply identities where the arguments of $B_3$ are shifted correspondingly. We've checked that we would arrive at the same final answer (\ref{I2dfinalfield}) given $0<\textrm{Re}(\tau)<1$. 
 }
\begin{equation}\label{Liidentity}
   \textrm{Li}_3(e^{2\pi i z}) -  \textrm{Li}_3(e^{-2\pi iz}) = \frac{4\pi^3 i}{3}B_3 (z) ,\quad 0<\textrm{Re}(z)<1,
\end{equation}
 where $B_3$ denotes the third Bernoulli polynomial, $B_3 (x) = x^3 - \frac{3}{2} x^2 + \frac{x}{2}. $
 Therefore, for the first integral in (\ref{I2dBHapp}), we have
\begin{equation}\label{res1}
\begin{aligned}
  &   - \int_0^1 dx\,\int_0^1 dy \, \log \theta_0\left(\sigma \left|\left(x-\frac{1}{3} \right)\tau+ \left(y+\frac{2}{3} \right)\sigma+\frac{1}{3}\right.-u_*\right)  \\
   = &   \frac{2\pi i ( 2\sigma + \tau - 6u_* - 1)}{9} -\frac{\pi i }{3\sigma \tau}\left[   - B_3 \left( \frac{2}{3} \tau + \frac{1}{3} - u_*\right) +B_3 \left(- \frac{1}{3} \tau + \frac{1}{3} - u_*\right)  \right]
\end{aligned}
\end{equation}
For the other three integrals in (\ref{I2dBHapp}), we follow exactly the same procedure, which leads to
\begin{equation}\label{res2}
\begin{aligned}
  &   - \int_0^1 dx\,\int_0^1 dy \,   \log \theta_0 \left(\sigma \left|\left(\frac{1}{3}-x \right) \tau + \left(\frac{1}{3}- y \right) \sigma + u_*   + \frac{2}{3} \right. \right)\\
  = & \frac{2\pi i ( 2\sigma + \tau - 6u_* - 1)}{9} 
   - \frac{\pi i }{3\sigma \tau} \left[ - B_3 \left( \frac{\tau}{3} + \frac{2}{3} + u_*\right) + B_3 \left( 
-\frac{2\tau}{3} + \frac{2}{3} + u_* \right) \right]
\end{aligned}
\end{equation}
\begin{equation}\label{res3}
 \begin{aligned}
     & \int_0^1 dx\,\int_0^1 dy \, \log \theta_0 (\sigma| -x\tau - y\sigma +u_*) =     \pi i (1 + 2u_* - \sigma - \tau) - \frac{\pi i }{3\sigma \tau} \left( B_3 (u_*) + B_3 (\tau - u_*)  \right) 
 \end{aligned}
\end{equation}
\begin{equation}\label{res4}
    \begin{aligned}
        &  \int_0^1 dx\,\int_0^1 dy \,\log \theta_0 \left(\sigma \left|\left( x -\frac{2}{3} \right) \tau + \left(y + \frac{1}{3} \right) \sigma - u_*   - \frac{1}{3} \right.\right) \\
        = & \frac{\pi i (-\sigma + \tau + 6u_* - 1)}{9}   - \frac{\pi i }{3\sigma \tau} \left[ - B_3 \left( - \frac{\tau}{3} + \frac{1}{3} + u_* \right) + B_3 \left( \frac{2\tau}{3} + \frac{1}{3} + u_* \right) \right].
    \end{aligned}
\end{equation}
Adding the results of (\ref{res1}) - (\ref{res4}) together, we have
\begin{equation}
\begin{aligned}
       \frac{ -\log  \,\langle \mathcal{D}\rangle  }{N} 
   & = \underset{u_*}{\textrm{ext}} \left[\frac{2 \pi i (2 - \sigma - 2\tau )}{9}  - \frac{ 2\pi i (\tau - 1)^2}{9\sigma }  + i \pi  \frac{u_*^2}{\sigma \tau}\right] .
\end{aligned}
\end{equation}
The extremum of $u_*$ sits at $u_*=0$. Note that since $\textrm{Re}\left(i/(\sigma\tau)\right) \propto  \textrm{Im}\left(\sigma \tau \right)>0$, $u_* = 0$ minimizes the real part of the action and therefore gives the dominant contribution. Setting $u_*=0$, we recover the final answer
\begin{equation}
     \frac{ -\log  \,\langle \mathcal{D}\rangle  }{N}  = -\frac{2\pi i (\sigma + \tau - 1)^2 }{9\sigma}.
\end{equation}
This agrees with our final answer (\ref{I2dfinalfield}) in the main text.

\section{Field theory analysis in the $\sigma = \tau$ limit}\label{sec:puzzle}

In the main part of this paper, we focused on the range of chemical potentials (\ref{tausigmarelation}), where the 2d index formula (\ref{I2dequal}) has a convergent Plethystic exponential expression, assuming that the ${\rm SU}(N)$ holonomies take real values. Within this range, we find agreement between the bulk and boundary analysis.

However, the range of chemical potential (\ref{tausigmarelation}) excludes the special limit $\sigma = \tau, \Delta_1 = \Delta_2 = \Delta_3 = (2\sigma - 1)/3$. If we naively apply the index formula (\ref{I2dequal}) outside the range (\ref{tausigmarelation}) and in particular to the $\sigma = \tau$ limit, we encounter subtleties that we are not sure how to resolve. Intriguingly, instead of the more intricate analysis for the black hole saddle, it is for the thermal AdS saddle that we appear to find an surprising answer. In this appendix we present the puzzling computation and discuss some possible interpretations. We leave a satisfying resolution to future study.

For convenience, here we write down the expression for the 2d index that we will evaluate (after specifying to the case $\sigma = \tau$):
\begin{equation}\label{I2dsigma=tau}
    \mathcal{I}_{2d} 
    =  \sum_{i=1}^N \prod_{j\neq i} \frac{\theta_0\left(\sigma|u_{ji} + \frac{\sigma}{3} + \frac{1}{3}  \right)\theta_0(\sigma|u_{ij} + \frac{2\sigma}{3}  + \frac{2}{3} )}{\theta_0(\sigma|u_{ji} - \frac{\sigma}{3}  - \frac{1}{3}) \theta_0(\sigma|u_{ij})}.
\end{equation}

\subsection{Thermal AdS saddle}
As explained in section \ref{sec:field theory}, the thermal AdS saddle corresponds to the eigenvalue distribution $u_j = \frac{j}{N}, j =0,...,N-1$. For the purpose of evaluating the large $N$ limit of $\mathcal{I}_{2d}$, we'd like to write the $\theta_0$ functions into Plethystic exponential form. Since all $u_i$ are real, in order to make $\theta_0$-functions have convergent Plethystic exponentials, we need to shift the argument of $\theta_0(\sigma|u_{ji}-\frac{\sigma}{3}-\frac{1}{3})$ using identity (\ref{theta0shift})
\begin{equation}
\label{eq:ThermalAdSshift}
    \theta_0\left(\sigma\left|u_{ji}-\frac{\sigma}{3}\right.-\frac{1}{3}\right)=-e^{2\pi i \left(u_{ji}-\frac{1}{3}\sigma-\frac{1}{3}\right)}\theta_0 \left(\sigma\left|u_{ji}+\frac{2\sigma}{3}\right.-\frac{1}{3}\right).
\end{equation}
After this shift on the argument, all $\theta_0$-functions appear in \eqref{I2dsigma=tau} have valid Plethystic exponentials. Based on the same argument made in section \ref{sec:fieldAdS}, Plethystic exponentials evaluated at the thermal AdS saddle always contribute $e^{\mathcal{O}(1)}$ factors to the 2d index. Therefore, the large-$N$ contributions purely come from the prefactor in \eqref{eq:ThermalAdSshift}. In this case, we find that 
\begin{equation}\label{nonzeroads}
    -\log  \,\langle \mathcal{D}\rangle  = -2\pi i N\left(\frac{1}{3}\sigma+\frac{1}{3}\right)+\mathcal{O}(1),
\end{equation}
which has a nonvanishing $O(N)$ term. It seems to suggest a nonzero brane action, despite being on the thermal AdS background. 
We are not aware of a possible interpretation of this nonzero action.

One possible resolution to this is that one should first evaluate the large $N$ limit of the field theory expression within the range (\ref{tausigmarelation}), then analytically continue the large $N$ action to the $\sigma = \tau$ limit. This would lead to an answer that is simply zero which agrees with what we would expect from gravity. However, it is unclear to us what the physical motivation of such a prescription is. Of course, another possibility is that (\ref{nonzeroads}) does have a gravity interpretation.

\subsection{Black hole saddle}
When $\sigma = \tau$,
the supersymmetric black hole solution is associated to the $(1,0)$ saddle of the matrix integral\cite{Cabo-Bizet:2019eaf}, which can also be derived from the roots of the Bethe-ansatz equation\cite{Benini:2018mlo,Benini:2018ywd}. 
The eigenvalue $u_j$ is distributed uniformly on the line in $\sigma$ direction, i.e. $u_j = \frac{j}{N}\sigma, j=0,...,N-1$. This distribution of $u_j$ is different from directly the distribution obtained by taking the collinear limit $\sigma\rightarrow \tau$ of the parallelogram ansatz used in Section \ref{sec:field theory}. In \cite{Choi:2021rxi}, the relation between the solutions of two ansatz was studied, and it was shown that the large-$N$ contribution from both approaches agree. In this part of discussion, we use the saddle-point distribution from BAEs for simplicity, and in the end we will see that the result is consistent with the collinear limit of the analytic continuation of the result in Section \ref{sec:field theory}. 
Note that the expression \eqref{I2dsigma=tau} is invariant under $u_i \rightarrow u_i+\sigma$ shift, all terms in the summation over $i$ should be identical. In the large-$N$ limit, we rewrite the 2d index into the following integral expression
\begin{equation}
    \begin{aligned}
        -\frac{\log  \,\langle \mathcal{D}\rangle  }{N} =& \int_0^1 dx \left[-\log \theta_0\left(\sigma\left|x\sigma+\frac{\sigma}{3}+\frac{1}{3}\right.\right)-\log \theta_0\left(\sigma\left|-x\sigma+\frac{2\sigma}{3}+\frac{2}{3}\right.\right)\right.\\&\left. +\log \theta_0\left(\sigma\left|x\sigma-\frac{\sigma}{3}-\frac{1}{3}\right.\right)+\log \theta_0\left(\sigma\left|-x\sigma\right.\right)\right].
    \end{aligned}
\end{equation}

In order to evaluate this integral, we could adopt the same procedure as we use in Appendix \ref{sec:direct} to write the $\theta_0$-functions into convergent Plethystic exponentials by shifting the arguments suitably. After performing the shifts, we decompose the integral into the following parts
\begin{equation}
    \begin{aligned}
        -\frac{\log  \,\langle \mathcal{D}\rangle  }{N} = & -\int_0^1 dx \log \frac{\theta_0\left(\sigma|x\sigma+\frac{1}{3}\right)\theta_0\left(\sigma| (1-x)\sigma +\frac{2}{3}\right)}{\theta_0\left(\sigma|x\sigma-\frac{1}{3}\right)\theta_0\left(\sigma|(1-x)\sigma\right)}\\
        &+2\pi i\int_{\frac{2}{3}}^1 dx\,\left(x\sigma-\frac{2\sigma}{3}-\frac{1}{6}\right)-2\pi i\int_{\frac{2}{3}}^1 dx\, \left(-x\sigma+\frac{2\sigma}{3}+\frac{1}{6}\right)\\
        &+2\pi i \int_{0}^{\frac{1}{3}}dx\,\left(x\sigma-\frac{\sigma}{3}+\frac{1}{6}\right)+2\pi i\int_0^1 dx\, \left(-x\sigma+\frac{1}{2}\right).
    \end{aligned}
\end{equation}
The first line admits the convergent Plethystic exponential expression, so we can integrate out $x$ and write the result in terms of dilogarithm functions
\begin{equation}
    \begin{aligned}
        -\int_0^1 dx \log \frac{\theta_0\left(\sigma|x\sigma+\frac{1}{3}\right)\theta_0\left(\sigma| (1-x)\sigma +\frac{2}{3}\right)}{\theta_0\left(\sigma|x\sigma-\frac{1}{3}\right)\theta_0\left(\sigma|(1-x)\sigma\right)} = &\frac{i}{2\pi \sigma}\left[\text{Li}_2(e^{\frac{4\pi i}{3}})+\text{Li}_2(e^{-\frac{4\pi i}{3}})-2\text{Li}_2(1)\right],
    \end{aligned}
\end{equation}
where we use the series expansion of dilogarithm $\text{Li}_2(z)=\sum_{n=1}^\infty \frac{z^n}{n^2}$. For dilogarithm function, we have the following useful identity
\begin{equation}
    \textrm{Li}_2 (e^{2\pi i z}) + \textrm{Li}_2 (e^{-2\pi i z}) = 2\pi^2 B_2 (z), \quad 0\leq \textrm{Re}(z) < 1,
\end{equation}
where $B_2 (z)$ is the second Bernoulli polynomial.
By applying this identity to the dilogarithms in the $-\log \mathcal{I}_{\text{2d}}$ expression, we obtain the following leading term in the large-$N$ expansion
\begin{equation}
\label{eq:I2dfinalsigmaeqtau}
    -\frac{\log \,\langle \mathcal{D}\rangle  }{N} = -\frac{2\pi i}{9\sigma}-\frac{8\pi i}{9}(\sigma-1)=-\frac{2\pi i(2\sigma-1)^2}{9\sigma},
\end{equation}
which agrees with the result derived in \eqref{I2dfinalfield} provided that $\sigma$ is equal to $\tau$. Therefore, in this case, the final expression \eqref{eq:I2dfinalsigmaeqtau} is still consistent with the D3-brane effective action \eqref{SD3final} evaluated in the black hole background.

\section{Defects on shifted chemical potential saddle points}\label{sec:shifted}

As we have mentioned in the main text, an important conceptual point in the recent discussion of the ${1\over 16}$-BPS index and its gravity dual is the importance of saddle points with shifted chemical potentials. These are gravity solutions with the same fugacities but different chemical potentials at infinity, and the sum over them is important in reproducing the expected periodicity of the field theory index. This is very clear if we wrote the field theory expression in terms of a trace. After transforming to a microcanonical ensemble in the charges, this periodicity property implies the correct discreteness of the charge spectrum (in the original superconformal index or in the case with a surface operator insertion). The existence of these extra saddles is thus a field theory motivated constraint on the ``sum over geometries'' as expected of a gravitational path integral. In cases such as the near extremal or near BPS limits of the gravity path integral, the sum over these shifted saddles is not suppressed, and they have important consequences for the low energy physics \cite{Heydeman:2020hhw,Boruch:2022tno,Iliesiu:2021are,Iliesiu:2022kny}.

Given the chemical potentials $\sigma, \tau, \Delta_a, a= 1,2,3$ on the field theory side, one can generally consider infinitely many gravity solutions with chemical potentials $\sigma_g, \, \tau_g, \, \Delta_{a,g} , \, a = 1,2,3$ shifted relative to the field theory ones by independent integers (while satisfying the constraint $\sigma_g +  \tau_g - \Delta_{1,g} - \Delta_{2,g} - \Delta_{3,g} = \pm1$). Here we used subscript $g$ to highlight that the chemical potentials showing up in gravity can be distinct from the ones at the boundary. Even though these gravity solutions formally exist, their stability and the convergence of the sum over them are generally unclear. More essential to our consideration here, the full characterization of these solutions on the field theory side, in particular, in the language of large $N$ saddle points for the unitary matrix integral  \cite{Choi:2021rxi} (see e.g. \cite{Aharony:2021zkr} for a different approach using Thermal Bethe Ansatz), is still under active research and have not been fully understood to our best knowledge. 

It is natural to wonder about how the defect/D3-brane behaves on these other saddles. We will restrict our attention to the case where
\begin{equation}\label{shiftedgrav}
    \sigma_g = \sigma  + 3n_1, \quad \tau_g = \tau + 3n_2, \quad \Delta_{1,g} = \Delta_{2,g} =\Delta_{3,g} = \frac{\sigma + \tau - 1}{3} + n_1 + n_2 , \quad n_1, n_2 \in \mathbb{Z},
\end{equation}
so that the black hole solution still has the same R-charges and can be described by the same metric as in Section \ref{sec:grav}.\footnote{Notice that not all of these gravity solutions are stable. For the case with $\sigma=\tau$, it was shown in \cite{Aharony:2021zkr} that the action can be lower by adding wrapped Euclidean D3 branes unless $\sigma_g = \tau_g$. We thank Ofer Aharony and Ohad Mamroud for comments on this point.} The only difference is that the chemical potentials in Section \ref{sec:grav} are now replaced by the ones with subscript $g$ and the rest of the analysis goes through. Therefore, following (\ref{SD3final}), the action of the D3-brane on the shifted gravity saddles, after regularization, is
\begin{equation}\label{Sshifted}
      I_{D3, \textrm{E, finite}} = -2\pi i N \frac{(\sigma_g + \tau_g - 1)^2}{9 \sigma_g} =-2\pi i N \frac{(\sigma  + \tau + 3(n_1 + n_2)- 1)^2}{9 (\sigma + 3n_1 )} .
\end{equation}

On the field theory side, \cite{Choi:2021rxi} proposed a family of large $N$ saddle points that naturally correspond to (\ref{shiftedgrav}). These saddle points are characterized by a uniform parallelogram distribution of holonomies $u_i$, similar to what we considered in (\ref{sec:fieldBH}), but with two edges of the parallelogram given by $\sigma + 3n_1, \tau + 3n_2$. It was claimed in \cite{Choi:2021rxi} that this new parallelogram saddle would satisfy the large $N$ saddle point equations of the unitary matrix model, with the shifted chemical potentials, namely $\sigma + 3n_1, \tau + 3n_2, \Delta_a + n_1 + n_2 , a= 1,2,3$ satisfying the same inequalities that the chemical potentials should satisfy for the original saddle point.\footnote{See (2.41) of \cite{Choi:2021rxi} for the detailed form of the inequalitites. We thank Seok Kim for correspondence on these points.} Therefore, we have
\begin{equation}\label{I2dexpshift}
    \langle \mathcal{D}\rangle  
    =  \sum_{i=1}^N \prod_{j\neq i} \frac{\theta_0\left(\sigma|u_{ji} + \frac{2(\sigma + 3n_1)}{3} - \frac{\tau+3n_2}{3} + \frac{1}{3}  \right)\theta_0\left(\sigma|u_{ij} + \frac{\sigma + 3n_1}{3} + \frac{\tau + 3n_2}{3} + \frac{2}{3} \right)}{\theta_0\left(\sigma|u_{ji} + \frac{\sigma + 3n_1}{3} - \frac{2(\tau + 3n_2)}{3} - \frac{1}{3}\right)\theta_0(\sigma|u_{ij})}, 
\end{equation}
where $u_{i} = u_{\hat{i},\hat{j}} = \frac{\hat{i}}{\sqrt{N}} (\sigma + 3n_1) + \frac{\hat{j}}{\sqrt{N}} (\tau + 3n_2) $. One might be puzzled by the explicit shift of the chemical potentials in (\ref{I2dexpshift}), but since we have $\theta_0 (\sigma| x + 1) = \theta_0 (\sigma| x ) $, the shifts in (\ref{I2dexpshift}) should not affect the final answer, given that $n_1, n_2 \in \mathbb{Z}$. This is of course true, but the benefit of (\ref{I2dexpshift}) is to highlight that the shifted chemical potentials $\sigma + 3n_1 , ...$ should be within the same range as the chemical potentials for which the discussion in Section \ref{sec:fieldBH} would be valid.  The calculation then follows exactly the discussion in Section \ref{sec:fieldBH}, with the chemical potentials replaced by the shifted ones. From (\ref{I2dfinalfield}), this yields a final answer
\begin{equation}
    - \log  \,\langle \mathcal{D}\rangle   = -2\pi i N \frac{(\sigma  + \tau + 3(n_1 + n_2)- 1)^2}{9 (\sigma + 3n_1 )} ,
\end{equation}
which agrees with (\ref{Sshifted}). 

However, we want to highlight an unsatisfying feature of this discussion. Generally, after we fix the values of $\sigma , \tau , \Delta_a$ which lie within the inequalities in \cite{Choi:2021rxi}, the shifted chemical potentials $\sigma + 3n_1 , \tau + 3n_2 , \Delta_a + n_1 + n_2$ will not satisfy the same inequalities and therefore we cannot apply the parallelogram ansatz. To make it more clear, in the above field theory calculation, we are not considering the shifted saddles for the \emph{same} chemical potentials, but rather shifted saddles for correspondingly \emph{different} chemical potentials. This appears to be different from the gravity picture where we can consider both the original saddle and the shifted ones for a fixed set of chemical potential. Therefore, the results in this appendix shouldn't be taken as a match between field theory side and the gravity side, but rather as only preliminary evidence. Nonetheless, we would like to emphasize that the difficulty of relating the two is already present at the level of the original BPS index, and not particularly related to the added defect. It would also be useful to understand whether one can apply different techniques, such as the Thermodynamic Bethe Ansatz \cite{Aharony:2021zkr} to compute the 4d-2d index.

\section{Conventions and relations for $\theta$-functions}\label{sec:convention}
The Plethystic exponential, which usually appears in the index calculation, is defined as follows:
\begin{equation}
\label{eqn:plethystic}
    \textrm{PE}\left( \frac{a}{1-q}\right) \equiv \exp\left[ \sum_{n=1}^\infty \frac{1}{n} \frac{a^n}{1-q^n} \right].
\end{equation}
Provided the above definition, we can associate the Plethystic exponential to the $q$-Pochhammer symbol
\begin{equation}\label{Pochhammer}
    (a;q)_\infty \equiv \prod_{k=0}^\infty (1-aq^k) = \frac{1}{  \textrm{PE}\left( \frac{a}{1-q}\right)  } 
\end{equation}
We introduce one useful special function $\theta_0$ in terms of the product of two $q$-Pochhammer symbols
\begin{equation}
\label{eqn:theta0}
    \theta_0 (\tau|u) = \left(e^{2\pi i u}; e^{2\pi i \tau}\right)_\infty  \left(e^{2\pi i (\tau - u)}; e^{2\pi i \tau} \right)_\infty  
\end{equation}
By massaging the two Plethystic exponentials in \eqref{eqn:theta0} together, we obtain a series expansion for $\log{\theta_0}$
\begin{equation}
    \log \theta_0(\tau | u) = -i\sum_{n=1}^{\infty}\frac{\cos{n\pi (2u-\tau)}}{n \sin{n\pi \tau}}.
\end{equation}

Note that the above Plethystic exponential expression does not always have a convergent exponent. The formula \eqref{eqn:plethystic} is only valid when $|q|,|a|<1$. Therefore, the $\theta_0$-function can be rewritten as a Plethystic exponential only if $\text{Im}(\tau)>0,\, 0<\text{Im}(u)<\text{Im}(\tau)$. The same criterion also works for the series expansion of $\log{\theta_0}$.  

The $\theta_0$-function has the following quasi-periodicity property:
\begin{equation}\label{theta0shift}
    \theta_0(\tau|u)=\theta_0(\tau|u+1)=-e^{2\pi i u}\theta_0(\tau|u+\tau).
\end{equation}
Moreover, the $\theta_0$-function possesses a few modular properties under the transformations $\tau \rightarrow \tau+1$ ($T$) and $u \rightarrow \frac{u}{\tau},\,\tau\rightarrow -\frac{1}{\tau}$ ($S$)
\begin{align}
    \theta_0(\tau|u) =& \theta_0(\tau+1|u), \\
    \theta_0(\tau|u) =& e^{\pi i B_{2,2}(u|\tau,-1)}\theta_0 \left(-\frac{1}{\tau} \left|\frac{u}{\tau} \right. \right),
\end{align}
where $B_{2,2}(u|\tau,-1)$ is the multiple Bernoulli polynomial defined as follows:
\begin{equation}
    B_{2,2}(u|\tau,-1) =-\frac{B_2(u+1)}{\tau}+B_1(u+1)-\frac{\tau}{6}.
\end{equation}
The $B_1(x)=x-\frac{1}{2}$ and $B_2(x) = x^2-x+\frac{1}{6} $ are ordinary Bernoulli polynomials.

Besides the $\theta_0$-function, we introduce the ordinary Jacobi Theta function $\theta_1$, which can be obtained from the $\theta_0$-function via the following identity:
\begin{equation}
    \theta_1(\tau|u) =ie^{-i\pi(u-\tau /4)}(q;q) \theta_0(\tau | u).
\end{equation}
The Jacobi theta function $\theta_1$ has the following double quasi-periodicity:
\begin{equation}
    \theta_1(\tau | u+a+b\tau) = (-1)^{a+b}e^{-2\pi ib u-i\pi b^2\tau}\theta_1(\tau|u).
\end{equation}

Here we summarize the zeroes and poles of $\theta_0(\tau|u)$, which is useful in performing the contour integral over the worldvolume gauge field. $\theta_0(\sigma|u)$ has no poles, while it has zeroes at $u = a +b\tau$. The derivative of $\theta_0(\tau|u)$ near $u=0$ is
\begin{equation}
    \theta_0(\tau|u)'|_{u = 0} = -2\pi i (q;q)^2.
\end{equation}
According to the residue theorem, we have the following contour integral
\begin{equation}
    \oint_{u=0} \frac{du}{\theta_0(\tau|u)} = -\frac{1}{(q;q)^2}.
\end{equation}
Another important identity of $\theta_0(\tau|u)$ is
\begin{equation}
    \theta_0(\tau|u) = -e^{2\pi i u}\theta_0(\tau|-u).
\end{equation}
This property can be derived directly from the definition of $\theta_0(\tau|u)$ in \eqref{eqn:theta0}.

\bibliographystyle{./JHEP}
\bibliography{references.bib}

\end{document}